\documentclass[longauth,pdfencoding=auto,psdextra]{aa}

\usepackage{graphicx}
\usepackage{natbib}
\usepackage{scalerel}
\usepackage{comment}
\usepackage{multirow}

\usepackage{glossaries}
\newacronym{dr1}{DR1}{Data Release 1}
\newacronym{q1}{Q1}{Quick Data Release 1}
\newacronym{dlnn}{DLNN}{Deep Learning Neural Network}
\newacronym{GBDT}{GBDT}{Gradient-Boosted Decision-Trees}
\newacronym{ml}{ML}{machine learning}
\newacronym{ms}{SFMS}{star-forming main-sequence}
\newacronym{sfg}{SFGs}{star-forming galaxies}
\newacronym{sfr}{SFR}{star-formation rates}
\newacronym{snr}{S/N}{signal-to-noise ratio}
\newacronym{sed}{SED}{spectral energy distribution}
\newacronym{pp}{PPs}{physical parameters}
\newacronym{pdf}{PDFs}{posterior distribution functions}
\newacronym{nn}{NNs}{nearest-neighbours}
\newacronym{UV}{UV}{ultraviolet}
\newacronym{NIR}{NIR}{Near-Infrared}
\newacronym{EWS}{EWS}{Euclid Wide Survey}
\newacronym{EDF}{EDFs}{Euclid Deep Fields}
\newacronym{EAF}{EAFs}{Euclid Auxiliary Fields}
\newacronym{sSFR}{sSFR}{specific \gls{sfr}}

\newcommand{\nnpz}{\texttt{NNPZ}}
\newcommand{\phosphoros}{\texttt{Phosphoros}}

\newcommand{\Mstarwun}{\logten(M_{\ast}/M_\odot)}
\newcommand{\sfrwun}{\logten\,(\mathrm{SFR}/M_\odot\,\mathrm{yr}^{-1})}

\usepackage[table]{xcolor}

\usepackage{orcidlink}
\newcommand{\orcid}[1]{\,\orcidlink{#1}}
\usepackage{txfonts}
\usepackage{hyperref}
\hypersetup{
    colorlinks=true,
    linkcolor=blue,
    filecolor=magenta,      
    urlcolor=blue,
    citecolor=blue
}
\makeatletter
\renewcommand*\aa@pageof{, page \thepage{} of \pageref*{LastPage}}
\makeatother

\usepackage[utf8]{inputenc}

\renewcommand{\linenumbers}[0]{}

\usepackage{euclid}

\begin{document}
\title{Euclid Data Release 1 (DR1)}
\subtitle{The PHZ processing function for the physical properties of galaxies.}

\author{Euclid Collaboration: A.~Enia\orcid{0000-0002-0200-2857}\thanks{\email{andrea.enia@cea.fr}}\inst{\ref{aff1},\ref{aff2}}
\and W.~G.~Hartley\orcid{0000-0002-4563-5101}\inst{\ref{aff3}}
\and M.~Bolzonella\orcid{0000-0003-3278-4607}\inst{\ref{aff2}}
\and F.~Dubath\orcid{0000-0002-6533-2810}\inst{\ref{aff3}}
\and S.~Paltani\orcid{0000-0002-8108-9179}\inst{\ref{aff3}}
\and L.~Pozzetti\orcid{0000-0001-7085-0412}\inst{\ref{aff2}}
\and C.~Saulder\orcid{0000-0002-0408-5633}\inst{\ref{aff4},\ref{aff5}}
\and M.~Tucci\inst{\ref{aff3}}
\and N.~Morisset\inst{\ref{aff3}}
\and D.~Scott\orcid{0000-0002-6878-9840}\inst{\ref{aff6}}
\and M.~Salvato\orcid{0000-0001-7116-9303}\inst{\ref{aff4}}
\and H.~Hildebrandt\orcid{0000-0002-9814-3338}\inst{\ref{aff7}}
\and J.~G.~Sorce\orcid{0000-0002-2307-2432}\inst{\ref{aff8},\ref{aff9}}
\and G.~Zamorani\orcid{0000-0002-2318-301X}\inst{\ref{aff2}}
\and B.~Altieri\orcid{0000-0003-3936-0284}\inst{\ref{aff10}}
\and A.~Amara\inst{\ref{aff11}}
\and S.~Andreon\orcid{0000-0002-2041-8784}\inst{\ref{aff12}}
\and N.~Auricchio\orcid{0000-0003-4444-8651}\inst{\ref{aff2}}
\and C.~Baccigalupi\orcid{0000-0002-8211-1630}\inst{\ref{aff13},\ref{aff14},\ref{aff15},\ref{aff16}}
\and M.~Baldi\orcid{0000-0003-4145-1943}\inst{\ref{aff17},\ref{aff2},\ref{aff18}}
\and A.~Balestra\orcid{0000-0002-6967-261X}\inst{\ref{aff19}}
\and S.~Bardelli\orcid{0000-0002-8900-0298}\inst{\ref{aff2}}
\and P.~Battaglia\orcid{0000-0002-7337-5909}\inst{\ref{aff2}}
\and A.~N.~Belikov\inst{\ref{aff20},\ref{aff21}}
\and A.~Biviano\orcid{0000-0002-0857-0732}\inst{\ref{aff14},\ref{aff13}}
\and E.~Branchini\orcid{0000-0002-0808-6908}\inst{\ref{aff22},\ref{aff23},\ref{aff12}}
\and M.~Brescia\orcid{0000-0001-9506-5680}\inst{\ref{aff24},\ref{aff25}}
\and S.~Camera\orcid{0000-0003-3399-3574}\inst{\ref{aff26},\ref{aff27},\ref{aff28}}
\and V.~Capobianco\orcid{0000-0002-3309-7692}\inst{\ref{aff28}}
\and C.~Carbone\orcid{0000-0003-0125-3563}\inst{\ref{aff29}}
\and J.~Carretero\orcid{0000-0002-3130-0204}\inst{\ref{aff30},\ref{aff31}}
\and F.~J.~Castander\orcid{0000-0001-7316-4573}\inst{\ref{aff32},\ref{aff33}}
\and M.~Castellano\orcid{0000-0001-9875-8263}\inst{\ref{aff34}}
\and G.~Castignani\orcid{0000-0001-6831-0687}\inst{\ref{aff2}}
\and S.~Cavuoti\orcid{0000-0002-3787-4196}\inst{\ref{aff25},\ref{aff35}}
\and K.~C.~Chambers\orcid{0000-0001-6965-7789}\inst{\ref{aff36}}
\and A.~Cimatti\inst{\ref{aff37}}
\and C.~Colodro-Conde\inst{\ref{aff38}}
\and G.~Congedo\orcid{0000-0003-2508-0046}\inst{\ref{aff39}}
\and C.~J.~Conselice\orcid{0000-0003-1949-7638}\inst{\ref{aff40}}
\and L.~Conversi\orcid{0000-0002-6710-8476}\inst{\ref{aff41},\ref{aff10}}
\and Y.~Copin\orcid{0000-0002-5317-7518}\inst{\ref{aff42}}
\and A.~Costille\inst{\ref{aff43}}
\and F.~Courbin\orcid{0000-0003-0758-6510}\inst{\ref{aff44},\ref{aff45},\ref{aff46}}
\and H.~M.~Courtois\orcid{0000-0003-0509-1776}\inst{\ref{aff47}}
\and M.~Cropper\orcid{0000-0003-4571-9468}\inst{\ref{aff48}}
\and H.~Degaudenzi\orcid{0000-0002-5887-6799}\inst{\ref{aff3}}
\and G.~De~Lucia\orcid{0000-0002-6220-9104}\inst{\ref{aff14}}
\and H.~Dole\orcid{0000-0002-9767-3839}\inst{\ref{aff9}}
\and X.~Dupac\inst{\ref{aff10}}
\and S.~Escoffier\orcid{0000-0002-2847-7498}\inst{\ref{aff49}}
\and M.~Farina\orcid{0000-0002-3089-7846}\inst{\ref{aff50}}
\and R.~Farinelli\inst{\ref{aff2}}
\and S.~Farrens\orcid{0000-0002-9594-9387}\inst{\ref{aff51}}
\and F.~Faustini\orcid{0000-0001-6274-5145}\inst{\ref{aff34}}
\and S.~Ferriol\inst{\ref{aff42}}
\and F.~Finelli\orcid{0000-0002-6694-3269}\inst{\ref{aff2},\ref{aff52}}
\and P.~Fosalba\orcid{0000-0002-1510-5214}\inst{\ref{aff33},\ref{aff32}}
\and N.~Fourmanoit\orcid{0009-0005-6816-6925}\inst{\ref{aff49}}
\and M.~Frailis\orcid{0000-0002-7400-2135}\inst{\ref{aff14}}
\and E.~Franceschi\orcid{0000-0002-0585-6591}\inst{\ref{aff2}}
\and M.~Fumana\orcid{0000-0001-6787-5950}\inst{\ref{aff29}}
\and L.~Gabarra\orcid{0000-0002-8486-8856}\inst{\ref{aff53}}
\and S.~Galeotta\orcid{0000-0002-3748-5115}\inst{\ref{aff14}}
\and K.~George\orcid{0000-0002-1734-8455}\inst{\ref{aff54}}
\and B.~Gillis\orcid{0000-0002-4478-1270}\inst{\ref{aff39}}
\and C.~Giocoli\orcid{0000-0002-9590-7961}\inst{\ref{aff2},\ref{aff18}}
\and J.~Gracia-Carpio\orcid{0000-0003-4689-3134}\inst{\ref{aff4}}
\and A.~Grazian\orcid{0000-0002-5688-0663}\inst{\ref{aff19}}
\and F.~Grupp\inst{\ref{aff4},\ref{aff5}}
\and S.~Gwyn\orcid{0000-0001-8221-8406}\inst{\ref{aff55}}
\and S.~V.~H.~Haugan\orcid{0000-0001-9648-7260}\inst{\ref{aff56}}
\and S.~Hemmati\orcid{0000-0003-2226-5395}\inst{\ref{aff57}}
\and H.~Hoekstra\orcid{0000-0002-0641-3231}\inst{\ref{aff58}}
\and W.~Holmes\orcid{0009-0007-8554-4646}\inst{\ref{aff59}}
\and F.~Hormuth\inst{\ref{aff60}}
\and A.~Hornstrup\orcid{0000-0002-3363-0936}\inst{\ref{aff61},\ref{aff62}}
\and M.~Huertas-Company\orcid{0000-0002-1416-8483}\inst{\ref{aff38},\ref{aff63},\ref{aff64}}
\and K.~Jahnke\orcid{0000-0003-3804-2137}\inst{\ref{aff65}}
\and M.~Jhabvala\inst{\ref{aff66}}
\and B.~Joachimi\orcid{0000-0001-7494-1303}\inst{\ref{aff67}}
\and S.~Kermiche\orcid{0000-0002-0302-5735}\inst{\ref{aff49}}
\and A.~Kiessling\orcid{0000-0002-2590-1273}\inst{\ref{aff59}}
\and B.~Kubik\orcid{0009-0006-5823-4880}\inst{\ref{aff42}}
\and M.~K\"ummel\orcid{0000-0003-2791-2117}\inst{\ref{aff5}}
\and M.~Kunz\orcid{0000-0002-3052-7394}\inst{\ref{aff68}}
\and H.~Kurki-Suonio\orcid{0000-0002-4618-3063}\inst{\ref{aff69},\ref{aff70}}
\and A.~M.~C.~Le~Brun\orcid{0000-0002-0936-4594}\inst{\ref{aff71}}
\and V.~Le~Brun\orcid{0000-0002-5027-1939}\inst{\ref{aff43}}
\and D.~Le~Mignant\orcid{0000-0002-5339-5515}\inst{\ref{aff43}}
\and P.~Liebing\orcid{0009-0005-8043-7150}\inst{\ref{aff48}}
\and S.~Ligori\orcid{0000-0003-4172-4606}\inst{\ref{aff28}}
\and P.~B.~Lilje\orcid{0000-0003-4324-7794}\inst{\ref{aff56}}
\and V.~Lindholm\orcid{0000-0003-2317-5471}\inst{\ref{aff69},\ref{aff70}}
\and I.~Lloro\orcid{0000-0001-5966-1434}\inst{\ref{aff72}}
\and M.~Magliocchetti\orcid{0000-0001-9158-4838}\inst{\ref{aff50}}
\and G.~Mainetti\orcid{0000-0003-2384-2377}\inst{\ref{aff73}}
\and D.~Maino\inst{\ref{aff74},\ref{aff29},\ref{aff75}}
\and O.~Mansutti\orcid{0000-0001-5758-4658}\inst{\ref{aff14}}
\and S.~Marcin\inst{\ref{aff76}}
\and O.~Marggraf\orcid{0000-0001-7242-3852}\inst{\ref{aff77}}
\and M.~Martinelli\orcid{0000-0002-6943-7732}\inst{\ref{aff34},\ref{aff78}}
\and N.~Martinet\orcid{0000-0003-2786-7790}\inst{\ref{aff43}}
\and F.~Marulli\orcid{0000-0002-8850-0303}\inst{\ref{aff79},\ref{aff2},\ref{aff18}}
\and R.~J.~Massey\orcid{0000-0002-6085-3780}\inst{\ref{aff80}}
\and D.~C.~Masters\orcid{0000-0001-5382-6138}\inst{\ref{aff81}}
\and N.~Mauri\orcid{0000-0001-8196-1548}\inst{\ref{aff37},\ref{aff18}}
\and E.~Medinaceli\orcid{0000-0002-4040-7783}\inst{\ref{aff2}}
\and S.~Mei\orcid{0000-0002-2849-559X}\inst{\ref{aff82},\ref{aff83}}
\and M.~Melchior\inst{\ref{aff84}}
\and M.~Meneghetti\orcid{0000-0003-1225-7084}\inst{\ref{aff2},\ref{aff18}}
\and E.~Merlin\orcid{0000-0001-6870-8900}\inst{\ref{aff19}}
\and G.~Meylan\orcid{0000-0001-6503-0209}\inst{\ref{aff85}}
\and J.~J.~Mohr\orcid{0000-0002-6875-2087}\inst{\ref{aff54}}
\and P.~Monaco\orcid{0000-0003-2083-7564}\inst{\ref{aff86},\ref{aff14},\ref{aff15},\ref{aff13}}
\and A.~Mora\orcid{0000-0002-1922-8529}\inst{\ref{aff87}}
\and M.~Moresco\orcid{0000-0002-7616-7136}\inst{\ref{aff79},\ref{aff2}}
\and C.~Moretti\orcid{0000-0003-3314-8936}\inst{\ref{aff14},\ref{aff13},\ref{aff15}}
\and L.~Moscardini\orcid{0000-0002-3473-6716}\inst{\ref{aff79},\ref{aff2},\ref{aff18}}
\and R.~Nakajima\orcid{0009-0009-1213-7040}\inst{\ref{aff77}}
\and C.~Neissner\orcid{0000-0001-8524-4968}\inst{\ref{aff88},\ref{aff31}}
\and S.-M.~Niemi\orcid{0009-0005-0247-0086}\inst{\ref{aff89}}
\and C.~Padilla\orcid{0000-0001-7951-0166}\inst{\ref{aff88}}
\and F.~Pasian\orcid{0000-0002-4869-3227}\inst{\ref{aff14}}
\and W.~J.~Percival\orcid{0000-0002-0644-5727}\inst{\ref{aff90},\ref{aff91},\ref{aff92}}
\and V.~Pettorino\orcid{0000-0002-4203-9320}\inst{\ref{aff89}}
\and A.~Pezzotta\orcid{0000-0003-0726-2268}\inst{\ref{aff12}}
\and S.~Pires\orcid{0000-0002-0249-2104}\inst{\ref{aff51}}
\and G.~Polenta\orcid{0000-0003-4067-9196}\inst{\ref{aff93}}
\and M.~Poncet\inst{\ref{aff94}}
\and L.~A.~Popa\inst{\ref{aff95}}
\and F.~Raison\orcid{0000-0002-7819-6918}\inst{\ref{aff4}}
\and A.~Renzi\orcid{0000-0001-9856-1970}\inst{\ref{aff96},\ref{aff97},\ref{aff2}}
\and J.~Rhodes\orcid{0000-0002-4485-8549}\inst{\ref{aff59}}
\and G.~Riccio\inst{\ref{aff25}}
\and I.~Risso\orcid{0000-0003-2525-7761}\inst{\ref{aff22},\ref{aff23},\ref{aff12}}
\and F.~Rizzo\orcid{0000-0002-9407-585X}\inst{\ref{aff14}}
\and E.~Romelli\orcid{0000-0003-3069-9222}\inst{\ref{aff14}}
\and M.~Roncarelli\orcid{0000-0001-9587-7822}\inst{\ref{aff2}}
\and B.~Rusholme\orcid{0000-0001-7648-4142}\inst{\ref{aff57}}
\and R.~Saglia\orcid{0000-0003-0378-7032}\inst{\ref{aff5},\ref{aff4}}
\and Z.~Sakr\orcid{0000-0002-4823-3757}\inst{\ref{aff98},\ref{aff99},\ref{aff100}}
\and A.~G.~S\'anchez\orcid{0000-0003-1198-831X}\inst{\ref{aff4}}
\and D.~Sapone\orcid{0000-0001-7089-4503}\inst{\ref{aff101}}
\and B.~Sartoris\orcid{0000-0003-1337-5269}\inst{\ref{aff4},\ref{aff14}}
\and M.~Schirmer\orcid{0000-0003-2568-9994}\inst{\ref{aff65}}
\and P.~Schneider\orcid{0000-0001-8561-2679}\inst{\ref{aff77}}
\and T.~Schrabback\orcid{0000-0002-6987-7834}\inst{\ref{aff102}}
\and M.~Scodeggio\inst{\ref{aff29}}
\and A.~Secroun\orcid{0000-0003-0505-3710}\inst{\ref{aff49}}
\and E.~Sihvola\orcid{0000-0003-1804-7715}\inst{\ref{aff103}}
\and C.~Sirignano\orcid{0000-0002-0995-7146}\inst{\ref{aff96},\ref{aff97}}
\and G.~Sirri\orcid{0000-0003-2626-2853}\inst{\ref{aff18}}
\and L.~Stanco\orcid{0000-0002-9706-5104}\inst{\ref{aff97}}
\and P.~Tallada-Cresp\'{i}\orcid{0000-0002-1336-8328}\inst{\ref{aff30},\ref{aff31}}
\and A.~N.~Taylor\inst{\ref{aff39}}
\and H.~I.~Teplitz\orcid{0000-0002-7064-5424}\inst{\ref{aff81}}
\and I.~Tereno\orcid{0000-0002-4537-6218}\inst{\ref{aff104},\ref{aff105}}
\and N.~Tessore\orcid{0000-0002-9696-7931}\inst{\ref{aff48}}
\and S.~Toft\orcid{0000-0003-3631-7176}\inst{\ref{aff106},\ref{aff107}}
\and R.~Toledo-Moreo\orcid{0000-0002-2997-4859}\inst{\ref{aff108},\ref{aff109}}
\and F.~Torradeflot\orcid{0000-0003-1160-1517}\inst{\ref{aff31},\ref{aff30}}
\and I.~Tutusaus\orcid{0000-0002-3199-0399}\inst{\ref{aff32},\ref{aff33},\ref{aff99}}
\and J.~Valiviita\orcid{0000-0001-6225-3693}\inst{\ref{aff69},\ref{aff70}}
\and T.~Vassallo\orcid{0000-0001-6512-6358}\inst{\ref{aff14},\ref{aff54}}
\and Y.~Wang\orcid{0000-0002-4749-2984}\inst{\ref{aff57}}
\and J.~Weller\orcid{0000-0002-8282-2010}\inst{\ref{aff5},\ref{aff4}}
\and A.~Zacchei\orcid{0000-0003-0396-1192}\inst{\ref{aff14},\ref{aff13}}
\and F.~M.~Zerbi\orcid{0000-0002-9996-973X}\inst{\ref{aff12}}
\and I.~A.~Zinchenko\orcid{0000-0002-2944-2449}\inst{\ref{aff110}}
\and E.~Zucca\orcid{0000-0002-5845-8132}\inst{\ref{aff2}}
\and V.~Allevato\orcid{0000-0001-7232-5152}\inst{\ref{aff25}}
\and M.~Ballardini\orcid{0000-0003-4481-3559}\inst{\ref{aff111},\ref{aff112},\ref{aff2}}
\and S.~Borgani\orcid{0000-0001-6151-6439}\inst{\ref{aff86},\ref{aff13},\ref{aff14},\ref{aff15},\ref{aff113}}
\and E.~Bozzo\orcid{0000-0002-8201-1525}\inst{\ref{aff3}}
\and C.~Burigana\orcid{0000-0002-3005-5796}\inst{\ref{aff114},\ref{aff52}}
\and R.~Cabanac\orcid{0000-0001-6679-2600}\inst{\ref{aff99}}
\and M.~Calabrese\orcid{0000-0002-2637-2422}\inst{\ref{aff115},\ref{aff29}}
\and A.~Cappi\inst{\ref{aff116},\ref{aff2}}
\and F.~Caro\orcid{0009-0003-1053-0507}\inst{\ref{aff34}}
\and T.~Castro\orcid{0000-0002-6292-3228}\inst{\ref{aff117},\ref{aff14},\ref{aff13}}
\and D.~Di~Ferdinando\inst{\ref{aff18}}
\and J.~A.~Escartin~Vigo\inst{\ref{aff4}}
\and G.~Fabbian\orcid{0000-0002-3255-4695}\inst{\ref{aff9}}
\and J.~Garc\'ia-Bellido\orcid{0000-0002-9370-8360}\inst{\ref{aff98}}
\and T.~Gasparetto\orcid{0000-0002-7913-4866}\inst{\ref{aff34}}
\and V.~Gautard\inst{\ref{aff1}}
\and C.~M.~Gutierrez\orcid{0000-0001-7854-783X}\inst{\ref{aff38},\ref{aff118}}
\and A.~Loureiro\orcid{0000-0002-4371-0876}\inst{\ref{aff119},\ref{aff120}}
\and J.~Macias-Perez\orcid{0000-0002-5385-2763}\inst{\ref{aff121}}
\and C.~Mancini\orcid{0000-0002-4297-0561}\inst{\ref{aff29}}
\and R.~Maoli\orcid{0000-0002-6065-3025}\inst{\ref{aff122},\ref{aff34}}
\and R.~B.~Metcalf\orcid{0000-0003-3167-2574}\inst{\ref{aff79},\ref{aff2}}
\and M.~P\"ontinen\orcid{0000-0001-5442-2530}\inst{\ref{aff69}}
\and E.~Sarpa\orcid{0000-0002-1256-655X}\inst{\ref{aff14}}
\and V.~Scottez\orcid{0009-0008-3864-940X}\inst{\ref{aff123},\ref{aff124}}
\and M.~Sereno\orcid{0000-0003-0302-0325}\inst{\ref{aff2},\ref{aff18}}
\and M.~Tenti\orcid{0000-0002-4254-5901}\inst{\ref{aff18}}
\and M.~Viel\orcid{0000-0002-2642-5707}\inst{\ref{aff13},\ref{aff14},\ref{aff16},\ref{aff15},\ref{aff113}}
\and M.~Wiesmann\orcid{0009-0000-8199-5860}\inst{\ref{aff56}}
\and J.~A.~Acevedo~Barroso\orcid{0000-0002-9654-1711}\inst{\ref{aff59}}
\and R.~Adam\orcid{0009-0000-5380-1109}\inst{\ref{aff116}}
\and Y.~Akrami\orcid{0000-0002-2407-7956}\inst{\ref{aff98},\ref{aff125}}
\and I.~T.~Andika\orcid{0000-0001-6102-9526}\inst{\ref{aff5}}
\and G.~Angora\orcid{0000-0002-0316-6562}\inst{\ref{aff25},\ref{aff111}}
\and S.~Anselmi\orcid{0000-0002-3579-9583}\inst{\ref{aff97},\ref{aff96},\ref{aff126}}
\and M.~Archidiacono\orcid{0000-0003-4952-9012}\inst{\ref{aff74},\ref{aff75}}
\and G.~Aric\`o\orcid{0000-0002-2802-2928}\inst{\ref{aff18}}
\and F.~Atrio-Barandela\orcid{0000-0002-2130-2513}\inst{\ref{aff127}}
\and E.~Aubourg\orcid{0000-0002-5592-023X}\inst{\ref{aff82},\ref{aff128}}
\and M.~Baes\orcid{0000-0002-3930-2757}\inst{\ref{aff129}}
\and L.~Bazzanini\orcid{0000-0003-0727-0137}\inst{\ref{aff111},\ref{aff2}}
\and D.~Bertacca\orcid{0000-0002-2490-7139}\inst{\ref{aff96},\ref{aff19},\ref{aff97}}
\and M.~Bethermin\orcid{0000-0002-3915-2015}\inst{\ref{aff130}}
\and F.~Beutler\orcid{0000-0003-0467-5438}\inst{\ref{aff39}}
\and L.~Bisigello\orcid{0000-0003-0492-4924}\inst{\ref{aff19}}
\and A.~Blanchard\orcid{0000-0001-8555-9003}\inst{\ref{aff99}}
\and L.~Blot\orcid{0000-0002-9622-7167}\inst{\ref{aff131},\ref{aff71}}
\and H.~B\"ohringer\orcid{0000-0001-8241-4204}\inst{\ref{aff4},\ref{aff54},\ref{aff132}}
\and M.~Bonici\orcid{0000-0002-8430-126X}\inst{\ref{aff90},\ref{aff29}}
\and M.~L.~Brown\orcid{0000-0002-0370-8077}\inst{\ref{aff40}}
\and S.~Bruton\orcid{0000-0002-6503-5218}\inst{\ref{aff133}}
\and A.~Calabro\orcid{0000-0003-2536-1614}\inst{\ref{aff34}}
\and B.~Camacho~Quevedo\orcid{0000-0002-8789-4232}\inst{\ref{aff13},\ref{aff16},\ref{aff14}}
\and C.~S.~Carvalho\inst{\ref{aff105}}
\and F.~Cogato\orcid{0000-0003-4632-6113}\inst{\ref{aff79},\ref{aff2}}
\and T.~E.~Collett\orcid{0000-0001-5564-3140}\inst{\ref{aff134}}
\and A.~R.~Cooray\orcid{0000-0002-3892-0190}\inst{\ref{aff135}}
\and P.~Corcho-Caballero\orcid{0000-0001-6327-7080}\inst{\ref{aff20}}
\and B.~Csizi\orcid{0000-0003-3227-6581}\inst{\ref{aff102}}
\and O.~Cucciati\orcid{0000-0002-9336-7551}\inst{\ref{aff2}}
\and H.~Dannerbauer\orcid{0000-0001-7147-3575}\inst{\ref{aff38},\ref{aff118}}
\and T.~de~Boer\orcid{0000-0001-5486-2747}\inst{\ref{aff36}}
\and G.~Desprez\orcid{0000-0001-8325-1742}\inst{\ref{aff20}}
\and A.~D\'iaz-S\'anchez\orcid{0000-0003-0748-4768}\inst{\ref{aff136}}
\and S.~Di~Domizio\orcid{0000-0003-2863-5895}\inst{\ref{aff22},\ref{aff23}}
\and J.~M.~Diego\orcid{0000-0001-9065-3926}\inst{\ref{aff137}}
\and V.~Duret\orcid{0009-0009-0383-4960}\inst{\ref{aff49}}
\and Y.~Fang\orcid{0000-0002-0334-6950}\inst{\ref{aff5}}
\and A.~Farina\orcid{0009-0000-3420-929X}\inst{\ref{aff12},\ref{aff23}}
\and A.~M.~N.~Ferguson\orcid{0000-0001-7934-1278}\inst{\ref{aff39}}
\and A.~Finoguenov\orcid{0000-0002-4606-5403}\inst{\ref{aff69}}
\and A.~Fontana\orcid{0000-0003-3820-2823}\inst{\ref{aff34}}
\and A.~Franco\orcid{0000-0002-4761-366X}\inst{\ref{aff138},\ref{aff139},\ref{aff140}}
\and Y.~Fu\orcid{0000-0002-0759-0504}\inst{\ref{aff58},\ref{aff20}}
\and K.~Ganga\orcid{0000-0001-8159-8208}\inst{\ref{aff82}}
\and R.~Gavazzi\orcid{0000-0002-5540-6935}\inst{\ref{aff43},\ref{aff141}}
\and E.~Gaztanaga\orcid{0000-0001-9632-0815}\inst{\ref{aff32},\ref{aff33},\ref{aff134}}
\and Z.~Ghaffari\orcid{0000-0002-6467-8078}\inst{\ref{aff14},\ref{aff13}}
\and F.~Giacomini\orcid{0000-0002-3129-2814}\inst{\ref{aff18}}
\and F.~Gianotti\orcid{0000-0003-4666-119X}\inst{\ref{aff2}}
\and G.~Gozaliasl\orcid{0000-0002-0236-919X}\inst{\ref{aff142},\ref{aff69}}
\and A.~Gruppuso\orcid{0000-0001-9272-5292}\inst{\ref{aff2},\ref{aff18}}
\and M.~Guidi\orcid{0000-0001-9408-1101}\inst{\ref{aff17},\ref{aff2}}
\and A.~Hall\orcid{0000-0002-3139-8651}\inst{\ref{aff39}}
\and N.~A.~Hatch\orcid{0000-0001-5600-0534}\inst{\ref{aff143}}
\and C.~Hern\'andez-Monteagudo\orcid{0000-0001-5471-9166}\inst{\ref{aff118},\ref{aff38}}
\and J.~Hjorth\orcid{0000-0002-4571-2306}\inst{\ref{aff144}}
\and M.~Jauzac\orcid{0000-0003-1974-8732}\inst{\ref{aff145},\ref{aff80},\ref{aff146},\ref{aff147},\ref{aff99}}
\and J.~J.~E.~Kajava\orcid{0000-0002-3010-8333}\inst{\ref{aff148},\ref{aff149},\ref{aff150}}
\and Y.~Kang\orcid{0009-0000-8588-7250}\inst{\ref{aff102}}
\and V.~Kansal\orcid{0000-0002-4008-6078}\inst{\ref{aff151},\ref{aff152}}
\and D.~Karagiannis\orcid{0000-0002-4927-0816}\inst{\ref{aff111},\ref{aff153}}
\and J.~Kim\orcid{0000-0003-2776-2761}\inst{\ref{aff53}}
\and C.~C.~Kirkpatrick\inst{\ref{aff103}}
\and A.~Kov\'acs\orcid{0000-0002-5825-579X}\inst{\ref{aff154},\ref{aff155}}
\and I.~Kova{\v{c}}i{\'{c}}\orcid{0000-0001-6751-3263}\inst{\ref{aff129}}
\and K.~Koyama\orcid{0000-0001-6727-6915}\inst{\ref{aff134}}
\and S.~Kruk\orcid{0000-0001-8010-8879}\inst{\ref{aff10}}
\and M.~C.~Lam\orcid{0000-0002-9347-2298}\inst{\ref{aff39}}
\and M.~Lattanzi\orcid{0000-0003-1059-2532}\inst{\ref{aff112}}
\and L.~Legrand\orcid{0000-0003-0610-5252}\inst{\ref{aff156},\ref{aff9}}
\and M.~Lembo\orcid{0000-0002-5271-5070}\inst{\ref{aff141}}
\and G.~Leroy\orcid{0009-0004-2523-4425}\inst{\ref{aff145},\ref{aff80}}
\and G.~F.~Lesci\orcid{0000-0002-4607-2830}\inst{\ref{aff79},\ref{aff2}}
\and J.~Lesgourgues\orcid{0000-0001-7627-353X}\inst{\ref{aff157}}
\and L.~Linke\orcid{0000-0002-2622-8113}\inst{\ref{aff102}}
\and S.~J.~Liu\orcid{0000-0001-7680-2139}\inst{\ref{aff50}}
\and G.~Maggio\orcid{0000-0003-4020-4836}\inst{\ref{aff14}}
\and F.~Mannucci\orcid{0000-0002-4803-2381}\inst{\ref{aff158}}
\and F.~R.~Marleau\orcid{0000-0002-1442-2947}\inst{\ref{aff102}}
\and C.~J.~A.~P.~Martins\orcid{0000-0002-4886-9261}\inst{\ref{aff159},\ref{aff160}}
\and M.~Miluzio\inst{\ref{aff10},\ref{aff161}}
\and G.~Morgante\inst{\ref{aff2}}
\and S.~Nadathur\orcid{0000-0001-9070-3102}\inst{\ref{aff134}}
\and K.~Naidoo\orcid{0000-0002-9182-1802}\inst{\ref{aff134},\ref{aff65}}
\and A.~Navarro-Alsina\orcid{0000-0002-3173-2592}\inst{\ref{aff77}}
\and A.~Nersesian\orcid{0000-0001-6843-409X}\inst{\ref{aff162},\ref{aff129}}
\and S.~Nesseris\orcid{0000-0002-0567-0324}\inst{\ref{aff98}}
\and L.~Nicastro\orcid{0000-0001-8534-6788}\inst{\ref{aff2}}
\and F.~Oppizzi\orcid{0000-0003-3904-8370}\inst{\ref{aff23},\ref{aff97},\ref{aff96}}
\and F.~Pace\orcid{0000-0001-8039-0480}\inst{\ref{aff26},\ref{aff27},\ref{aff28},\ref{aff163}}
\and D.~Paoletti\orcid{0000-0003-4761-6147}\inst{\ref{aff2},\ref{aff52}}
\and G.~Parimbelli\orcid{0000-0002-2539-2472}\inst{\ref{aff32},\ref{aff16}}
\and F.~Passalacqua\orcid{0000-0002-8606-4093}\inst{\ref{aff97}}
\and K.~Paterson\orcid{0000-0001-8340-3486}\inst{\ref{aff65}}
\and C.~Pattison\orcid{0000-0003-3272-2617}\inst{\ref{aff134}}
\and R.~Pello\orcid{0000-0003-0858-6109}\inst{\ref{aff43}}
\and A.~Pisani\orcid{0000-0002-6146-4437}\inst{\ref{aff49}}
\and D.~Potter\orcid{0000-0002-0757-5195}\inst{\ref{aff164}}
\and G.~W.~Pratt\inst{\ref{aff51}}
\and S.~Quai\orcid{0000-0002-0449-8163}\inst{\ref{aff79},\ref{aff2}}
\and M.~Radovich\orcid{0000-0002-3585-866X}\inst{\ref{aff19}}
\and F.~Ricci\orcid{0000-0001-5742-5980}\inst{\ref{aff165},\ref{aff34}}
\and G.~Rodighiero\orcid{0000-0002-9415-2296}\inst{\ref{aff96},\ref{aff19}}
\and W.~Roster\orcid{0000-0002-9149-6528}\inst{\ref{aff4}}
\and M.~Sahl\'en\orcid{0000-0003-0973-4804}\inst{\ref{aff166}}
\and D.~B.~Sanders\orcid{0000-0002-1233-9998}\inst{\ref{aff36}}
\and A.~Schneider\orcid{0000-0001-7055-8104}\inst{\ref{aff164}}
\and D.~Sciotti\orcid{0009-0008-4519-2620}\inst{\ref{aff34},\ref{aff78}}
\and E.~Sellentin\orcid{0009-0002-2655-3458}\inst{\ref{aff167},\ref{aff58}}
\and S.~Serjeant\orcid{0000-0002-0517-7943}\inst{\ref{aff168}}
\and F.~Shankar\orcid{0000-0001-8973-5051}\inst{\ref{aff169}}
\and L.~C.~Smith\orcid{0000-0002-3259-2771}\inst{\ref{aff170}}
\and S.~A.~Stanford\orcid{0000-0003-0122-0841}\inst{\ref{aff171}}
\and I.~Szapudi\orcid{0000-0003-2274-0301}\inst{\ref{aff36}}
\and M.~Talia\orcid{0000-0003-4352-2063}\inst{\ref{aff79},\ref{aff2}}
\and K.~Tanidis\orcid{0000-0001-9843-5130}\inst{\ref{aff172}}
\and C.~Tao\orcid{0000-0001-7961-8177}\inst{\ref{aff49}}
\and F.~Tarsitano\orcid{0000-0002-5919-0238}\inst{\ref{aff173},\ref{aff174},\ref{aff3}}
\and G.~Testera\orcid{0000-0003-2970-766X}\inst{\ref{aff23}}
\and R.~Teyssier\orcid{0000-0001-7689-0933}\inst{\ref{aff175}}
\and S.~Tosi\orcid{0000-0002-7275-9193}\inst{\ref{aff22},\ref{aff12},\ref{aff23}}
\and A.~Troja\orcid{0000-0003-0239-4595}\inst{\ref{aff14}}
\and C.~Uhlemann\orcid{0000-0001-7831-1579}\inst{\ref{aff176},\ref{aff177}}
\and C.~Valieri\inst{\ref{aff18}}
\and A.~Venhola\orcid{0000-0001-6071-4564}\inst{\ref{aff178}}
\and D.~Vergani\orcid{0000-0003-0898-2216}\inst{\ref{aff2}}
\and P.~Vielzeuf\orcid{0000-0003-2035-9339}\inst{\ref{aff49}}
\and S.~Vinciguerra\orcid{0009-0005-4018-3184}\inst{\ref{aff43}}
\and M.~von~Wietersheim-Kramsta\orcid{0000-0003-4986-5091}\inst{\ref{aff80},\ref{aff145}}
\and N.~A.~Walton\orcid{0000-0003-3983-8778}\inst{\ref{aff170}}
\and L.~Wang\orcid{0000-0002-6736-9158}\inst{\ref{aff179},\ref{aff20}}
\and J.~R.~Weaver\orcid{0000-0003-1614-196X}\inst{\ref{aff180}}
\and A.~H.~Wright\orcid{0000-0001-7363-7932}\inst{\ref{aff7}}}
										   
\institute{CEA Saclay, DFR/IRFU, Service d'Astrophysique, Bat. 709, 91191 Gif-sur-Yvette, France\label{aff1}
\and
INAF-Osservatorio di Astrofisica e Scienza dello Spazio di Bologna, Via Piero Gobetti 93/3, 40129 Bologna, Italy\label{aff2}
\and
Department of Astronomy, University of Geneva, ch. d'Ecogia 16, 1290 Versoix, Switzerland\label{aff3}
\and
Max Planck Institute for Extraterrestrial Physics, Giessenbachstr. 1, 85748 Garching, Germany\label{aff4}
\and
Universit\"ats-Sternwarte M\"unchen, Fakult\"at f\"ur Physik, Ludwig-Maximilians-Universit\"at M\"unchen, Scheinerstr.~1, 81679 M\"unchen, Germany\label{aff5}
\and
Department of Physics and Astronomy, University of British Columbia, Vancouver, BC V6T 1Z1, Canada\label{aff6}
\and
Ruhr University Bochum, Faculty of Physics and Astronomy, Astronomical Institute (AIRUB), German Centre for Cosmological Lensing (GCCL), 44780 Bochum, Germany\label{aff7}
\and
Univ. Lille, CNRS, Centrale Lille, UMR 9189 CRIStAL, 59000 Lille, France\label{aff8}
\and
Universit\'e Paris-Saclay, CNRS, Institut d'astrophysique spatiale, 91405, Orsay, France\label{aff9}
\and
ESAC/ESA, Camino Bajo del Castillo, s/n., Urb. Villafranca del Castillo, 28692 Villanueva de la Ca\~nada, Madrid, Spain\label{aff10}
\and
School of Mathematics and Physics, University of Surrey, Guildford, Surrey, GU2 7XH, UK\label{aff11}
\and
INAF-Osservatorio Astronomico di Brera, Via Brera 28, 20122 Milano, Italy\label{aff12}
\and
IFPU, Institute for Fundamental Physics of the Universe, via Beirut 2, 34151 Trieste, Italy\label{aff13}
\and
INAF-Osservatorio Astronomico di Trieste, Via G. B. Tiepolo 11, 34143 Trieste, Italy\label{aff14}
\and
INFN, Sezione di Trieste, Via Valerio 2, 34127 Trieste TS, Italy\label{aff15}
\and
SISSA, International School for Advanced Studies, Via Bonomea 265, 34136 Trieste TS, Italy\label{aff16}
\and
Dipartimento di Fisica e Astronomia, Universit\`a di Bologna, Via Gobetti 93/2, 40129 Bologna, Italy\label{aff17}
\and
INFN-Sezione di Bologna, Viale Berti Pichat 6/2, 40127 Bologna, Italy\label{aff18}
\and
INAF-Osservatorio Astronomico di Padova, Via dell'Osservatorio 5, 35122 Padova, Italy\label{aff19}
\and
Kapteyn Astronomical Institute, University of Groningen, PO Box 800, 9700 AV Groningen, The Netherlands\label{aff20}
\and
ATG Europe BV, Huygensstraat 34, 2201 DK Noordwijk, The Netherlands\label{aff21}
\and
Dipartimento di Fisica, Universit\`a di Genova, Via Dodecaneso 33, 16146, Genova, Italy\label{aff22}
\and
INFN-Sezione di Genova, Via Dodecaneso 33, 16146, Genova, Italy\label{aff23}
\and
Department of Physics "E. Pancini", University Federico II, Via Cinthia 6, 80126, Napoli, Italy\label{aff24}
\and
INAF-Osservatorio Astronomico di Capodimonte, Via Moiariello 16, 80131 Napoli, Italy\label{aff25}
\and
Dipartimento di Fisica, Universit\`a degli Studi di Torino, Via P. Giuria 1, 10125 Torino, Italy\label{aff26}
\and
INFN-Sezione di Torino, Via P. Giuria 1, 10125 Torino, Italy\label{aff27}
\and
INAF-Osservatorio Astrofisico di Torino, Via Osservatorio 20, 10025 Pino Torinese (TO), Italy\label{aff28}
\and
INAF-IASF Milano, Via Alfonso Corti 12, 20133 Milano, Italy\label{aff29}
\and
Centro de Investigaciones Energ\'eticas, Medioambientales y Tecnol\'ogicas (CIEMAT), Avenida Complutense 40, 28040 Madrid, Spain\label{aff30}
\and
Port d'Informaci\'{o} Cient\'{i}fica, Campus UAB, C. Albareda s/n, 08193 Bellaterra (Barcelona), Spain\label{aff31}
\and
Institute of Space Sciences (ICE, CSIC), Campus UAB, Carrer de Can Magrans, s/n, 08193 Barcelona, Spain\label{aff32}
\and
Institut d'Estudis Espacials de Catalunya (IEEC),  Edifici RDIT, Campus UPC, 08860 Castelldefels, Barcelona, Spain\label{aff33}
\and
INAF-Osservatorio Astronomico di Roma, Via Frascati 33, 00078 Monteporzio Catone, Italy\label{aff34}
\and
INFN -- Sezione di Napoli, Via Cinthia 6, 80126, Napoli, Italy\label{aff35}
\and
Institute for Astronomy, University of Hawaii, 2680 Woodlawn Drive, Honolulu, HI 96822, USA\label{aff36}
\and
Dipartimento di Fisica e Astronomia "Augusto Righi" - Alma Mater Studiorum Universit\`a di Bologna, Viale Berti Pichat 6/2, 40127 Bologna, Italy\label{aff37}
\and
Instituto de Astrof\'{\i}sica de Canarias, E-38205 La Laguna, Tenerife, Spain\label{aff38}
\and
Institute for Astronomy, University of Edinburgh, Royal Observatory, Blackford Hill, Edinburgh EH9 3HJ, UK\label{aff39}
\and
Jodrell Bank Centre for Astrophysics, Department of Physics and Astronomy, University of Manchester, Oxford Road, Manchester M13 9PL, UK\label{aff40}
\and
European Space Agency/ESRIN, Largo Galileo Galilei 1, 00044 Frascati, Roma, Italy\label{aff41}
\and
Universit\'e Lyon 1, CNRS/IN2P3, IP2I Lyon, Villeurbanne, France\label{aff42}
\and
Aix-Marseille Universit\'e, CNRS, CNES, LAM, Marseille, France\label{aff43}
\and
Institut de Ci\`{e}ncies del Cosmos (ICCUB), Universitat de Barcelona (IEEC-UB), Mart\'{i} i Franqu\`{e}s 1, 08028 Barcelona, Spain\label{aff44}
\and
Instituci\'o Catalana de Recerca i Estudis Avan\c{c}ats (ICREA), Passeig de Llu\'{\i}s Companys 23, 08010 Barcelona, Spain\label{aff45}
\and
Institut de Ciencies de l'Espai (IEEC-CSIC), Campus UAB, Carrer de Can Magrans, s/n Cerdanyola del Vall\'es, 08193 Barcelona, Spain\label{aff46}
\and
Universit\'e Lyon 1, CNRS/IN2P3, IUF, IP2I Lyon, Villeurbanne, France \label{aff47}
\and
Mullard Space Science Laboratory, University College London, Holmbury St Mary, Dorking, Surrey RH5 6NT, UK\label{aff48}
\and
Aix-Marseille Universit\'e, CNRS/IN2P3, CPPM, Marseille, France\label{aff49}
\and
INAF-Istituto di Astrofisica e Planetologia Spaziali, via del Fosso del Cavaliere, 100, 00100 Roma, Italy\label{aff50}
\and
Universit\'e Paris-Saclay, Universit\'e Paris Cit\'e, CEA, CNRS, AIM, 91191, Gif-sur-Yvette, France\label{aff51}
\and
INFN-Bologna, Via Irnerio 46, 40126 Bologna, Italy\label{aff52}
\and
Department of Physics, University of Oxford, Keble Road, Oxford OX1 3RH, UK\label{aff53}
\and
University Observatory, LMU Faculty of Physics, Scheinerstr.~1, 81679 Munich, Germany\label{aff54}
\and
National Research Council, Herzberg Astronomy and Astrophysics Research Centre, 5071 W. Saanich Rd. Victoria, BC, V9E 2E7, Canada\label{aff55}
\and
Institute of Theoretical Astrophysics, University of Oslo, P.O. Box 1029 Blindern, 0315 Oslo, Norway\label{aff56}
\and
Caltech/IPAC, 1200 E. California Blvd., Pasadena, CA 91125, USA\label{aff57}
\and
Leiden Observatory, Leiden University, Einsteinweg 55, 2333 CC Leiden, The Netherlands\label{aff58}
\and
Jet Propulsion Laboratory, California Institute of Technology, 4800 Oak Grove Drive, Pasadena, CA, 91109, USA\label{aff59}
\and
Felix Hormuth Engineering, Goethestr. 17, 69181 Leimen, Germany\label{aff60}
\and
Technical University of Denmark, Elektrovej 327, 2800 Kgs. Lyngby, Denmark\label{aff61}
\and
Cosmic Dawn Center (DAWN), Denmark\label{aff62}
\and
Universit\'e PSL, Observatoire de Paris, Sorbonne Universit\'e, CNRS, LERMA, 75014, Paris, France\label{aff63}
\and
Universit\'e Paris-Cit\'e, 5 Rue Thomas Mann, 75013, Paris, France\label{aff64}
\and
Max-Planck-Institut f\"ur Astronomie, K\"onigstuhl 17, 69117 Heidelberg, Germany\label{aff65}
\and
NASA Goddard Space Flight Center, Greenbelt, MD 20771, USA\label{aff66}
\and
Department of Physics and Astronomy, University College London, Gower Street, London WC1E 6BT, UK\label{aff67}
\and
Universit\'e de Gen\`eve, D\'epartement de Physique Th\'eorique and Centre for Astroparticle Physics, 24 quai Ernest-Ansermet, CH-1211 Gen\`eve 4, Switzerland\label{aff68}
\and
Department of Physics, P.O. Box 64, University of Helsinki, 00014 Helsinki, Finland\label{aff69}
\and
Helsinki Institute of Physics, Gustaf H{\"a}llstr{\"o}min katu 2, University of Helsinki, 00014 Helsinki, Finland\label{aff70}
\and
Laboratoire d'etude de l'Univers et des phenomenes eXtremes, Observatoire de Paris, Universit\'e PSL, Sorbonne Universit\'e, CNRS, 92190 Meudon, France\label{aff71}
\and
SKAO, Jodrell Bank, Lower Withington, Macclesfield SK11 9FT, UK\label{aff72}
\and
Centre de Calcul de l'IN2P3/CNRS, 21 avenue Pierre de Coubertin 69627 Villeurbanne Cedex, France\label{aff73}
\and
Dipartimento di Fisica "Aldo Pontremoli", Universit\`a degli Studi di Milano, Via Celoria 16, 20133 Milano, Italy\label{aff74}
\and
INFN-Sezione di Milano, Via Celoria 16, 20133 Milano, Italy\label{aff75}
\and
University of Applied Sciences and Arts of Northwestern Switzerland, School of Computer Science, 5210 Windisch, Switzerland\label{aff76}
\and
Universit\"at Bonn, Argelander-Institut f\"ur Astronomie, Auf dem H\"ugel 71, 53121 Bonn, Germany\label{aff77}
\and
INFN-Sezione di Roma, Piazzale Aldo Moro, 2 - c/o Dipartimento di Fisica, Edificio G. Marconi, 00185 Roma, Italy\label{aff78}
\and
Dipartimento di Fisica e Astronomia "Augusto Righi" - Alma Mater Studiorum Universit\`a di Bologna, via Piero Gobetti 93/2, 40129 Bologna, Italy\label{aff79}
\and
Department of Physics, Institute for Computational Cosmology, Durham University, South Road, Durham, DH1 3LE, UK\label{aff80}
\and
Infrared Processing and Analysis Center, California Institute of Technology, Pasadena, CA 91125, USA\label{aff81}
\and
Universit\'e Paris Cit\'e, CNRS, Astroparticule et Cosmologie, 75013 Paris, France\label{aff82}
\and
CNRS-UCB International Research Laboratory, Centre Pierre Bin\'etruy, IRL2007, CPB-IN2P3, Berkeley, USA\label{aff83}
\and
University of Applied Sciences and Arts of Northwestern Switzerland, School of Engineering, 5210 Windisch, Switzerland\label{aff84}
\and
Institute of Physics, Laboratory of Astrophysics, Ecole Polytechnique F\'ed\'erale de Lausanne (EPFL), Observatoire de Sauverny, 1290 Versoix, Switzerland\label{aff85}
\and
Dipartimento di Fisica - Sezione di Astronomia, Universit\`a di Trieste, Via Tiepolo 11, 34131 Trieste, Italy\label{aff86}
\and
Telespazio UK S.L. for European Space Agency (ESA), Camino bajo del Castillo, s/n, Urbanizacion Villafranca del Castillo, Villanueva de la Ca\~nada, 28692 Madrid, Spain\label{aff87}
\and
Institut de F\'{i}sica d'Altes Energies (IFAE), The Barcelona Institute of Science and Technology, Campus UAB, 08193 Bellaterra (Barcelona), Spain\label{aff88}
\and
European Space Agency/ESTEC, Keplerlaan 1, 2201 AZ Noordwijk, The Netherlands\label{aff89}
\and
Waterloo Centre for Astrophysics, University of Waterloo, Waterloo, Ontario N2L 3G1, Canada\label{aff90}
\and
Department of Physics and Astronomy, University of Waterloo, Waterloo, Ontario N2L 3G1, Canada\label{aff91}
\and
Perimeter Institute for Theoretical Physics, Waterloo, Ontario N2L 2Y5, Canada\label{aff92}
\and
Space Science Data Center, Italian Space Agency, via del Politecnico snc, 00133 Roma, Italy\label{aff93}
\and
Centre National d'Etudes Spatiales -- Centre spatial de Toulouse, 18 avenue Edouard Belin, 31401 Toulouse Cedex 9, France\label{aff94}
\and
Institute of Space Science, Str. Atomistilor, nr. 409 M\u{a}gurele, Ilfov, 077125, Romania\label{aff95}
\and
Dipartimento di Fisica e Astronomia "G. Galilei", Universit\`a di Padova, Via Marzolo 8, 35131 Padova, Italy\label{aff96}
\and
INFN-Padova, Via Marzolo 8, 35131 Padova, Italy\label{aff97}
\and
Instituto de F\'isica Te\'orica UAM-CSIC, Campus de Cantoblanco, 28049 Madrid, Spain\label{aff98}
\and
Institut de Recherche en Astrophysique et Plan\'etologie (IRAP), Universit\'e de Toulouse, CNRS, UPS, CNES, 14 Av. Edouard Belin, 31400 Toulouse, France\label{aff99}
\and
Universit\'e St Joseph; Faculty of Sciences, Beirut, Lebanon\label{aff100}
\and
Departamento de F\'isica, FCFM, Universidad de Chile, Blanco Encalada 2008, Santiago, Chile\label{aff101}
\and
Universit\"at Innsbruck, Institut f\"ur Astro- und Teilchenphysik, Technikerstr. 25/8, 6020 Innsbruck, Austria\label{aff102}
\and
Department of Physics and Helsinki Institute of Physics, Gustaf H\"allstr\"omin katu 2, University of Helsinki, 00014 Helsinki, Finland\label{aff103}
\and
Departamento de F\'isica, Faculdade de Ci\^encias, Universidade de Lisboa, Edif\'icio C8, Campo Grande, PT1749-016 Lisboa, Portugal\label{aff104}
\and
Instituto de Astrof\'isica e Ci\^encias do Espa\c{c}o, Faculdade de Ci\^encias, Universidade de Lisboa, Tapada da Ajuda, 1349-018 Lisboa, Portugal\label{aff105}
\and
Cosmic Dawn Center (DAWN)\label{aff106}
\and
Niels Bohr Institute, University of Copenhagen, Jagtvej 128, 2200 Copenhagen, Denmark\label{aff107}
\and
Universidad Polit\'ecnica de Cartagena, Departamento de Electr\'onica y Tecnolog\'ia de Computadoras,  Plaza del Hospital 1, 30202 Cartagena, Spain\label{aff108}
\and
European University of Technology EUt+, European Union\label{aff109}
\and
Astronomisches Rechen-Institut, Zentrum f\"ur Astronomie der Universit\"at Heidelberg, M\"onchhofstr. 12-14, 69120 Heidelberg, Germany\label{aff110}
\and
Dipartimento di Fisica e Scienze della Terra, Universit\`a degli Studi di Ferrara, Via Giuseppe Saragat 1, 44122 Ferrara, Italy\label{aff111}
\and
Istituto Nazionale di Fisica Nucleare, Sezione di Ferrara, Via Giuseppe Saragat 1, 44122 Ferrara, Italy\label{aff112}
\and
ICSC - Centro Nazionale di Ricerca in High Performance Computing, Big Data e Quantum Computing, Via Magnanelli 2, Bologna, Italy\label{aff113}
\and
INAF, Istituto di Radioastronomia, Via Piero Gobetti 101, 40129 Bologna, Italy\label{aff114}
\and
Astronomical Observatory of the Autonomous Region of the Aosta Valley (OAVdA), Loc. Lignan 39, I-11020, Nus (Aosta Valley), Italy\label{aff115}
\and
Universit\'e C\^{o}te d'Azur, Observatoire de la C\^{o}te d'Azur, CNRS, Laboratoire Lagrange, Bd de l'Observatoire, CS 34229, 06304 Nice cedex 4, France\label{aff116}
\and
Department of Mathematical Physics, Institute of Physics, University of S\~ao Paulo, R. do Mat\~ao 1371, 05508-090, S\~ao Paulo, SP, Brazil\label{aff117}
\and
Universidad de La Laguna, Dpto. Astrof\'\i sica, E-38206 La Laguna, Tenerife, Spain\label{aff118}
\and
Oskar Klein Centre for Cosmoparticle Physics, Department of Physics, Stockholm University, Stockholm, SE-106 91, Sweden\label{aff119}
\and
Astrophysics Group, Blackett Laboratory, Imperial College London, London SW7 2AZ, UK\label{aff120}
\and
Univ. Grenoble Alpes, CNRS, Grenoble INP, LPSC-IN2P3, 53, Avenue des Martyrs, 38000 Grenoble, France\label{aff121}
\and
Dipartimento di Fisica, Sapienza Universit\`a di Roma, Piazzale Aldo Moro 2, 00185 Roma, Italy\label{aff122}
\and
Institut d'Astrophysique de Paris, 98bis Boulevard Arago, 75014, Paris, France\label{aff123}
\and
ICL, Junia, Universit\'e Catholique de Lille, LITL, 59000 Lille, France\label{aff124}
\and
CERCA/ISO, Department of Physics, Case Western Reserve University, 10900 Euclid Avenue, Cleveland, OH 44106, USA\label{aff125}
\and
Laboratoire Univers et Th\'eorie, Observatoire de Paris, Universit\'e PSL, Universit\'e Paris Cit\'e, CNRS, 92190 Meudon, France\label{aff126}
\and
Departamento de F{\'\i}sica Fundamental. Universidad de Salamanca. Plaza de la Merced s/n. 37008 Salamanca, Spain\label{aff127}
\and
IRFU, CEA, Universit\'e Paris-Saclay 91191 Gif-sur-Yvette Cedex, France\label{aff128}
\and
Universiteit Gent, Department of Physics and Astronomy, Proeftuinstraat 86 N3, 9000 Ghent, Belgium
\label{aff129}
\and
Universit\'e de Strasbourg, CNRS, Observatoire astronomique de Strasbourg, UMR 7550, 67000 Strasbourg, France\label{aff130}
\and
Center for Data-Driven Discovery, Kavli IPMU (WPI), UTIAS, The University of Tokyo, Kashiwa, Chiba 277-8583, Japan\label{aff131}
\and
Max-Planck-Institut f\"ur Physik, Boltzmannstr. 8, 85748 Garching, Germany\label{aff132}
\and
California Institute of Technology, 1200 E California Blvd, Pasadena, CA 91125, USA\label{aff133}
\and
Institute of Cosmology and Gravitation, University of Portsmouth, Portsmouth PO1 3FX, UK\label{aff134}
\and
Department of Physics \& Astronomy, University of California Irvine, Irvine CA 92697, USA\label{aff135}
\and
Departamento F\'isica Aplicada, Universidad Polit\'ecnica de Cartagena, Campus Muralla del Mar, 30202 Cartagena, Murcia, Spain\label{aff136}
\and
Instituto de F\'isica de Cantabria, Edificio Juan Jord\'a, Avenida de los Castros, 39005 Santander, Spain\label{aff137}
\and
Department of Mathematics and Physics E. De Giorgi, University of Salento, Via per Arnesano, CP-I93, 73100, Lecce, Italy\label{aff138}
\and
INFN, Sezione di Lecce, Via per Arnesano, CP-193, 73100, Lecce, Italy\label{aff139}
\and
INAF-Sezione di Lecce, c/o Dipartimento Matematica e Fisica, Via per Arnesano, 73100, Lecce, Italy\label{aff140}
\and
Institut d'Astrophysique de Paris, UMR 7095, CNRS, and Sorbonne Universit\'e, 98 bis boulevard Arago, 75014 Paris, France\label{aff141}
\and
Department of Computer Science, Aalto University, PO Box 15400, Espoo, FI-00 076, Finland\label{aff142}
\and
School of Physics and Astronomy, University of Nottingham, University Park, Nottingham NG7 2RD, UK\label{aff143}
\and
DARK, Niels Bohr Institute, University of Copenhagen, Jagtvej 155, 2200 Copenhagen, Denmark\label{aff144}
\and
Department of Physics, Centre for Extragalactic Astronomy, Durham University, South Road, Durham, DH1 3LE, UK\label{aff145}
\and
Astrophysics Research Centre, University of KwaZulu-Natal, Westville Campus, Durban 4041, South Africa\label{aff146}
\and
School of Mathematics, Statistics \& Computer Science, University of KwaZulu-Natal, Westville Campus, Durban 4041, South Africa\label{aff147}
\and
Department of Physics and Astronomy, Vesilinnantie 5, University of Turku, 20014 Turku, Finland\label{aff148}
\and
Finnish Centre for Astronomy with ESO (FINCA), Quantum, Vesilinnantie 5, University of Turku, 20014 Turku, Finland\label{aff149}
\and
Serco for European Space Agency (ESA), Camino bajo del Castillo, s/n, Urbanizacion Villafranca del Castillo, Villanueva de la Ca\~nada, 28692 Madrid, Spain\label{aff150}
\and
ARC Centre of Excellence for Dark Matter Particle Physics, Melbourne, Australia\label{aff151}
\and
Centre for Astrophysics \& Supercomputing, Swinburne University of Technology,  Hawthorn, Victoria 3122, Australia\label{aff152}
\and
Department of Physics and Astronomy, University of the Western Cape, Bellville, Cape Town, 7535, South Africa\label{aff153}
\and
MTA-CSFK Lend\"ulet Large-Scale Structure Research Group, Konkoly-Thege Mikl\'os \'ut 15-17, H-1121 Budapest, Hungary\label{aff154}
\and
Konkoly Observatory, HUN-REN CSFK, MTA Centre of Excellence, Budapest, Konkoly Thege Mikl\'os {\'u}t 15-17. H-1121, Hungary\label{aff155}
\and
Brazilian Center for Research in Physics (CBPF), Dr. Xavier Sigaud st. 150, zip 22290-180, Rio de Janeiro, RJ, Brazil\label{aff156}
\and
Institute for Theoretical Particle Physics and Cosmology (TTK), RWTH Aachen University, 52056 Aachen, Germany\label{aff157}
\and
INAF-Osservatorio Astrofisico di Arcetri, Largo E. Fermi 5, 50125, Firenze, Italy\label{aff158}
\and
Centro de Astrof\'{\i}sica da Universidade do Porto, Rua das Estrelas, 4150-762 Porto, Portugal\label{aff159}
\and
Instituto de Astrof\'isica e Ci\^encias do Espa\c{c}o, Universidade do Porto, CAUP, Rua das Estrelas, PT4150-762 Porto, Portugal\label{aff160}
\and
HE Space for European Space Agency (ESA), Camino bajo del Castillo, s/n, Urbanizacion Villafranca del Castillo, Villanueva de la Ca\~nada, 28692 Madrid, Spain\label{aff161}
\and
STAR Institute, University of Li{\`e}ge, Quartier Agora, All\'ee du six Ao\^ut 19c, 4000 Li\`ege, Belgium\label{aff162}
\and
Instituto de Astrof\'isica e Ci\^encias do Espa\c{c}o, Faculdade de Ci\^encias, Universidade de Lisboa, Campo Grande, 1749-016 Lisboa, Portugal\label{aff163}
\and
Department of Astrophysics, University of Zurich, Winterthurerstrasse 190, 8057 Zurich, Switzerland\label{aff164}
\and
Department of Mathematics and Physics, Roma Tre University, Via della Vasca Navale 84, 00146 Rome, Italy\label{aff165}
\and
Theoretical astrophysics, Department of Physics and Astronomy, Uppsala University, Box 524, 751 20 Uppsala, Sweden\label{aff166}
\and
Mathematical Institute, University of Leiden, Einsteinweg 55, 2333 CA Leiden, The Netherlands\label{aff167}
\and
School of Physical Sciences, The Open University, Milton Keynes, MK7 6AA, UK\label{aff168}
\and
School of Physics \& Astronomy, University of Southampton, Highfield Campus, Southampton SO17 1BJ, UK\label{aff169}
\and
Institute of Astronomy, University of Cambridge, Madingley Road, Cambridge CB3 0HA, UK\label{aff170}
\and
Department of Physics and Astronomy, University of California, Davis, CA 95616, USA\label{aff171}
\and
Center for Astrophysics and Cosmology, University of Nova Gorica, Nova Gorica, Slovenia\label{aff172}
\and
Kobayashi-Maskawa Institute for the Origin of Particles and the Universe, Nagoya University, Chikusa-ku, Nagoya, 464-8602, Japan\label{aff173}
\and
Institute for Particle Physics and Astrophysics, Dept. of Physics, ETH Zurich, Wolfgang-Pauli-Strasse 27, 8093 Zurich, Switzerland\label{aff174}
\and
Department of Astrophysical Sciences, Peyton Hall, Princeton University, Princeton, NJ 08544, USA\label{aff175}
\and
Fakult\"at f\"ur Physik, Universit\"at Bielefeld, Postfach 100131, 33501 Bielefeld, Germany\label{aff176}
\and
School of Mathematics, Statistics and Physics, Newcastle University, Herschel Building, Newcastle-upon-Tyne, NE1 7RU, UK\label{aff177}
\and
Space physics and astronomy research unit, University of Oulu, Pentti Kaiteran katu 1, FI-90014 Oulu, Finland\label{aff178}
\and
SRON Netherlands Institute for Space Research, Landleven 12, 9747 AD, Groningen, The Netherlands\label{aff179}
\and
MIT Kavli Institute for Astrophysics and Space Research, Massachusetts Institute of Technology, Cambridge, MA 02139, USA\label{aff180}}    

\abstract{\Euclid is a European Space Agency mission mapping the extragalactic sky, aiming to provide the most-accurate 3D view of the Universe at optical and near-infrared wavelengths, by measuring the photometric redshifts of billions of galaxies. Within the Euclid Consortium, this task is executed by the PHZ Organisation Unit. PHZ takes as input the multi-wavelength photometry of \Euclid and ground-based data, and returns as output the source classification, photometric redshifts for the \Euclid core science (i.e., cosmology), and photometric redshifts plus physical parameters and absolute magnitudes for non-cosmological science. As part of the first Euclid Data Release (DR1), we describe the details of the DR1 pipeline for non-cosmological science, along with the released data products, and their validation. The non-cosmological science data products are obtained with \nnpz, a nearest-neighbour algorithm developed within PHZ that infers the properties of each target galaxy from its closest matches, in colour--magnitude space, within a reference sample of galaxies with known redshifts and physical parameters. We built this reference sample from \Euclid observations of COSMOS, one of the Euclid Auxiliary Fields, labelling each galaxy with redshifts and physical parameters derived from the COSMOS2020 multi-wavelength photometry, consistently with the \Euclid core-science results. Combined with the adjustments and priors required to generalise the results across the entire extragalactic sky, this approach yields high-quality and internally consistent data products. We tested the performance using both internal metrics and comparisons to ancillary data sets (e.g., COSMOS2020, 3D-HST, and DESI) and scaling relations (e.g., the star-forming main sequence); the results are in line with expectations and pre-DR1 simulations for both the Euclid Wide Survey and Euclid Deep Fields.}


   \titlerunning{PHZ processing function in DR1 for physical properties of galaxies}
   \authorrunning{Euclid Collaboration: Enia et al.}
   
   \maketitle
%
   
\section{\label{sc:Intro}Introduction}
\Euclid is a European Space Agency mission designed to map the large-scale distribution of matter in the Universe. The spacecraft carries two instruments: the visible imaging 
instrument \citep[VIS,][]{EuclidSkyVIS} and the \gls{NIR} Spectrometer and Photometer 
\citep[NISP,][]{EuclidSkyNISP}, which together provide high-resolution optical imaging and \gls{NIR} imaging and spectroscopy over approximately $\num{11000}\,\mathrm{deg}^2$ of the extragalactic sky, the \gls{EWS}. At the end of the nominal mission, the \gls{EWS} will reach a uniform depth of $24.5$ AB magnitudes in the visible band and fainter than $24.0$ in the \gls{NIR} bands \citep{Scaramella-EP1}. Although \Euclid was primarily designed to constrain dark energy and dark matter through weak gravitational lensing and galaxy clustering, the resulting data set is a substantial resource for non-cosmological science as well \citep{EuclidSkyOverview}.

Achieving the mission's full scientific potential requires photometric redshifts -- or photo-$z$s -- for billions of galaxies. These will be used both to reconstruct the three-dimensional distribution of matter and to trace galaxy evolution across a wide range of cosmic time. To meet the required precision, \Euclid incorporates multi-band photometry from ground-based surveys, including the Dark Energy Survey \citep[DES,][]{2015AJ....150..150F, 2016MNRAS.460.1270D}, the Ultraviolet Near Infrared Optical Northern Survey \citep[UNIONS,][]{2025AJ....170..324G}, and, in future, the Legacy Survey of Space and Time \citep[LSST,][]{2008SerAJ.176....1I}.

The processing of these data is managed by the Euclid Science Ground Segment (SGS), organised into self-contained processing functions (PFs). The PHZ PF is responsible for photometric redshifts, \gls{pp} of galaxies, source classification, \gls{sed} estimation, and tomographic binning. The \gls{pp} -- stellar masses, \gls{sfr}, metallicities, and others -- are central to the non-cosmological science cases that aim to reconstruct the assembly history of the Universe. For this purpose, PHZ developed a machine-learning algorithm, the Nearest-Neighbour Photometric Redshifts \citep[\nnpz,][]{2018PASJ...70S...9T, EP-Enia, Q1-TP005, Q1-SP031}. \nnpz\ estimates multi-dimensional posterior distributions of photo-$z$s and \gls{pp} by searching, in colour--magnitude space, for the nearest neighbours of each target galaxy within a pre-defined reference sample of objects with known properties.
\begin{table*}[]
    \centering
    \caption{Filters for the Euclid DR1, with associated observed depths.}\label{tab:filters}
    \begin{tabular}{lccccccc}
        \hline
        \hline
        \noalign{\vskip 1pt}
        Band &  $\lambda_{\rm eff}$ [\micron] & EWS & EDF-N & EDF-F & EDF-S & PV-005 & CDFS \\
         &  & $\sim 1900\,\mathrm{deg}^2$ & $\sim 20\,\mathrm{deg}^2$ & $\sim 10\,\mathrm{deg}^2$ & $\sim 20\,\mathrm{deg}^2$  & $\sim 1\,\mathrm{deg}^2$ & $\sim 0.1\,\mathrm{deg}^2$  \\
        \hline
        \noalign{\vskip 1pt}
        $u_\sfont{CFHT/MegaCam}$ & $0.372$ & \phantom{0}$23.2$\tablefootmark{a} & $ $ & $ $ & $ $ & $ $ \\
        $r_\sfont{CFHT/MegaCam}$ & $0.640$ & $24.0$ & $    $ & $    $ & $    $  & $    $  & $    $ \\
        $i_\sfont{PAN-STARRS}$   & $0.755$ & $23.6$ & $    $ & $    $ & $    $  & $    $  & $    $ \\
        $g_\sfont{HSC}$          & $0.480$ & $24.5$ & $26.4$ & $26.4$ & $    $  & $27.2$  & $26.8$ \\
        $r_\sfont{HSC}$          & $0.640$ & $    $ & $26.1$ & $    $ & $    $  & $26.5$  & $    $ \\
        $r2_\sfont{HSC}$         & $0.640$ & $    $ & $26.0$ & $25.4$ & $    $  & $26.5$  & $26.6$ \\
        $i_\sfont{HSC}$          & $0.755$ & $    $ & $24.9$ & $25.0$ & $    $  & $26.4$  & $25.2$ \\
        $i2_\sfont{HSC}$         & $0.755$ & $    $ & $25.5$ & $25.7$ & $    $  & $24.8$  & $26.2$ \\
        $z_\sfont{HSC}$          & $0.891$ & $23.0$ & $24.9$ & $24.6$ & $    $  & $26.3$  & $25.0$ \\
        $g_\sfont{Decam}$        & $0.473$ & $24.7$ & $    $ & $26.1$ & $24.7$  & $26.6$  & $26.8$ \\
        $r_\sfont{Decam}$        & $0.642$ & $24.3$ & $    $ & $25.8$ & $24.3$  & $25.9$  & $26.9$ \\
        $i_\sfont{Decam}$        & $0.784$ & $23.8$ & $    $ & $25.5$ & $23.7$  & $25.7$  & $26.7$ \\
        $z_\sfont{Decam}$        & $0.926$ & $23.0$ & $    $ & $25.1$ & $23.0$  & $24.2$  & $26.3$ \\
        VIS/\IE                  & $0.715$ & $24.8$ & $26.2$ & $24.9$ & $25.8$  & $26.2$  & $26.4$ \\
        NISP/\YE                 & $1.081$ & $23.2$ & $24.5$ & $23.4$ & $24.3$  & $24.7$  & $24.8$ \\ 
        NISP/\JE                 & $1.367$ & $23.3$ & $24.7$ & $23.5$ & $24.4$  & $24.8$  & $25.0$ \\
        NISP/\HE                 & $1.771$ & $23.3$ & $24.7$ & $23.5$ & $24.4$  & $24.8$  & $25.0$ \\
        \hline
    \end{tabular}
    \tablefoot{Reported magnitudes are the $10\,\sigma$ observed median depths of the observing tiles for an extended source in a $2 \times \mathrm{FWHM}$ diameter aperture. As a rule-of-thumb, the $5\,\sigma$ median depths are obtained by adding 0.75 mag.\\
    \tablefoottext{a}{The $u_\sfont{CFHT/MegaCam}$ band for the \gls{EWS} in the northern sky is only available for roughly half of the \gls{EWS}-N, and not for the EDF-N.}}
\end{table*}

In this paper, we describe the PHZ pipeline for photometric redshifts and galaxy \gls{pp} used for the first $\sim 1900\,\mathrm{deg}^2$ of the \citet{DR1cite}. The accuracy and precision of the released data products are validated against large spectroscopic samples from external surveys, demonstrating their reliability for a broad range of astrophysical applications. This paper is structured as follows. In Sect.\,\ref{sc:Data}, we describe the DR1 data and the ancillary data sets used to evaluate performance. In Sect.\,\ref{sc:Pipeline}, we describe the PHZ pipeline for recovering photo-$z$s and \gls{pp} for non-cosmological science, and the products stored in the Euclid Archive System \citep[EAS,][]{DR1-TP001}. In Sect.\,\ref{sc:Validation}, we validate the results against spectroscopic redshifts and \ha-estimated SFRs, while checking the relation between stellar masses and SFR, the \gls{ms}. In Sect.\,\ref{sc:Conclusions}, we present our conclusions.

All the results described in this work assume a flat cosmological-constant-dominated cold dark matter ($\Lambda$CDM) cosmology with $H_0 = 67.74\,\kmsMpc$, $\Omega_{\rm m} = 0.3089$, and $\Omega_{\Lambda} = 0.6911$ \citep{2016A&A...594A..13P}, and assume a \citet{2003PASP..115..763C} initial mass function (IMF). All magnitudes are given in the AB photometric system \citep{1983ApJ...266..713O}.

\section{\label{sc:Data}Data}
A comprehensive view of all the scientific objectives of the \Euclid mission is presented in \cite{EuclidSkyOverview}. A detailed description of \gls{dr1} is available in \citet{DR1-TP001}; here we provide a brief summary.

Euclid \gls{dr1} is composed of approximately $1900\,\mathrm{deg}^2$ of the extragalactic sky, broadly divided into the \gls{EWS}, the \gls{EDF}, and the \gls{EAF}. The \gls{EDF} are composed of three fields: EDF-North (EDF-N), EDF-Fornax (EDF-F), and EDF-South (EDF-S), while the \gls{EAF} are a series of well-known extragalactic fields \citep[further details available in][]{DR1-TP001, DR1-TP027}.\footnote{See also \url{https://www.cosmos.esa.int/web/euclid/euclid-dr1-expected-coverage} for the exact sky coverage and position of the fields in \gls{dr1}.} \Euclid observed all the \gls{dr1} footprint in four photometric bands, one in the visible \citep[$\IE$,][]{EuclidSkyVIS}, and three in the \gls{NIR} \citep[$\YE$, $\JE$, and $\HE$,][]{EuclidSkyNISP}. \Euclid observations are complemented by ground-based observations carried out with multiple instruments to cover the wavelength range between $0.3\,\micron$ and $1.8\,\micron$ by the Ultraviolet Near-Infrared Optical Northern Survey \citep[UNIONS,][]{2025AJ....170..324G} in the northern hemisphere; in the southern hemisphere, \Euclid is complemented with observations made by the Dark Energy Survey \citep[DES,][]{2015AJ....150..150F, 2016MNRAS.460.1270D}. In the \gls{EDF} and \gls{EAF}, other deeper ground-based observations are available, namely the Hawaii $20\,\mathrm{deg}^2$ survey \citep[H20,][]{EP-McPartland}. Euclid \gls{dr1} will also contain \Euclid NISP spectroscopic redshifts and spectral line measurements \citep{DR1-TP023, DR1-TP024}.

While the \gls{EWS} covers the sky with a series of single reference observation sequences \citep[ROS, in Figure\,1 of][]{DR1-TP027}, in contiguous $\ang{;32} \times \ang{;32}$ “tiles”, the \gls{EAF} and \gls{EDF} have been observed multiple times with varying depth in tiles of $\ang{;17} \times \ang{;17}$: EDF-F has been observed two times, with an expected improvement in depth of $0.4$ magnitudes with respect to the \gls{EWS}; EDF-S ten times ($1\,\mathrm{mag}$ deeper), and EDF-N eleven times (the same). The depths of the different \gls{EAF} depend instead on different factors (e.g., zodiacal background). All the inputs for the PHZ pipeline come from the MER processing function, described in \citet{DR1-TP009}. Each catalogue contains all the objects detected in $\IE$ \citep{DR1-TP002} or in a NIR-stacked detection image that is the combination of $\YE$, $\JE$, and $\HE$ \citep{DR1-TP003}. There are $\sim 9000$ tiles in \gls{dr1}, with a total of $\num{880955708}$ unique detected sources, $\num{30809240}$ in the \gls{EDF} and $\num{850146468}$ in the \gls{EWS}.

For the photometry, for each detected source in a given tile, there are three main sets of measurements: aperture-, model-, and template-fitting photometry. Within PHZ, we are interested in the total emission of the source, while preserving the measurement of colours belonging to the same physical region of each galaxy. For VIS and NIR-stacked images we used the total flux computed within a Kron aperture with the A-PHOT algorithm \citep{2015A&A...582A..15M, 2016A&A...595A..97M}. For all the other bands, the total flux is obtained by scaling with a colour term,
\begin{equation}
    F_{\rm band}^{\rm total} = F_{\rm detection}^{\rm total} \times \frac{F_{\rm band}^{\rm aperture}}{F_{\rm detection}^{\rm aperture}}
\end{equation}
where $F^{\rm aperture}_{\rm band}$ is the aperture flux measured in the band of interest, and $F^{\rm aperture}_{\rm detection}$ is the aperture flux measured in the detection image, whether $I_{\rm E}$ or the NIR-stacked image. Both aperture fluxes are measured on images degraded to the worst resolution among the bands considered. As input to PHZ, we used the flux measured in a circular aperture of diameter $2 \times {\rm FWHM}$, converted to total flux. In Table\;\ref{tab:filters} we report the filters used in this work, with the observed $10\,\sigma$ depths for an extended source in an aperture whose diameter is twice the FWHM from the lowest resolution band between the optical and \Euclid ones \citep[further details on the photometry in][]{DR1-TP006, DR1-TP009, DR1-TP010}. 

\begin{figure*}
    \centering
    \includegraphics[width=\hsize]{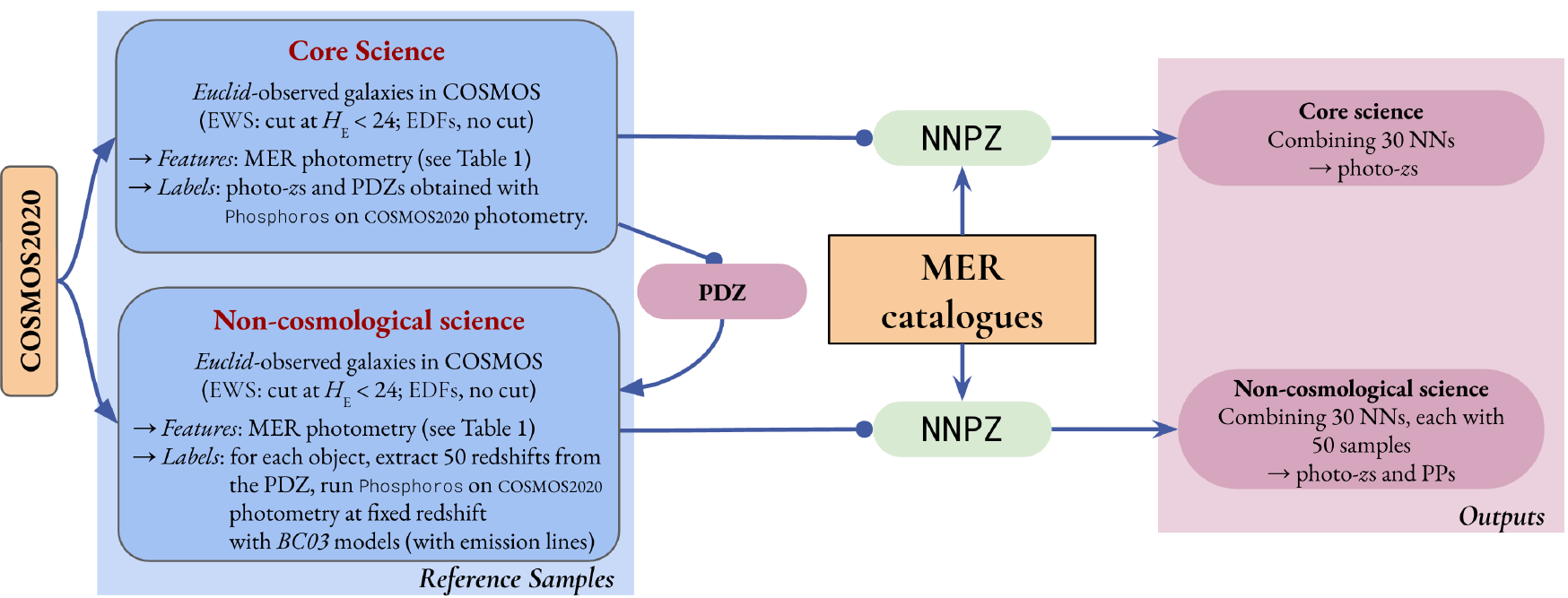}
    \caption{Schematic view of the PHZ pipeline for the core and non-cosmological science, where the \gls{pp} are measured. The pipeline inputs are within orange boxes, outputs are in pink, and specific tasks in green. Both reference samples are built running \phosphoros\ on COSMOS2020 photometry. The output PDZs obtained when finding the labels for the core-science reference sample are used as an input for building the labels for the non-cosmological science reference sample: $50$ redshifts are randomly extracted from the PDZ, and used to run \phosphoros\ at fixed redshift to infer the \gls{pp} and absolute magnitudes. After the \nnpz\ run, the 30 NNs are used to infer the outputs both for core-science (only photo-$z$s) and non-cosmological science, where each NN will have $50$ samples attached from the \phosphoros\ run described above. Further details available in \citet{DR1-TP011} and Sect.\,\ref{sc:Pipeline}.}
    \label{fig:pipeline2}
\end{figure*}

When reporting the results in Sect.\,\ref{sc:Validation}, we applied a series of quality cuts to make the analysis more robust, removing low-quality sources. These are:
\begin{itemize}
    \item \texttt{SPURIOUS$\_$FLAG} $= 0$;
    \item \texttt{DET\_QUALITY\_FLAG} $< 4$;
    \item \texttt{MUMAX\_MINUS\_MAG} $> -2.6$.
\end{itemize}
This selection removes low-quality objects from the sample, as well as stars and compact objects, e.g., QSOs. For details on the meaning of the flags, see \citet{DR1-TP009}. 

We further cleaned our sample looking for reliable extragalactic sources with both the PHZ source classification of the DR1 data products and the \nnpz\ flag on the recovered photo-$z$ and \gls{pp}, imposing:
\begin{itemize}
    \item \texttt{PHZ$\_$CLASSIFICATION} $= 2$;
    \item \texttt{PHYS$\_$PARAM$\_$FLAGS} $= 0$.
\end{itemize}
The first condition selects sources that have been classified as galaxies, while with the second we select only those sources with at least a nearest-neighbour in the reference sample with a weight higher than $10^{-3}$ \citep[see][and Sect.\,\ref{sc:Pipeline} for further details on the pipeline]{DR1-TP011}.

Additionally, we applied an $\HE$ magnitude cut that limits the signal-to-noise ratio (S/N, measured in an aperture of $2\times\,\mathrm{FWHM}$ diameter) to five. We warn the reader that the above described selection was chosen only for the validation of the \gls{dr1} products, and it is not intended as the blueprint for the different scientific studies that can be done with \gls{dr1} data. To give an example, for statistical studies that require a rigorous definition of the survey area, the \Euclid photometric visibility mask PhotoZ \citep[VMPZ, see][]{DR1-TP028} are better suited. We also note that the above selection, by requiring \texttt{PHZ\_CLASSIFICATION = 2} and removing compact, non-stellar sources (via \texttt{MUMAX\_MINUS\_MAG} cut), effectively restricts our validation sample to extended objects only. We therefore assumed that all extended sources are galaxies, ignoring the fact that some of them are in fact AGNs. The performance of the PHZ pipeline for AGN and QSO-classified objects is not discussed in this work.


In order to assess the quality of the photometric redshifts and \gls{pp} (e.g., stellar masses or SFRs), we compared them with the available literature compilations of spectroscopic redshifts and (when available) high-quality \gls{pp}, whether because they have been measured with a large number of photometric points or with spectral line information available. The \gls{EDF} are the part of the Euclid \gls{dr1} with the most coverage from other extragalactic spectroscopic surveys. A full list of references for high-quality spec-$z$ galaxies that matches with the \Euclid observations is available in \citet{DR1-TP011}, \citet{Q1-SP031}, or in Table\,5 of \citet{Q1-TP005}. We compared these with the results of the Deep Pipeline for the \gls{EDF} and of the single EWS-like passes of the same fields, in order to assess the quality of the results for the \gls{EWS}. Despite being the deepest, with ten ROSs, only a handful of reliable spectroscopic redshifts are available within the EDF-S; therefore, in the following sections we do not show for EDF-S the full set of tests that we performed for the \gls{EAF}, EDF-N, and EDF-F, and we limit our report of the EDF-S metrics to Table\,\ref{tab:phzpp_metrics}.

Given the technical details on how the PHZ results for non-cosmological science have been measured, a field of extreme importance is the Cosmic Evolution Survey (COSMOS). The latest release \citep[COSMOS2020,][]{2022ApJS..258...11W} was used within PHZ to build the reference samples \citep[described in Sect.\,\ref{sc:Pipeline} and][]{DR1-TP011}. In this work, we also used it to compare photo-$z$s and \gls{pp} with what can be seen as the best quality available stellar masses and SFRs, both for validating the reference sample, and as part of the Performance Verification 005 field (see Sect.\,\ref{sc:aux}), an EAF covering the innermost part of COSMOS. In COSMOS, we also compared our results with the compilation of available spectroscopic redshifts in \citet{2026ApJS..282....6K} with redshift quality $Q_{\rm f} = 4$. For the EAF Chandra Deep Field South (CDFS), we compared them with the 3D-HST compilation of spectroscopic redshifts and \gls{pp} \citep{2013arXiv1305.2140V, 2016ApJS..225...27M}.

Finally, for a limited sample of approximately $\num{5000}$ sources within $0 < z < 0.6$, we compared the measured SFRs with those reported in the value-added catalogue of \gls{pp} for the Dark Energy Spectroscopic Instrument \citep[DESI,][]{2016arXiv161100036D, 2024A&A...691A.308S}. Specifically, we used SFRs derived from both the \ha\ and \hb\ emission lines following \citet{1998ApJ...498..541K}. Although these estimates are not as reliable as spectroscopic redshifts for establishing the ‘ground truth’, they are sufficiently accurate to provide an indication of the quality of the recovered SFRs, at least in the South.

\section{\label{sc:Pipeline}OU-PHZ pipeline for physical parameters}
A more complete overview of the full OU-PHZ pipeline for \gls{dr1} is given in \citet{DR1-TP011}. In this section, we complement that with a specific focus on photometric redshifts and \gls{pp} for non-cosmological science, with galaxies being the main focus. In fact, as shown in Figure 1 of \citet{DR1-TP011}, the PHZ production pipeline is broadly divided into two branches: the core-science branch \citep[e.g., cosmology and weak-lensing, described in Sections 7 and 8 of][]{EuclidSkyOverview} and the non-cosmological science branch (e.g., extragalactic astrophysics, see Section 9 of the \Euclid overview paper). Both branches are intertwined, relying on the same inputs and -- for \gls{pp} -- from outputs of the core-science branch; as such, at least a short summary of both is needed to gain full comprehension of the PHZ data products for our purposes.
\begin{table}
\centering
\caption{Summary of all the PHZ pipeline output. The meaning of these parameters is described in Sect.\,\ref{sc:Pipeline} and in \citet{DR1-TP011}.}\label{tab:PHZout}
\renewcommand{\arraystretch}{1.2}
    \begin{tabular}{cp{4.8cm}}
    \hline
    \hline
    \noalign{\vskip 1pt}
    File Name & Description. \\
    \hline
    \noalign{\vskip 1pt}
    \texttt{PHZCAT} & Core science results: photo-$z$s and PDZs. \\
    \texttt{PPSAMPLINGINDEX} & Indices, weights, and scaling factors of the 30 NNs. \\
    \texttt{PPSAMPLINGFILE} & Non-cosmological science results: 50 samples from the combined \gls{pdf} of all the NNs. \\
    \texttt{PHYSPARAM} & Non-cosmological science results: median, mode, and percentiles of all the quantities in Table\,\ref{tab:DMDR1_PPs}. \\
    \texttt{PHYSPARAMNIR} & \phosphoros\ results on NISP-detected galaxies; only photo-$z$s and PDZs. \\
    \texttt{PHYSPARAMQSO} & \phosphoros\ results on QSO-classified objects; only photo-$z$s and PDZs. \\
    \texttt{CLASSCAT} & Results from the classification pipeline for galaxies. \\
    \texttt{STARCLASS} & Results from the classification pipeline for stars. \\
    \hline
    \end{tabular}
\end{table}
\begin{table}[t]
\centering
\caption{Summary of physical parameters provided in the Euclid \gls{dr1} PHZ pipeline output.}
\label{tab:DMDR1_PPs}
    \begin{tabular}{ll}
    \hline
    \hline
    \noalign{\vskip 1pt}
    Property & Units / Scale \\
    \hline
    \noalign{\vskip 1pt}
    Stellar Population & \\
    \texttt{STELLARMASS} & $\Mstarwun$ \\
    \texttt{SFR} & $\sfrwun$ \\
    \texttt{MASSFORMED} & $\Mstarwun$ \\
    \texttt{SFHAGE} & Gyr \\
    \texttt{SFHTIMESCALE} & Gyr \\
    \texttt{STELLARMETALLICITY} & $Z/Z_{\odot}$ \\
    \hline
    \noalign{\vskip 1pt}
    Dust & \\
    \texttt{AV} & mag \\
    \hline
    \noalign{\vskip 1pt}
    Absolute Magnitudes & \\
    \texttt{M\_VIS} & mag \\
    \texttt{M\_[FUV, NUV, U, B, V]} & mag \\
    \texttt{M\_[R, I, Z]} & mag \\
    \texttt{M\_[Y, J, H, K$_{\mathtt{S}}$]} & mag \\ 
    \hline
    \end{tabular}
\tablefoot{Each parameter includes associated median, mode, and $68\%$ confidence intervals derived from the \gls{pdf}.}
\end{table}

Given the large number of detected sources (see Sect.\,\ref{sc:Data}), \gls{ml} methods have been extensively tested and then used within PHZ to speed up the computational process, while achieving a performance comparable to template fitting methods \citep[see e.g.,][]{Desprez-EP10, EP-Enia}. As for the non-cosmological pipeline for \gls{q1}, the \gls{dr1} science data products were obtained using a \gls{nn} algorithm, called \nnpz. First described in \citet{2018PASJ...70S...9T}, and later developed within PHZ, \nnpz\ works by searching for each target galaxy a $k$ number of \gls{nn} in the features space (i.e., magnitude and colour) from a previously defined reference sample. In the \Euclid pipeline, the number of \gls{nn} was set to $30$, the optimal compromise between a reliable inference of photo-$z$s and \gls{pp}, and the computational cost of storing the \gls{nn} information. The \gls{nn} have their own weights -- measured from the $\chi^2$ distance between the reference and the target -- and scaling parameters from the \gls{sed} amplitude. For each target source, the $30$ neighbours are combined to infer its labels, i.e., the median (or the mode) of the (weighted) distribution of \gls{nn}; these are ultimately the inferred photo-$z$s, \gls{pp}, and absolute magnitudes of the target galaxies. The same approach is employed in the core-science branch of the PHZ pipeline, where the photometric redshifts are the lone labels attached to the reference sample, and thus the only quantity measured in the process. This is a key difference from what was done for \gls{q1}, when the core-science photometric redshifts were measured using \phosphoros\ \citep{Q1-TP005}.

Another difference from \gls{q1} is in the definition and construction of the reference sample. In \gls{q1}, the reference sample was built from a grid of templates and exponentially delayed star-formation histories (see Section\,6.1 of \citealt{Q1-TP005}, or Section\,3 of \citealt{Q1-SP031}). For \gls{dr1}, observations of the \gls{EAF} are available, thus making it possible to build the reference sample from a catalogue of real, observed galaxies, rather than a grid of templates. For both branches, we built the reference sample from \Euclid observations of COSMOS, one of the \gls{EAF}. There are in fact two different versions of the reference sample, depending on whether it is used for the processing of the \gls{EWS} or the processing for the \gls{EDF}. In practice, we built the samples by matching the Euclid \gls{dr1} observations of COSMOS with the COSMOS2020 catalogue -- including \Euclid, DES, and HSC bands -- cut in magnitude at a depth similar to that of the \Euclid data to which they were applied. For the \gls{EWS}, the reference sample is made of $\sim 375\,\mathrm{k}$ galaxies cut at $\HE < 24$ or $\IE < 25$; for the \gls{EDF}, the reference sample is made of $\sim 630\,\mathrm{k}$ galaxies, without any cut in magnitude (or, cut at the implicit UltraVISTA $H$-band limit in COSMOS).

In both cases, the reference sample labels (i.e., the photo-$z$s) were not taken directly from the COSMOS2020 catalogue, but were measured from a dedicated \phosphoros\ run on the available COSMOS2020 photometry \citep[from the Classic catalogue, see Table\,3 in][]{DR1-TP011}, ranging from $u$ to $K$ bands, and then cross-matched with the \Euclid MER catalogue in the same field, to attach the \Euclid observations as features in the reference sample. In building the core-science branch reference sample, the \phosphoros\ templates were obtained from an interpolation of 33 SED-models, including emission lines, looking for the best-fit photometric redshifts \citep[more details in][]{DR1-TP011}, with a volume prior applied. The outcome of the run is a reference sample whose features are the \Euclid, DES, UNIONS, and HSC bands (for the \gls{EDF}), and the \phosphoros-measured photo-$z$s as associated labels, and it is used in \nnpz\ to evaluate the core-science results.

We constructed the non-cosmological science reference sample as follows. In order to associate the \gls{pp} and absolute magnitudes with each reference source along with photo-$z$s, we drew $50$ random redshifts from the PDZ of the above mentioned \phosphoros\ run, and executed another dedicated run of \phosphoros\ at fixed redshift on the COSMOS2020 photometries -- this time also including the first two channels of IRAC, see Table\,3 in \citet{DR1-TP011} -- obtaining a set of \gls{pp} and absolute magnitudes. As such, each reference source has $50$ different realisations with their own photometric redshifts and correlated \gls{pp} and absolute magnitudes. This procedure ensures a high level of internal consistency between the core science products and the non‑cosmological science results. The models used in \phosphoros\ are a set of \citet[][2016 version\footnote{\url{http://www.bruzual.org/bc03/Updated_version_2016/}}]{2003MNRAS.344.1000B} templates using the MILES stellar library, with added emission lines; following what has been done in \citet{Q1-SP031}, and addressing some issues already found in \gls{q1} \citep[see Section\,6.1.1 of][]{Q1-TP005}, we removed models with unrealistically young ages -- $\log_{10}(T_0/\mathrm{yr}) < 8$ -- and implemented a complex grid of templates in steps of $\diff z = 0.01$, accounting for a redshift-dependent age prior.

One of the main setbacks of such an \gls{ml}-based approach is the risk of overfitting the available reference data: in \Euclid, it would mean replicating what is found in $2\,\deg^2$ of COSMOS on the whole $1900\,\deg^2$ of the extragalactic sky in \gls{dr1}. For example, the existence of large-scale structures in COSMOS must be smoothed out when searching for the redshifts of galaxies in the rest of the sky. It might also be possible that extremely massive galaxies not present in COSMOS with $\log_{10} (M_\ast/M_\odot) \geq 12.5$ \citep[e.g., Section\,10.3 in][]{EuclidSkyOverview} do actually exist in the whole $\sim \num{11000}\,\deg^2$ of the survey. Different measures were taken within PHZ to mitigate and eliminate these potential problems.

The first is a weight\footnote{This is another weight, not to be confounded with the $\chi^2$ weight assigned to the $30$ \gls{nn}, to which is it multiplied when inferring the photo-$z$s and \gls{pp}.} that is applied to all the objects in the reference sample, acting as a multiplicative prior for each NN. This slightly underweights some COSMOS-related redshift structures while favouring others. The weighting scheme is applied both for the core and the non-cosmological science part of the pipeline. Further details can be found in \citet{DR1-TP011}.
\begin{figure}
    \centering
    \includegraphics[width=\hsize]{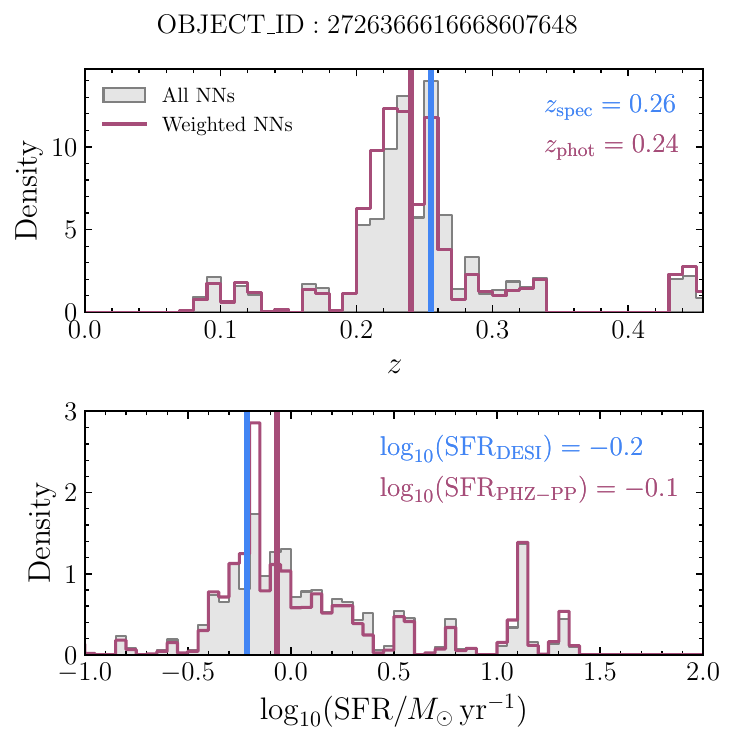}
    \caption{Example of the \nnpz\ output for a galaxy in EDF-N. The grey histogram is the distribution of all the NN, while the purple one is the one weighted with the product of the $\chi^2$-distance-inferred weight and the intrinsic one to smooth out cosmic structures in COSMOS (see Sect.\,\ref{sc:Pipeline}). In blue we report the DESI measured spectroscopic redshift and SFR obtained from observations of \ha\ and \hb\ emission lines, and in pink, the (median) PHZ inferred value.}
    \label{fig:out_ex}
\end{figure}

The second is a factor that accounts for the \gls{sed} amplitude scalings between the observed and reference sample colours. For each NN, there is an associated scaling factor, which enters the evaluation of \gls{pp}. For example, a galaxy with an important scaling between the reference sample's combined \gls{nn} $\HE$ magnitudes and its own observed $\HE$ magnitude will see its stellar mass properly scaled with respect to the one coming straight from the combination of the weighted \gls{nn}, allowing \nnpz\ to find objects with extremely high stellar masses that are not actually present in the reference sample. In \gls{q1}, the scaling factor was left free to vary, generating some troubling cases where wrong redshift attributions (particularly low-$z$ galaxies misplaced at high-$z$) artificially inflated the stellar masses to unphysical values $\Mstarwun > 14$. For \gls{dr1}, the allowed values for the scaling factor are not completely free, but follow a lognormal distribution, substantially decreasing the presence of such outliers.

The pipeline takes as input the tile-by-tile source catalogues generated by the MER PF. These include all objects detected in the $\IE$ images or in the NIR-stacked images combining the $\YE$, $\JE$, and $\HE$ data \citep[further details available in][and Sect.\,\ref{sc:Data} for how these have been ingested in PHZ]{DR1-TP009}. What is inside the final data product of the non-cosmological science PHZ pipeline is dictated by the Euclid \gls{dr1} Data Model (DMDR1).\footnote{Available here: \url{http://st-dm.pages.euclid-sgs.uk/data-product-doc/dmdr1/index.html}} The DMDR1 guarantees uniformity and compatibility among the different Level 3 (LE3) science products, strictly defining the content of the \gls{pp} catalogue. In addition to dictating the particular \gls{pp} to be estimated (e.g., stellar mass, SFR, and dust attenuation), it also specifies the statistical estimators needed for each, which are the median, mode, and 68\% confidence intervals obtained from the \gls{pdf}. These outputs are stored in all the files named \texttt{PHYSPARAM}. Also, files containing the indices of the reference sample \gls{nn}, weights, and scaling factor for each galaxy are available (\texttt{PPSAMPLINGINDEX}); finally, files with $50$ samples for each parameter, extracted from the full \gls{pdf}, are also available in the SAS (\texttt{PPSAMPLINGFILE}, only for the Deep Processing). A summary of all the PHZ release output files is given in Table\,\ref{tab:PHZout}, while an overview of the \gls{pp} released in compliance with the DMDR1 is summarised in Table\,\ref{tab:DMDR1_PPs}. Among the less immediately obvious parameters, \texttt{SFHAGE} refers to the age of the galaxy (i.e., the time elapsed since the onset of star formation), \texttt{SFHTIMESCALE} is the characteristic timescale of the star-formation history, usually identified as the $\tau$ parameter, and \texttt{MASSFORMED} is the total stellar mass ever formed. In Fig.\,\ref{fig:out_ex} we show an example of an \nnpz\ output for a galaxy in the EDF-N, limited to $z$ and $\mathrm{SFR}$. The distributions of the unweighted and weighted \gls{nn} are reported there, where we also report a comparison with the known $z_{\rm spec}$ and $\mathrm{SFR}$ from \ha\ and \hb\ measurements in DESI.
\begin{figure*}
    \centering
    \includegraphics[width=\hsize]{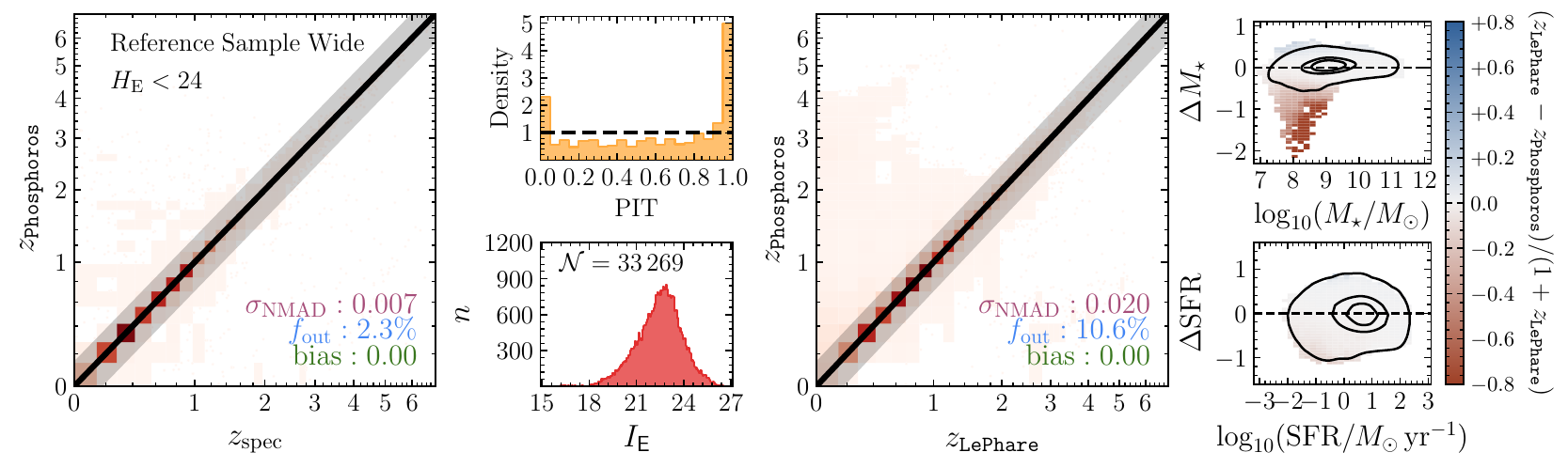}
    \includegraphics[width=\hsize]{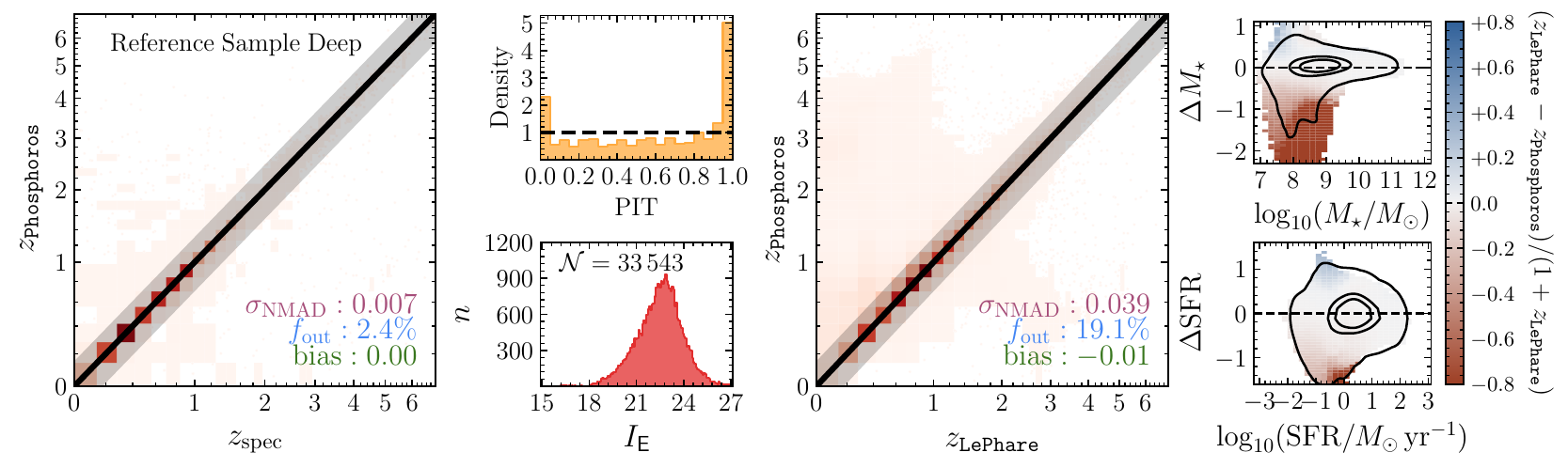}
    \caption{Comparison between the reference sample for the \gls{EWS} (first row) and \gls{EDF} (second row) used in \gls{dr1}. In the first column, the comparison is made with the subsample of reference sample sources that matched with the compilation of spectroscopic redshifts in COSMOS by \citet{2026ApJS..282....6K}. In the third column, the comparison is instead with the best-fit \texttt{LePhare} results for all the reference sample sources that matched with the COSMOS2020 Classic catalogue \citep{2022ApJS..258...11W}. The black line is the 1:1 relation, while the shaded grey area is the region beyond which a prediction is considered to be an outlier. The reported metrics are $\sigma_{\rm NMAD}$ (purple), the outlier fraction, $f_{\rm out}$ (blue), and the bias (green), all defined in Sect.\,\ref{sc:metrics}. The PIT (second columns) is measured as in Eq.~(\ref{eq:pit}) for redshifts. The fourth column shows the difference \texttt{LePhare} minus \texttt{Phosphoros} photometric redshifts, normalized to the \texttt{LePhare} photo-$z$s, while $\Delta M_\ast$ is the difference between the \texttt{LePhare} and \phosphoros\ best-fit measured logarithmic masses (see Sect.\,\ref{sc:Pipeline} for further details), and similarly for SFRs. The contours are the area containing $86\%$, $39\%$ (corresponding to the $2\sigma$ and $1\sigma$ levels for a 2D histogram), and $20\%$ of the sample.}
    \label{fig:RefSample_C2020}
\end{figure*}

\section{\label{sc:Validation}Validation}
In this section we describe and validate the \gls{dr1} data products. We do this by comparing with well-known, state-of-the-art ancillary data, and quantifying the performance using a common set of metrics.

\subsection{Metrics for quality assessment}\label{sc:metrics}
Following what has already been implemented in \citet{Q1-SP031}, to quantify the performance we define a uniform set of metrics, with some slight differences when referring to redshifts or \gls{pp}. Here we report their definitions and refer the reader to \citet{EP-Enia} for a full discussion on the chosen thresholds and the catastrophic outlier definition.

We define the normalised median absolute deviation as
\begin{equation}
{\rm NMAD} = 1.48\,\times\,{\rm median}
\begin{cases}
 \,\dfrac{|z_{\rm pred} - z_{\rm test}|}{1+z_{\rm test}} - b\:, & \text{for redshifts,} \\
 \,|y_{\rm pred} - y_{\rm test}| - b\:, & \text{for \gls{pp},}
\end{cases}
\end{equation}
with $b$ being the model bias (see below). The outlier fraction is defined as
\begin{equation}
f_{\rm out}\,=
    \begin{cases}
    \,\dfrac{|z_{\rm pred} - z_{\rm test}|}{1+z_{\rm test}} > 0.15\:, & \text{for redshifts,} \\
    \,|y_{\rm pred} - y_{\rm test}| > t_{\rm out}\:, & \text{for \gls{pp},} 
    \end{cases}
\end{equation}
with $t_{\rm out} = 0.4$ for stellar masses and $t_{\rm out} = 0.8$ for \gls{sfr}. We define the bias as
\begin{equation}
b\,={\rm median}
    \begin{cases}
    \,\left( \dfrac{z_{\rm pred} - z_{\rm test}}{1+z_{\rm test}} \right)\:, & \text{for redshifts,} \\
    \,(y_{\rm pred} - y_{\rm test})\:, & \text{for \gls{pp}.}
    \end{cases}
\end{equation}
In all of those cases, we compare a set of true values $z_{\rm test}$ and $y_{\rm test}$, with predicted values $z_{\rm pred}$ and $y_{\rm pred}$ (on a logarithmic scale for \gls{pp}).

Finally, we also quantify the information on the quality of the \gls{pdf} with the probability integral transform \citep[PIT,][]{10.2307/2981683, 2018A&A...609A.111D}. PIT measures the overall distribution of the cumulative distribution functions of a parameter $\theta$, $p(\theta)$, measured at its true values $\theta_{\rm true}$
\begin{equation}\label{eq:pit}
    {\rm PIT} = \int_{-\infty}^{\theta_{\rm true}} p(\theta) \, \diff \theta\:,
\end{equation}
where $\theta$ might be either photometric redshifts (in that case, $\theta_{\rm true} = z_{\rm spec}$) or a generic physical parameter (e.g., SFRs, with $\theta_{\rm true}$ being SFR measured from all the spectral information coming from the \ha\ and \hb\ emission lines). When these correctly reproduce the PDF of the sources, the PIT is flat. A concave, U-shaped PIT signals an underestimation of the width of the PDFs (and thus, the associated errors); the opposite, a convex-shaped PIT, signals an overestimation. Outliers appear as peaks (i.e., PIT equal to $0$ or $1$) at the edges of the distribution. Biases in the recovered quantities show up in the PIT as an asymmetry with respect to a flat shape.
\begin{figure*}
    \centering
    \includegraphics[width=\hsize]{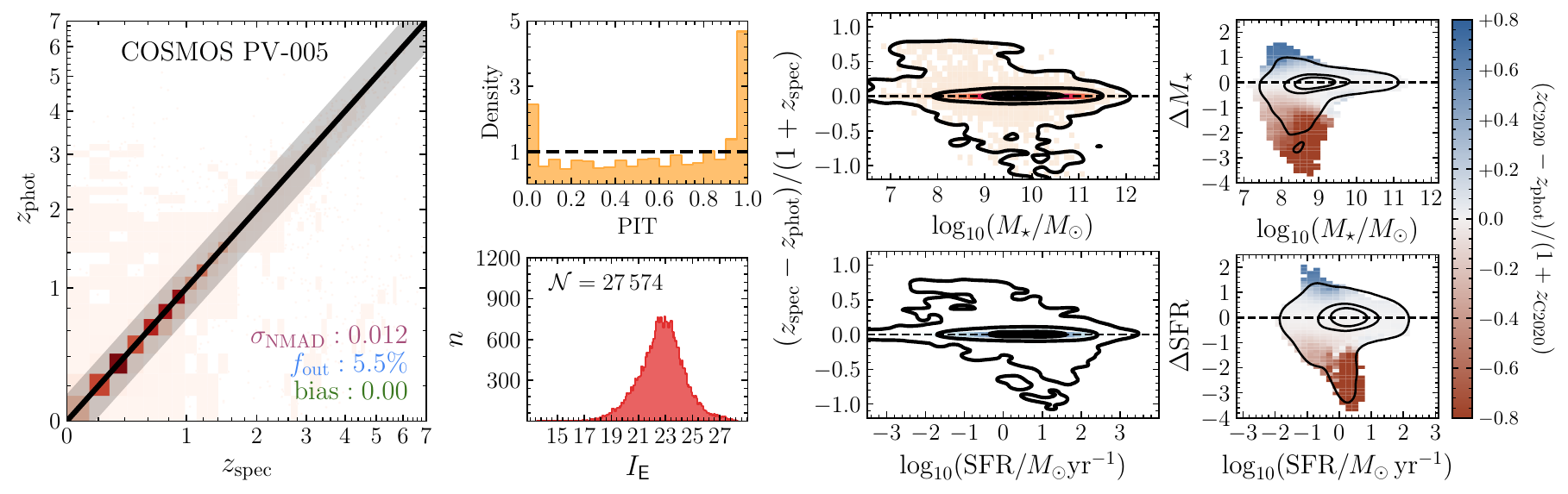}
    \includegraphics[width=\hsize]{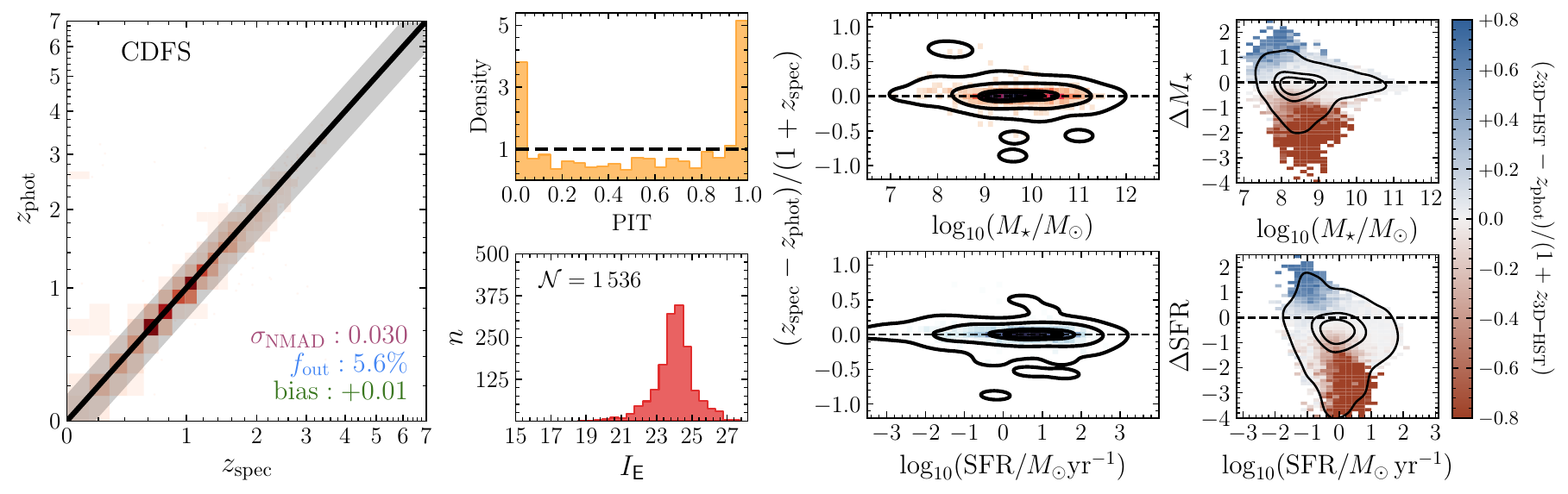}
        \caption{Results for the COSMOS PV-005 (first row) and CDFS (second row), two of the \gls{EAF}. \emph{First column}: comparison between the PHZ-PP measured photometric redshifts, and the compilation of spectroscopic redshifts in \citet{2026ApJS..282....6K} for COSMOS, and \citet{2012ApJ...758L..17B} for CDFS, with associated distribution of $\IE$ magnitudes (second column, bottom). Other plots and lines as in Fig.\,\ref{fig:RefSample_C2020}. \emph{Fourth column}: the difference between the \texttt{LePhare} best-fit measured masses in the COSMOS2020 Classic catalogue (first row) and one in 3D-HST (second row), and the \nnpz\ measured stellar masses, defined as $\Delta M_\ast$; similar for SFR. Everything is colour coded per the normalized difference between the parent catalogue (best) photometric redshift, and the \Euclid PHZ one. Contours are the same as in Fig.\,\ref{fig:RefSample_C2020}.}
    \label{fig:AuxFields}
\end{figure*}

\subsection{Reference sample and auxiliary fields}\label{sc:aux}
As a first necessary step to assess the quality of the PHZ data products, we must determine how well the reference sample was built, and how representative it is of the whole generic population. As reported in Sect.\,\ref{sc:Pipeline}, the reference sample for non-cosmological science was obtained from a \phosphoros\ run of all \Euclid observed galaxies in COSMOS2020 with BC03 models, for each galaxy at $50$ fixed redshifts, which are extracted from the PDZ of the \phosphoros\ run made to construct the core-science reference sample. These labels were then associated with each reference sample object, having the \Euclid observations of the field as features. There are two reference samples, one for the \gls{EDF}, with \num{628281} galaxies, and one for the \gls{EWS}, obtained from the former with a cut in $\HE < 24$ or $\IE < 25$,  with \num{374829} galaxies.

COSMOS is one of the most well-known fields in the extragalactic sky, with a considerable amount of multi-wavelength observations able to constrain photometric redshifts and \gls{pp} at the state-of-the-art level. We compared the labels we obtained for the reference sample with the compilation of spectroscopic redshifts in COSMOS by \citet{2026ApJS..282....6K} -- the unique sources catalogue -- as ground-truth for the redshifts, and the best-fit \texttt{LePhare} results from the COSMOS2020 Classic catalogue by \citet{2022ApJS..258...11W}, as the best available comparison for stellar masses and SFRs. As already reported in Sect.\,\ref{sc:Data}, we selected only sources with redshift quality $Q_{\rm f} = 4$. We caution that the spectroscopic samples used for validation are not representative of the full photometric sample: this quality flag cut preferentially retains secure redshifts for bright sources or sources with a high S/N, for which photometric redshift recovery is more reliable. Thus, the reported $\sigma_{\rm NMAD}$ and outlier fractions are likely mild underestimates of the true values for the full survey population, particularly at the faint end. This effect is reported in Table\,\ref{tab:phzpp_metrics}, where the degradation of performance with increasing $I_{\rm E}$ magnitude is evident.

The results are presented in Fig.\,\ref{fig:RefSample_C2020}. In the top panels, we report the results for the reference sample for the \gls{EWS}; in the bottom panel, the one for the \gls{EDF}. The leftmost panels always report the comparison between the \phosphoros-recovered photometric redshifts and the spectroscopic ones, with $\num{33269}$ matches for the \gls{EWS} and $\num{33543}$ for the \gls{EDF}. In both cases, the metrics are almost the same -- namely $\sigma_{\rm NMAD} = 0.007$, outlier fraction around $2.3\%$, no bias -- with an almost negligible difference of $0.1\%$ in the outlier fraction between the \gls{EWS} and the \gls{EDF}, a consequence of the magnitude cut removing a small fraction (about $200$) of the fainter objects with an incorrect redshift attribution. The PIT distributions, shown in the centre left panel of Fig.\,\ref{fig:RefSample_C2020}, are broadly flat for both fields, indicating that the photometric redshift PDFs are well-calibrated, with no significant signs of over or underestimation of the width of the PDFs, and only a mild excess at the edges of the distribution tracing the small fraction of catastrophic outliers already captured by $f_\mathrm{out}$.

In the right panels, we show a direct comparison with the COSMOS2020 catalogue, specifically the \texttt{LePhare} redshifts, where both axes show photo-$z$s estimates. The difference between \gls{EWS} and \gls{EDF} is expected, given the magnitude cut at $\HE < 24$ for the former, which reduces the number of objects almost by half from $\num{628281}$ to $\num{374829}$, removing mostly fainter galaxies, which are more difficult to place in redshift. The rightmost panels highlight the difference between the \texttt{LePhare}'s estimated stellar masses and SFRs and the same quantities in the reference sample (there called $\Delta M_\ast$ and $\Delta \mathrm{SFR}$), colour-coded per the median normalised redshift difference in the bin $(z_{\rm spec} - z_{\rm phot}) / (1 + z_{\rm spec})$. As the contours show, the vast majority ($> 95\%$) of the reference sample has stellar masses and SFRs in good accordance with the ones from the COSMOS2020 catalogue, within $0.2\mbox{--}0.4\,\mathrm{dex}$ for the stellar masses and $1\,\mathrm{dex}$ for SFRs. These checks assure us about the quality of the reference sample and its use in recovering the photo-$z$s and \gls{pp} for the \Euclid non-cosmological science results.
\begin{table*}[t]
\centering
\caption{The quality of the recovered photo-$z$s in the PHZ-PP pipeline, overall and in bins of \IE magnitude.}\label{tab:phzpp_metrics}
    \begin{tabular}{llcccc}
    \hline
    \hline
    \noalign{\vskip 1pt}
    Field & Metric & All & $15 < \IE \leq 21.5$ & $21.5 < \IE \leq 23$ & $23 < \IE \leq 25$ \\
    \hline
    \noalign{\vskip 1pt}
    \multirow{4}{*}{EDF-N} & $\mathcal{N}$ & \num{19291} & 9510 & 5884 & 3756 \\
     & $\sigma_{\rm NMAD}$ & 0.022 & 0.023 & 0.018 & 0.023 \\
     & $f_{\rm out}$ (\%) & 2.8 & 1.6 & 2.5 & 5.5 \\
     & bias & 0.00 & 0.00 & 0.00 & 0.00 \\
    \hline
    \multirow{4}{*}{EDF-S} & $\mathcal{N}$ & 74 & 49 & 25 & 0 \\
     & $\sigma_{\rm NMAD}$ & 0.023 & 0.020 & 0.027 & \dots \\
     & $f_{\rm out}$ (\%) & 6.8 & 6.1 & 8.0 & \dots \\
     & bias & +0.01 & +0.01 & +0.01 & \dots \\
    \hline
    \multirow{4}{*}{EDF-F} & $\mathcal{N}$ & \num{52582} & \num{23790} & \num{20966} & 6274 \\
     & $\sigma_{\rm NMAD}$ & 0.025 & 0.025 & 0.024 & 0.036 \\
     & $f_{\rm out}$ (\%) & 8.0 & 5.6 & 7.6 & 14.2 \\
     & bias & 0.00 & 0.00 & 0.00 & 0.00 \\
    \hline
    \multirow{4}{*}{CDFS} & $\mathcal{N}$ & 1536 & 43 & 182 & 1097 \\
     & $\sigma_{\rm NMAD}$ & 0.030 & 0.016 & 0.016 & 0.030 \\
     & $f_{\rm out}$ (\%) & 5.6 & 0.0 & 2.7 & 5.4 \\
     & bias & +0.01 & 0.00 & +0.01 & +0.01 \\
    \hline
    \multirow{4}{*}{COSMOS PV-005} & $\mathcal{N}$ & \num{27574} & 6013 & 9823 & \num{10186} \\
     & $\sigma_{\rm NMAD}$ & 0.012 & 0.010 & 0.009 & 0.015 \\
     & $f_{\rm out}$ (\%) & 5.5 & 1.8 & 2.6 & 6.6 \\
     & bias & 0.00 & 0.00 & 0.00 & 0.00 \\
    \hline
    \multirow{4}{*}{EWS-N} & $N$ & \num{19291} & 9510 & 5884 & 3756 \\
     & $\sigma_{\rm NMAD}$ & 0.021 & 0.017 & 0.022 & 0.034 \\
     & $f_{\rm out}$ (\%) & 3.7 & 0.4 & 3.8 & 9.4 \\
     & bias & +0.01 & +0.01 & +0.01 & 0.00 \\
    \hline
    \multirow{4}{*}{EWS-S} & $N$ & \num{52582} & \num{23790} & \num{20966} & 6274 \\
     & $\sigma_{\rm NMAD}$ & 0.027 & 0.024 & 0.025 & 0.048 \\
     & $f_{\rm out}$ (\%) & 7.6 & 4.4 & 7.2 & 16.2 \\
     & bias & 0.00 & 0.00 & +0.01 & 0.00 \\
    \hline
    \end{tabular}
\end{table*}

As a first test of the pipeline results, we compared them with observations of the \gls{EAF}, in particular the innermost part of COSMOS (called the Performance Verification Phase $5$ field, or PV-005) and the CDFS (a sub-field within the EDF-F). These fields have been observed multiple times, more than the \gls{EWS} or \gls{EDF} (see Table\,\ref{tab:filters}), which should result in better recovered photo-$z$s and \gls{pp}. In both cases we used ancillary data to quantify the goodness of the pipeline results, i.e., the previously reported COSMOS2020 catalogues \citep{2022ApJS..258...11W, 2026ApJS..282....6K} and the 3D-HST ones \citep{2013arXiv1305.2140V, 2016ApJS..225...27M}. As the main focus of this work are galaxies, we further cleaned the samples by removing those sources flagged as X-ray detected.
\begin{figure*}
    \centering
    \includegraphics[width=\hsize]{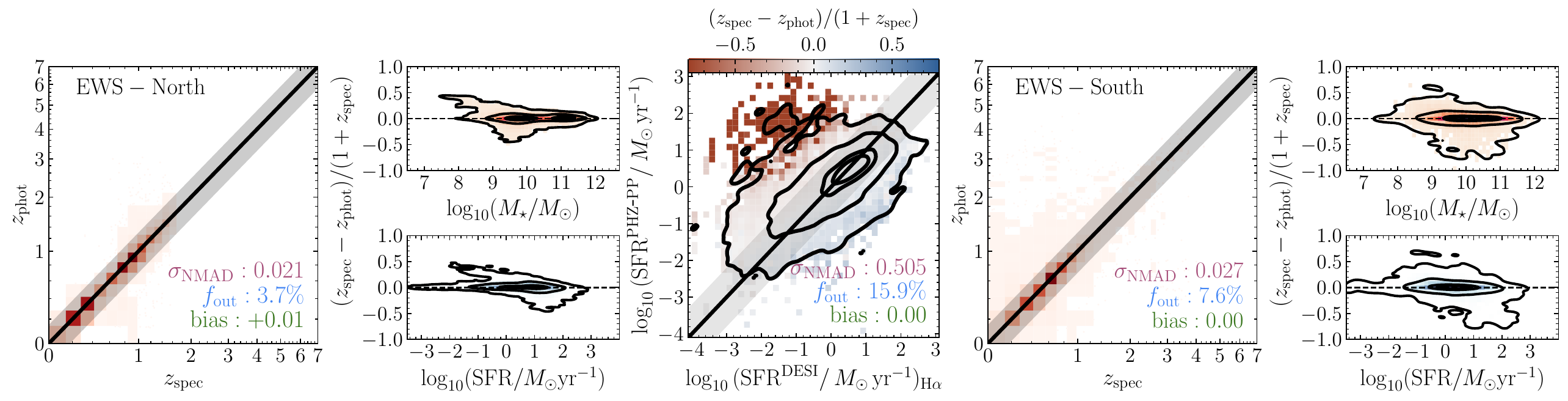}
    \caption{Results for the EWS-like processing of EDF-N (first, second, and third column), which can be considered as a proxy for the \gls{EWS} in the northern hemisphere, and EDF-F (fourth and fifth column), which can be considered as a proxy for the \gls{EWS} in the southern hemisphere. Contours are similar to those in Fig.\,\ref{fig:RefSample_C2020}.}
    \label{fig:EWS_WideLike}
\end{figure*}

The results are reported in Fig.\,\ref{fig:AuxFields}. For these fields, we also applied the photometric, classification, and quality cuts reported in Sect.\,\ref{sc:Data}. We once again find reassuring results, with $\sigma_{\rm NMAD}$ of $0.012$ (PV-005, on $\num{27574}$ galaxies with spec-$z$) and $0.03$ (CDFS, on $\num{1536}$ galaxies with spec-$z$), and small outlier fractions of $5.5\%$ and $5.6\%$. The small bias of $+0.01$ in the CDFS is probably just a statistical fluctuation due to the small number and distribution of spec-$z$ objects. The centre-right panels show, for the spec-$z$ subsamples, the normalised redshift difference as a function of the distribution of stellar masses and SFRs. For Q1, as a consequence of the specifics of the method to infer results for non-cosmological science, similar plots showed how the non-negligible fraction of objects with unrealistically high SFRs ($10^{3\mbox{--}4}\,M_\odot\,{\rm yr^{-1}}$) were all due to a wrong redshift attribution \citep[see e.g. Figure~3 of][]{Q1-SP031}. In the new results, the impact of these outliers is limited, if not absent (the log-normal scaling factor in \nnpz\ reduces the chances of finding such extreme objects), and the resulting \gls{pp} are indeed better constrained and occupy a more reasonable region of the \gls{pp} space.
\begin{figure*}
    \centering
    \includegraphics[width=\hsize]{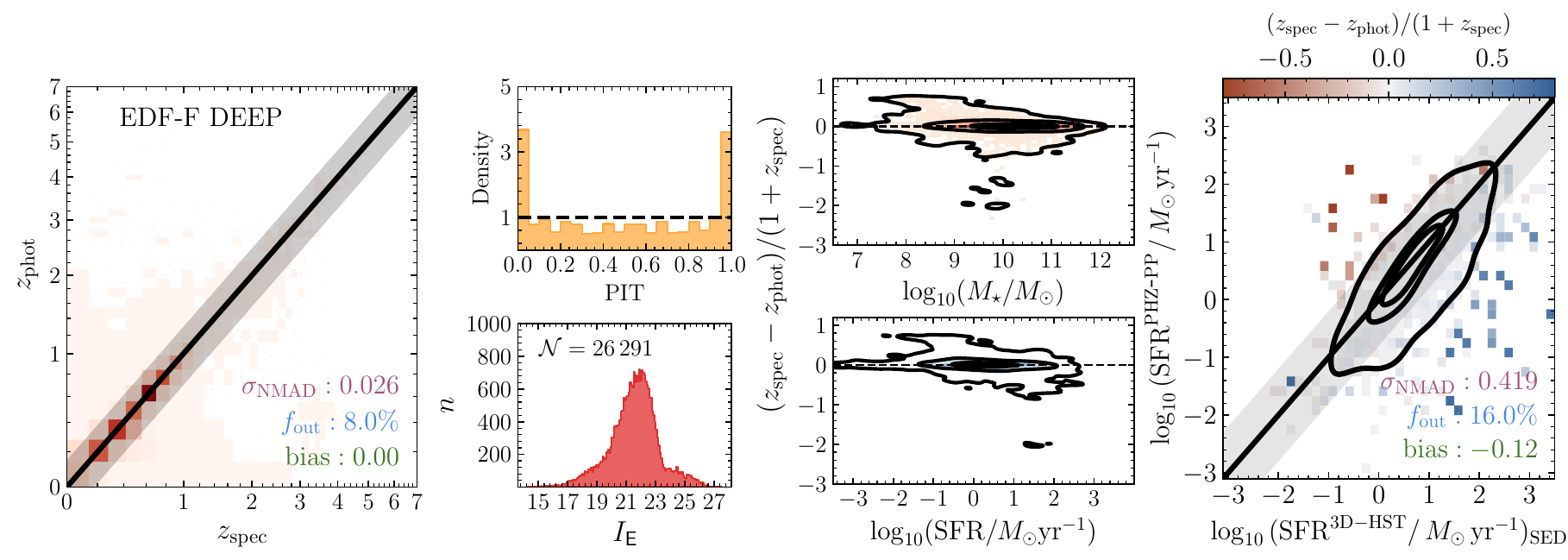}
    \includegraphics[width=\hsize]{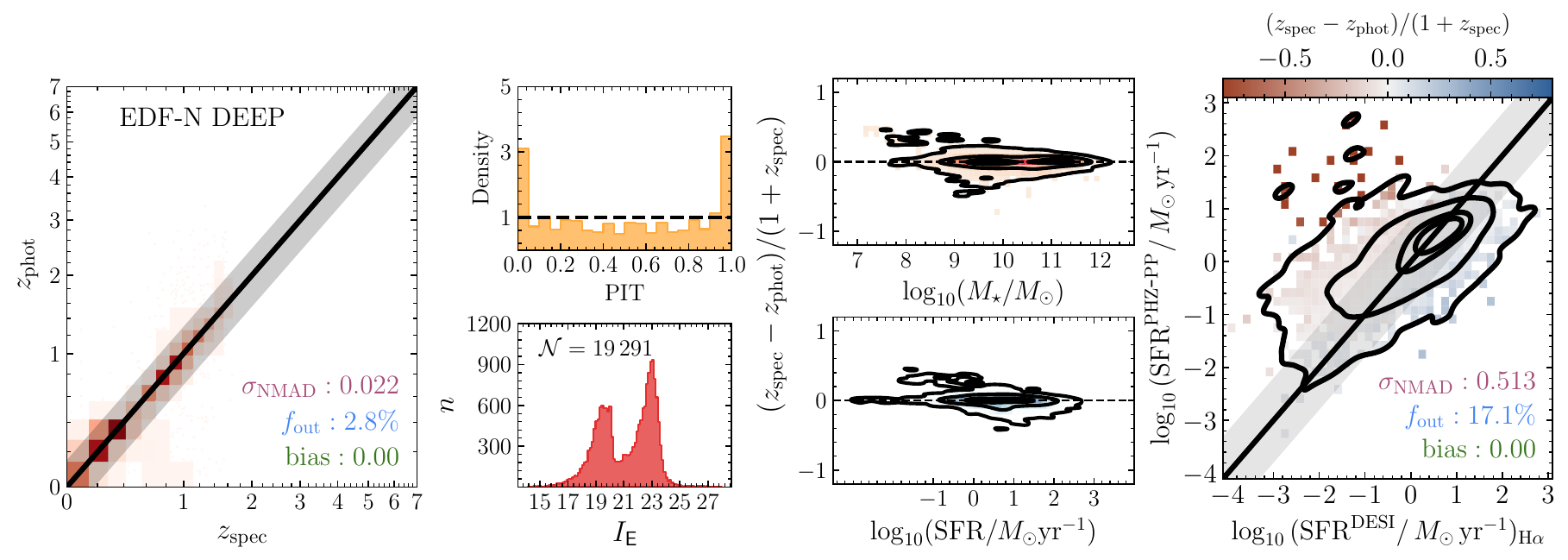}
    \caption{Results for the Deep processing of EDF-F (\emph{top panels}) and EDF-N (\emph{bottom panels}). \emph{Left}: comparison between the PHZ-PP measured photometric redshifts, and a compilation of spectroscopic redshifts (see Sect.\,\ref{sc:Data}). Other plots and lines as in Fig.\,\ref{fig:RefSample_C2020}. \emph{Right}: like the centre panel of Fig.\,\ref{fig:EWS_WideLike}; for EDF-F the comparison is with the SED-fitting inferred \gls{sfr} in GOODS-S in 3D-HST; for EDF-N, with the \ha\ and \hb\ detected lines in DESI.}
    \label{fig:DeepFields}
\end{figure*}

\subsection{Euclid Wide Survey}\label{sc:EWS}
While the \gls{EWS} covers the vast majority of the \Euclid observed extragalactic sky, most of the inference regarding the quality of the data products comes from the EWS-like processing of the fields with the most ancillary data for comparison, i.e., the EDF-N for the Northern hemisphere (\gls{EWS}-N), and the EDF-F for the Southern hemisphere (\gls{EWS}-S). The \gls{EDF} are obtained by stacking multiple ROSs on the same field (see Sect.\,\ref{sc:Data}, and also \citealt{DR1-TP001, DR1-TP009}); each single pass on the \gls{EDF} is effectively the same as an \gls{EWS} pass (thus EWS-like processing). Therefore, we inferred the quality of the data products for the \gls{EWS} by looking at the data quality of the single EWS-like processing of the EDF-F and EDF-N.

The results are shown in Fig.\,\ref{fig:EWS_WideLike}. The leftmost, centre-left, and centre panels are for the \gls{EWS}-N, while the centre-right and rightmost panel are for \gls{EWS}-S. The recovered photometric redshifts, in accord with the ones from the cosmological pipeline \citep{DR1-TP011}, show an important improvement with respect to the ones reported in Q1 \citep{Q1-TP005} -- $\sigma_{\rm NMAD}$ decreasing from about $0.03$ to $0.02$, and around $10\%$ fewer catastrophic outliers -- where the impossibility of building a reference sample from observations of true galaxies forced the use of a reference sample of synthetic models built from BC03 galaxies, with some issues especially in the recovery of the specific SFRs \citep[see Section\,6.1.1 of][]{Q1-TP005}.

Other than the available compilations of spectroscopic redshifts, we also compared the PHZ inferred SFRs with the ones coming from multiple observations of both \ha\ and \hb\ emission lines in DESI. These are available only for a limited number (around $\num{3000}$) of sources in the \gls{EWS}-N, but are the best available measurements to compare the SFRs to, and hence the closest possible ground truth values for this \gls{pp}. Accounting for how the SFRs were estimated -- that is, from broad-band photometry extending to $1.6\,\micron$ in the observed-frame -- the results are as expected. As reported in the centre panel of Fig.\,\ref{fig:EWS_WideLike}, a $\sigma_{\rm NMAD}$ of $0.505$ and an outlier fraction of $16\%$ (and no bias) are actually less than what was expected from numerical simulations \citep{EP-Enia}. The observed scatter with respect to the 1:1 relation is inherent due to the difficulties of inferring SFRs from the photometric data only and not just a consequence of a wrong redshift measurement; however, wrongly measured redshifts populate the vast majority of the catastrophic outliers found outside of a $0.8\,\mathrm{dex}$ cushion, especially from lower-$z$ galaxies misplaced to higher-$z$.

\subsection{Euclid Deep Fields}\label{sc:EDF}
There are three fields that compose the \gls{EDF}: the EDF-N ($20\,\mathrm{deg}^2$), near the Northern Ecliptic Pole; the EDF-F ($10\,\mathrm{deg}^2$), in the Fornax cluster which has been the subject of multiple multi-wavelength surveys in the past (e.g., GOODS-S); the EDF-S ($20\,\mathrm{deg}^2$), in an area of the Southern sky where the Galactic extinction is at its lowest level. While the plan is to observe all of them with (at least) $40$ ROSs, for DR1 there are eleven ROSs for EDF-N and ten for the EDF-S, and two ROSs for EDF-F.
\begin{figure*}
    \centering
    \includegraphics[width=\hsize]{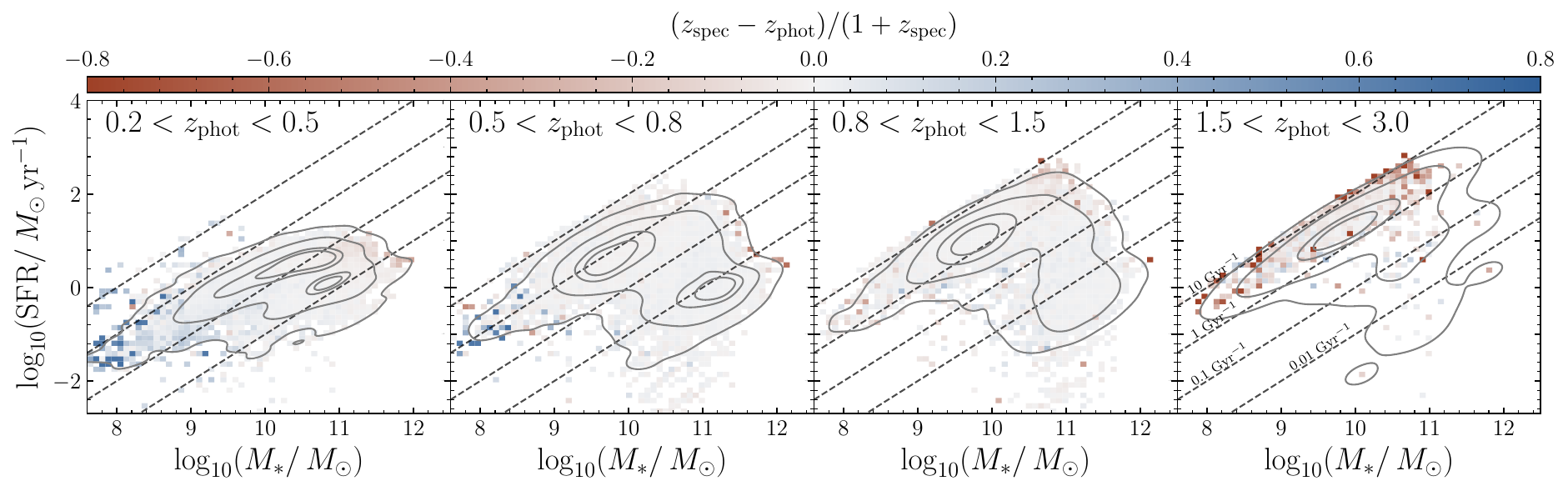}
    \includegraphics[width=\hsize]{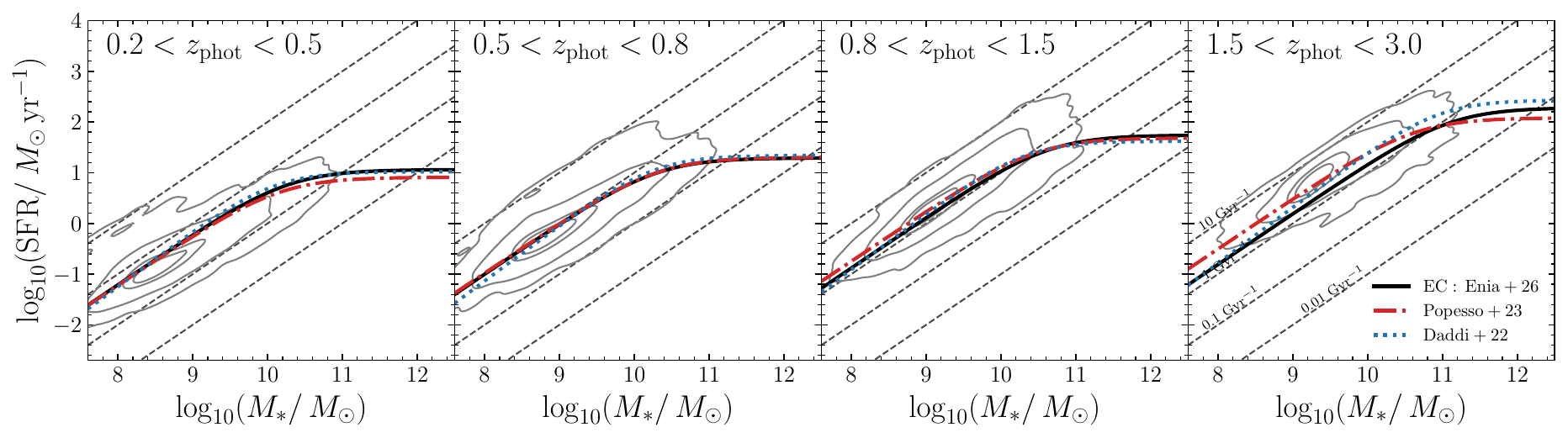}
    \caption{Insight into the \gls{ms} in the \gls{EDF}. \emph{Top}: $\mathrm{SFR}$--$M_\ast$ relation for the sample with available spectroscopic redshifts, colour-coded by the normalised redshift difference, in four different redshift bins. Blue squares highlight redshift underestimations, and the opposite for the red ones. \emph{Bottom}: $\mathrm{SFR}$--$M_\ast$ plot, compared with a compilation of published \gls{ms} determnations from \citet[][dashed black line]{Q1-SP031}, \citet[][dashed-dotted red line]{2023MNRAS.519.1526P}, and \citet[][dotted blue line]{2022A&A...661L...7D}, all evaluated at the median redshift of each redshift bin (or to the closest reported value). Dashed grey lines highlight loci of constant sSFR.}
    \label{fig:SFMS}
\end{figure*}

As reported in Sect.\,\ref{sc:EWS}, we used a compilation of ancillary data to infer the quality of the data products. The results are reported in Fig.\,\ref{fig:DeepFields}. The depth improvement with respect to the \gls{EWS} is reflected in the quality of the recovered photo-$z$s. For EDF-N, we find $\sigma_\mathrm{NMAD} = 0.022$ and an outlier fraction of $2.8\%$, with no bias, using a spectroscopic sample of $\num{19291}$ galaxies. Since EDF-F has been observed fewer times, the results are understandably poorer, with $\sigma_\mathrm{NMAD} = 0.026$ and an outlier fraction of $8.0\%$, for a larger spectroscopic sample of $\num{52582}$  galaxies; this is consistent with what is found for the EWS-like processing of the same field (see Sect.\,\ref{sc:EWS}) and reflects the larger diversity of galaxy types and redshift distributions sampled by the deeper observations, as well as the higher fraction of faint sources included in the spectroscopic comparison sample. For EDF-S, the very limited number of available spectroscopic redshifts ($74$ in total, with none fainter than $\IE = 23$, as reported in Sect.\,\ref{sc:Data}) prevents a thorough assessment of the pipeline performance; the metrics we recover -- $\sigma_\mathrm{NMAD} = 0.023$, $f_\mathrm{out} = 6.8\%$, and a positive bias of $+0.01$, almost negligible given the small number of spectroscopic redshifts available -- are nevertheless consistent with what is found for the other two \gls{EDF}. In all three fields, the PIT distributions are broadly flat, indicating well-calibrated photo-$z$ PDFs, with only minor signs of a slight overestimation of the PDF widths.

As in Sect.~\ref{sc:EWS}, we complemented the photo-$z$ validation with a comparison of the PHZ-inferred SFRs against external estimates. For EDF-F, we used the SED-fitting inferred SFRs from the 3D-HST catalogue in GOODS-S \citep{2016ApJS..225...27M}, while for EDF-N, we used the \ha\ and \hb-based SFRs from the DESI value-added catalogue \citep{2024A&A...691A.308S}. The results are shown in the rightmost panels of Fig.\,\ref{fig:DeepFields}. For EDF-F, we find $\sigma_\mathrm{NMAD} = 0.419$ and an outlier fraction of $16.0\%$, with a small negative bias of $-0.12$ dex with respect to the 3D-HST SED-fitting SFRs, while for EDF-N we find $\sigma_\mathrm{NMAD} = 0.513$, outlier fraction of $17.1\%$, and a bias $b = 0$ with respect to the DESI \ha\ measurements. These numbers are consistent with what is found for the EWS-like processing (Fig.\,\ref{fig:EWS_WideLike}), confirming that the dominant source of scatter in the recovered SFRs is the intrinsic difficulty of inferring this quantity from broad-band photometry alone, rather than a systematic effect introduced by the depth of the observations or the specific ancillary data set used for comparison. As already noted for the \gls{EWS}, the catastrophic outliers in the SFR comparison are predominantly associated with catastrophic failures in the photometric redshift recovery, with low-$z$ galaxies misplaced at higher redshift artificially boosting their inferred SFRs. We note that a complementary view of the \Euclid-inferred SFRs could be provided by the statistical comparison with far-infrared photometry from \textit{Herschel} and SCUBA-2 via stacking, as done in \citet{Q1-SP071}, which finds broadly consistent SFRs with respect to the \Euclid-inferred ones for \gls{ms} galaxies in Q1.
\begin{figure*}
    \centering
    \includegraphics[width=\hsize]{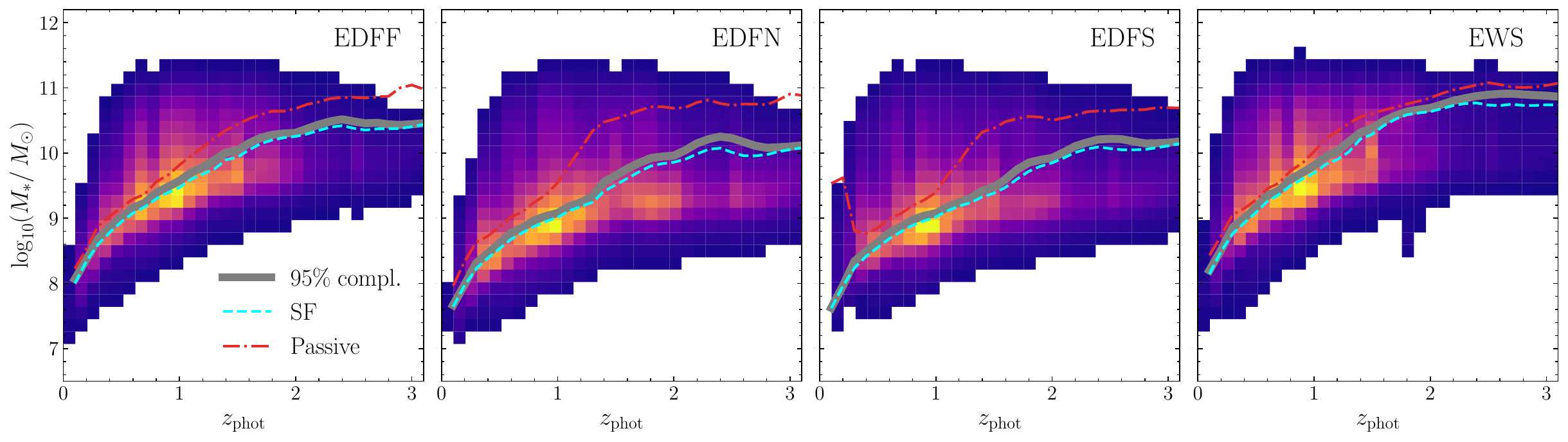}
    \caption{Stellar mass completeness for the \gls{dr1} \citep{2010A&A...523A..13P}, cut at each field depth in $\HE$ (Table\,\ref{tab:filters}), shown as the density of galaxies in the photometric redshifts versus stellar mass plane. The solid grey line is the $95\%$ stellar mass completeness limit, the red dashed-dotted line for passive galaxies, and the cyan dashed line for \gls{sfg}.}
    \label{fig:completeness}
\end{figure*}

For the sub-sample of sources with available spectroscopic redshifts, we tested the quality of the recovered PPs across different regions of the $M_*$--SFR plane as a function of the normalised redshift residual, in four redshift bins. We then extended the analysis to all sources in the EDFs satisfying the quality cuts of Sect.\,\ref{sc:Data}, comparing their distribution in the $M_*$--SFR plane with different published functional forms of the \gls{ms}. In both cases, we split the data into four redshift bins: $[0.2$, $0.5], [0.5$, $0.8], [0.8$, $1.5], [1.5$, $3.0]$. In the top panels of Fig.\,\ref{fig:SFMS}, we report the median distribution of spec-$z$ galaxies in bins within the $M_\ast$--SFR plane, coloured by their normalised redshift distribution $(z_{\rm spec} - z_{\rm phot})/(1 + z_{\rm spec})$. Ideally, when the colours start to diverge from white it means that the median normalised spectroscopic redshift difference in the bin is that of a catastrophic outlier (see definition in Sect.\,\ref{sc:Validation}). Red values indicate a redshift over-estimation, while blue values the opposite. The contours contain the area with $90\%$, $86\%$, $39\%$ (corresponding to the $3\,\sigma$, $2\,\sigma$, and $1\,\sigma$ levels for a 2D histogram), and $20\%$ of the sample. Thanks to the deeper data and especially the different definition of the reference sample and the scaling in \nnpz, the DR1 results show an important improvement with respect to the Q1 results, even in the case when the first two IRAC channels were added \citep[as visible in Figure\,4 in][]{Q1-SP031}. Where the majority of the (spectroscopic) sample lies, the quality of the recovered photometric redshift is extremely good, out to $z \sim 1.5$, still good out to $z \sim 3$, and with the signs of a (moderately) wrong redshift attribution only for high sSFRs.

In the lower panels, we show the $M_\ast$--SFR plane for the sample of \gls{sfg} selected with the $UVJ$ colours, in terms of contours highlighting the number density of the full sample of \gls{sfg} in DR1. There, we report three previously published \gls{ms} relations for comparison: Eq.\;(15) of \citet[][red dashed-dotted line]{2023MNRAS.519.1526P}, that is, a comprehensive compilation of 27 \gls{ms} relations from the literature fitted to the same functional form, in the mass range $8.5 < \Mstarwun < 11.5$; \citet[][dotted blue line]{2022A&A...661L...7D}, obtained from a stacking analysis of about $\num{400000}$ colour-selected \gls{sfg} in COSMOS \citep{2021A&A...647A.123D}, here also in the mass range $8.5 < \Mstarwun < 11.5$; and \citet[][black solid line]{Q1-SP031}, a fit of the \gls{ms} on the first Euclid Q1 data release, for more than eight million sources in the mass range $8.0 < \Mstarwun < 11.5$, and about $\num{35000}$ in the highest mass bin, which better constrains the bending of the \gls{ms}. We observe reasonable agreement between the \gls{ms} functional forms and the data, especially in the intermediate, more reliable redshift bins; going to higher redshifts, the DR1 results are a few dex ($\sim 0.1\mbox{--}0.2$) higher with respect to the \gls{ms} in \citet{2022A&A...661L...7D} and \citet{Q1-SP031}, and slightly more in accord with the one in \citet{2023MNRAS.519.1526P}.

\subsection{Stellar mass completeness}
In order to estimate the mass completeness of the \gls{dr1}, we followed the method described in \citet{2010A&A...523A..13P}. We selected galaxies close to the limiting $\HE$ magnitude of each field, measured for each galaxy the mass they would have at the limiting magnitude as 
\begin{equation}
\log_{10} (M_{\rm lim}/M_\odot) = \Mstarwun + 0.4(\HE - H_{\sfont{E},  \mathrm{lim}})\:.
\end{equation}
The $95$th percentile of the distribution of $M_{\rm lim} (z)$ for each redshift bin is the mass completeness, reported in Fig.\,\ref{fig:completeness} as grey lines. The classification into star-forming (dashed blue line) and passive galaxies (dash-dotted red line) was performed with the selection criteria based on the $UVJ$ colours \citep{2005ApJ...624L..81L, 2007ApJ...655...51W, 2010ApJ...713..738W}, but updated to the \Euclid rest-frame colours (Euclid Consortium: Mao et al. in prep.). Given the different field depths and multiwavelength coverage, we report the mass completeness separately for each field. The trend is broadly similar across all the \gls{EDF} -- with EDF-F having higher values by $\sim 0.25\,\mathrm{dex}$ -- with the $95\%$ completeness limit rising from $\log_{10}(M_\ast/M_\odot) \sim 8$ at $z \sim 0.1$ to $\sim 9.0\mbox{--}9.3$ at $z \sim 1$, and further increasing to $\log_{10}(M_\ast/M_\odot) \sim 10\mbox{--}10.3$ at $z \sim 2.5\mbox{--}3$. The completeness limit for passive galaxies is $\sim 0.4\mbox{--}0.6\,\mathrm{dex}$ higher than for star-forming ones at all redshifts, a consequence of their redder colours and higher mass-to-light ratios at fixed magnitude. A similar trend is observed for \gls{EWS}, although with higher values. We also checked if there were any differences between the EWS-S and EWS-N and found none. Finally, as a rule-of-thumb, rescaling the whole mass completeness to a less conservative $5\,\sigma$ limit -- which might be more appropriate depending on the scientific case -- means shifting the curves towards lower masses by $(0.30 \pm 0.02) \,\mathrm{dex}$, if considering a $\pm 0.05$ variation in the limiting magnitudes.

\section{\label{sc:Conclusions}Conclusions}

We have described the PHZ pipeline for the \Euclid DR1, in particular the non-cosmological science branch responsible for recovering photo-$z$s and \gls{pp} of galaxies. The pipeline is built around the nearest-neighbour algorithm \nnpz, which, for DR1, benefits from a fundamentally improved reference sample with respect to Q1: rather than relying on a grid of synthetic BC03 templates, the reference sample is now anchored to real \Euclid observations of galaxies in the COSMOS field, matched to the COSMOS2020 catalogue and labelled through a dedicated \phosphoros\ run. This approach ensures internal consistency between the core-science and non-cosmological science branches of the PHZ pipeline, since the photometric redshift probability distribution functions from the core-science run are directly propagated into the construction of the PP labels.

We constructed two versions of the reference sample -- one for the \gls{EWS} (approximately $\num{375000}$ galaxies, cut at $\HE < 24$) and one for the \gls{EDF} ($\num{630000}$ galaxies, with no magnitude cut) -- and introduced a set of additional measures to mitigate the risk of overfitting the sample variance of COSMOS and the occurrence of unphysical outliers in the recovered stellar masses, namely a source-weighting scheme and a lognormal prior on the SED amplitude scaling factor. The data products released through the pipeline cover the full extent of the DR1 footprint -- the \gls{EWS}, the three \gls{EDF}, and the \gls{EAF} -- and include photo-$z$s, stellar masses, SFRs, and other properties reported in Table\,\ref{tab:DMDR1_PPs}, alongside their associated PDFs (Table\,\ref{tab:PHZout}).

The validation results confirm the quality of the DR1 data products across all fields and processing depths. The reference sample itself shows excellent agreement with available spectroscopic redshifts, with $\sigma_\mathrm{NMAD} = 0.007$ and an outlier fraction of $2.3\%$, comparable to or slightly better than the state-of-the-art results from the COSMOS2020 catalogue. For the \gls{EAF} -- in particular the PV-005 observations of COSMOS and CDFS -- both observed more times than the EWS and reaching depths of $\IE \sim 26.2$ and $26.4$, respectively (Table\,\ref{tab:filters}) -- $\sigma_{\rm NMAD}$ of $0.012$ and $0.030$ are obtained respectively, with small outlier fractions and negligible biases, on spectroscopic samples of $\num{28118}$ and $1536$ galaxies reaching down to $\IE \sim 25 \mbox{--} 27$. For the \gls{EWS}, inferred from the EWS-like processing of the EDF-N and EDF-F (reaching $I_{\rm E} \sim 26.2$ and $24.9$, respectively, but processed at EWS depth of $I_{\rm E} \sim 24.8$), $\sigma_{\rm NMAD}$ of $0.021$ (North) and $0.027$ (South) are obtained, with outlier fractions of $3.7\%$ and $7.6\%$ -- a significant improvement with respect to the Q1 results (where $\sigma_{\rm NMAD}$ of around $0.03$ were reported), driven by the use of a real-galaxy reference sample. For the \gls{EDF}, despite being deeper than the EWS (at least for the EDF-N with $I_{\rm E} \sim 26.2$, while the EDF-F is more similar at $24.9$, compared to $\sim 24.8$ for the EWS), the spectroscopic comparison samples are correspondingly fainter, probing a more diverse and challenging galaxy population; the results are nevertheless consistent with the EWS findings, with $\sigma_{\rm NMAD}$ of $0.022$ and $0.025$ for EDF-N and EDF-F respectively. The comparison of SFRs with DESI \ha\  and \hb\ measurements yields a $\sigma_{\rm NMAD}$ of $\sim 0.5\,\mathrm{dex}$ and outlier fractions of $\sim 16\%$, results that are in line with expectations from numerical simulations and reflect the inherent difficulty of inferring SFRs from broad-band photometry alone. Finally, the distribution of \gls{sfg} in the $M_\ast$--SFR plane across the EDFs is shown to be in good agreement with well-known \gls{ms} relations, further confirming the overall reliability of the recovered \gls{pp}.

Euclid DR1 constitutes a major step forward for extragalactic astrophysics, with nearly $880$ million detected sources across more than $1900\,\mathrm{deg}^2$ of the extragalactic sky, and reliable photo-$z$s and \gls{pp} derived from a self-consistent, observation-based pipeline. It represents the largest homogeneous catalogue of galaxy physical properties ever assembled. Future data releases, benefiting from an ever-growing survey footprint, deeper observations of the EDFs, and continued improvements to the reference sample and the \nnpz\ pipeline, will further extend and refine these results, solidifying \Euclid's role as a cornerstone resource for the study of galaxy formation and evolution.

\begin{acknowledgements}
AnEn, MiBo, LuPo, SoFo, ViAl, MaMe, SaQu, DaVe, MaTa, acknowledge support from the ELSA project. "ELSA: Euclid Legacy Science Advanced analysis tools" (Grant Agreement no. 101135203) is funded by the European Union. Views and opinions expressed are however those of the author(s) only and do not necessarily reflect those of the European Union or Innovate UK. Neither the European Union nor the granting authority can be held responsible for them. UK participation is funded through the UK HORIZON guarantee scheme under Innovate UK grant 10093177.
MiBo, AnEn acknowledge the support from INAF Minigrant 2023 `ADIEU'.
\AckDatalabs
\AckEC \AckDRone
\end{acknowledgements}

\bibliography{Euclid, biblio, DR1}

@ARTICLE{DR1-TP001,
       author = {{Euclid Collaboration: Aussel}, H. and others},
        title = "{Euclid Data Release 1 (DR1). Data release overview}",
      journal = {arXiv e-prints},
         year = 2026,
        month = nov,
          eid = {arXiv:2611.16510},
        pages = {arXiv:2611.16510},
archivePrefix = {arXiv},
       eprint = {2611.16510},
 primaryClass = {astro-ph.CO},
       adsurl = {https://ui.adsabs.harvard.edu/abs/2026arXiv261116510E}
}

@ARTICLE{DR1-TP002,
       author = {{Euclid Collaboration: McCracken}, H.~J. and {Benson}, K. and {Cropper}, M. and others},
        title = "{Euclid Data Release 1 (DR1): VIS processing and data products}",
      journal = {A\&A, submitted},
         year = 2026,
        month = nov,
          eid = {arXiv:2611.15890},
        pages = {arXiv:2611.15890},
archivePrefix = {arXiv},
       eprint = {2611.15890},
 primaryClass = {astro-ph.CO},
       adsurl = {https://ui.adsabs.harvard.edu/abs/2026arXiv261115890E}
}

@ARTICLE{DR1-TP003,
       author = {{Euclid Collaboration: Polenta}, G. and {Frailis}, M. and {Alavi}, A. and others},
        title = "{Euclid Data Release 1 (DR1). Near-infrared processing and data products}",
      journal = {A\&A, submitted},
         year = 2026,
        month = nov,
          eid = {arXiv:2611.16119},
        pages = {arXiv:2611.16119},
archivePrefix = {arXiv},
       eprint = {2611.16119},
 primaryClass = {astro-ph.IM},
       adsurl = {https://ui.adsabs.harvard.edu/abs/2026arXiv261116119E}
}

@ARTICLE{DR1-TP006,
       author = {{Euclid Collaboration: Corcho-Caballero}, P. and {Balbinot}, E. and {de Jong}, J.~T.~A. and others},
       title = "{Euclid Data Release 1 (DR1). Colouring Euclid with external ground-based imaging data}",
      journal = {A\&A, submitted},
         year = 2026,
        month = nov,
          eid = {arXiv:2611.16533},
        pages = {arXiv:2611.16533},
archivePrefix = {arXiv},
       eprint = {2611.16533},
 primaryClass = {astro-ph.CO},
       adsurl = {https://ui.adsabs.harvard.edu/abs/2026arXiv261116533E}
}

@ARTICLE{DR1-TP009,
       author = {{Euclid Collaboration: K{\"u}mmel}, M. and {Romelli}, E. and {Gracia-Carpio}, J. and others},
       title = "{Euclid Data Release 1 (DR1): The MERge processing}",
      journal = {A\&A, submitted},
         year = 2026,
        month = nov,
          eid = {arXiv:2611.17832},
        pages = {arXiv:2611.17832},
archivePrefix = {arXiv},
       eprint = {2611.17832},
 primaryClass = {astro-ph.IM},
       adsurl = {https://ui.adsabs.harvard.edu/abs/2026arXiv261117832E}
}

@ARTICLE{DR1-TP010,
       author = {{Euclid Collaboration: Gasparetto}, T. and {Merlin}, E. and {Castellano}, M. and others},
       title = "{Euclid Data Release 1 (DR1). Multi-band photometry validation on COSMOS extragalactic sources}",
      journal = {A\&A, submitted},
         year = 2026,
        month = nov,
          eid = {arXiv:2611.16825},
        pages = {arXiv:2611.16825},
archivePrefix = {arXiv},
       eprint = {2611.16825},
 primaryClass = {astro-ph.CO},
       adsurl = {https://ui.adsabs.harvard.edu/abs/2026arXiv261116825E}
}

@ARTICLE{DR1-TP011,
       author = {{Euclid Collaboration: Hartley}, W.~G. and {Paltani}, S. and {Dubath}, F. and others},
        title = "{Euclid Data Release 1 (DR1): Photometric-redshift pipeline overview and validation}",
      journal = {A\&A, submitted},
         year = 2026,
        month = nov,
          eid = {arXiv:2611.15987},
        pages = {arXiv:2611.15987},
archivePrefix = {arXiv},
       eprint = {2611.15987},
 primaryClass = {astro-ph.CO},
       adsurl = {https://ui.adsabs.harvard.edu/abs/2026arXiv261115987E}
}

@ARTICLE{DR1-TP023,
        author ={{Euclid Collaboration: Scodeggio}, M. and {Copin}, Y. and others},
        title = "{Euclid Data Release 1 (DR1). From spectrograms to spectra: The SIR spectroscopic processing function}",
        journal = {in preparation},
        year = 2026
        }

@ARTICLE{DR1-TP024,
       author = {{Euclid Collaboration: Fumana}, M. and {Copin}, Y. and {Appleton}, P.~N. and others},
       title = "{Euclid Data Release 1 (DR1). SIR spectroscopic calibration}",
      journal = {A\&A, submitted},
         year = 2026,
        month = nov,
          eid = {arXiv:2611.17305},
        pages = {arXiv:2611.17305},
archivePrefix = {arXiv},
       eprint = {2611.17305},
 primaryClass = {astro-ph.IM},
       adsurl = {https://ui.adsabs.harvard.edu/abs/2026arXiv261117305E}
}

@ARTICLE{DR1-TP027,
       author = {{Euclid Collaboration: Tereno}, I. and {Dinis}, J. and {Dupac}, X. and others},
       title = "{Euclid Data Release 1 (DR1). The Euclid surveys at DR1}",
      journal = {arXiv preprint},
         year = 2026,
        month = nov,
          eid = {arXiv:2611.17040},
        pages = {arXiv:2611.17040},
archivePrefix = {arXiv},
       eprint = {2611.17040},
 primaryClass = {astro-ph.CO},
       adsurl = {https://ui.adsabs.harvard.edu/abs/2026arXiv261117040E}
}

@ARTICLE{DR1-TP028,
       author = {{Euclid Collaboration: Odier}, J. and {Macias-Perez}, J.~F. and {Alguero}, G. and others},
       title = "{Euclid Data Release 1 (DR1). Euclid photometric masks}",
      journal = {arXiv e-prints},
         year = 2026,
        month = nov,
          eid = {arXiv:2611.16602},
        pages = {arXiv:2611.16602},
archivePrefix = {arXiv},
       eprint = {2611.16602},
 primaryClass = {astro-ph.CO},
       adsurl = {https://ui.adsabs.harvard.edu/abs/2026arXiv261116602E}
}

@ARTICLE{Q1-SP071,
        author ={{Euclid Collaboration: Hill}, R. and {Abghari}, A. and {Scott}, D. and others},
        title = "{Euclid Quick Data Release (Q1). XXIII. The average far-infrared properties of Euclid-selected star-forming galaxies}",
      journal = {\aap},
         year = 2026,
        month = jun,
       volume = {711},
          eid = {A23},
        pages = {A23},
          doi = {10.1051/0004-6361/202557514},
       adsurl = {https://ui.adsabs.harvard.edu/abs/2026A&A...711A..23E}
}

@ARTICLE{Q1-TP005,
       author = {{Euclid Collaboration: Tucci}, M. and {Paltani}, S. and {Hartley}, W.~G. and others},
       title = "{Euclid Quick Data Release (Q1). V. Photometric redshifts and physical properties of galaxies through the PHZ processing function}",
      journal = {\aap},
         year = 2026,
        month = jun,
       volume = {711},
          eid = {A5},
        pages = {A5},
          doi = {10.1051/0004-6361/202554588},
       adsurl = {https://ui.adsabs.harvard.edu/abs/2026A&A...711A...5E}
}

@ARTICLE{Q1-SP031,
       author = {{Euclid Collaboration: Enia}, A. and {Pozzetti}, L. and {Bolzonella}, M. and others},
        title = "{Euclid Quick Data Release (Q1). XIV. A first view of the star-forming main sequence in the Euclid Deep Fields}",
      journal = {\aap},
         year = 2026,
        month = jun,
       volume = {711},
          eid = {A14},
        pages = {A14},
          doi = {10.1051/0004-6361/202554576},
       adsurl = {https://ui.adsabs.harvard.edu/abs/2026A&A...711A..14E}
}

@ARTICLE{EuclidSkyOverview,
author = {{Euclid Collaboration: Mellier}, Y. and {Abdurro'uf} and {Acevedo~Barroso}, J.A. and others},
	title = {Euclid - I. Overview of the Euclid mission},
	DOI= "10.1051/0004-6361/202450810",
	url= "https://doi.org/10.1051/0004-6361/202450810",
	journal = {A\&A},
	year = 2025,
	volume = 697,
	pages = "A1",
}

@ARTICLE{EuclidSkyVIS,
author = {{Euclid Collaboration: Cropper}, M. and {Al-Bahlawan}, A. and {Amiaux}, J. and others},
	title = {Euclid - II. The VIS instrument},
	DOI= "10.1051/0004-6361/202450996",
	url= "https://doi.org/10.1051/0004-6361/202450996",
	journal = {A\&A},
	year = 2025,
	volume = 697,
	pages = "A2",
}

@ARTICLE{EuclidSkyNISP,
author = {{Euclid Collaboration: Jahnke}, K. and {Gillard}, W. and {Schirmer}, M. and others},
	title = {Euclid - III. The NISP Instrument},
	DOI= "10.1051/0004-6361/202450786",
	url= "https://doi.org/10.1051/0004-6361/202450786",
	journal = {A\&A},
	year = 2025,
	volume = 697,
	pages = "A3",
}

@misc{DR1cite,
author = "{Euclid Data Release 1}",
howpublished = "\url{https://doi.org/10.57780/esa-d079c51}",
year = 2026
}

@ARTICLE{EP-McPartland,
       author = {{Euclid Collaboration: McPartland}, C.~J.~R. and {Zalesky}, L. and {Weaver}, J.~R. and others},
        title = "{Euclid preparation: LXIV. The Cosmic Dawn Survey (DAWN) of the Euclid Deep and Auxiliary Fields}",
      journal = {\aap},
         year = 2025,
        month = mar,
       volume = {695},
          eid = {A259},
        pages = {A259},
          doi = {10.1051/0004-6361/202451849},
archivePrefix = {arXiv},
       eprint = {2408.05275},
 primaryClass = {astro-ph.GA},
       adsurl = {https://ui.adsabs.harvard.edu/abs/2025A&A...695A.259E}
}

@ARTICLE{EP-Enia,
       author = {{Euclid Collaboration: Enia}, A. and {Bolzonella}, M. and {Pozzetti}, L. and others},
        title = "{Euclid preparation: LI. Forecasting the recovery of galaxy physical properties and their relations with template-fitting and machine-learning methods}",
      journal = {\aap},
         year = 2024,
        month = nov,
       volume = {691},
          eid = {A175},
        pages = {A175},
          doi = {10.1051/0004-6361/202451425},
archivePrefix = {arXiv},
       eprint = {2407.07940},
 primaryClass = {astro-ph.GA},
       adsurl = {https://ui.adsabs.harvard.edu/abs/2024A&A...691A.175E}
}

@ARTICLE{Desprez-EP10,
       author = {{Euclid Collaboration: Desprez}, G. and {Paltani}, S. and {Coupon}, J. and others},
        title = "{Euclid preparation. X. The Euclid photometric-redshift challenge}",
      journal = {\aap},
         year = 2020,
        month = dec,
       volume = {644},
          eid = {A31},
        pages = {A31},
          doi = {10.1051/0004-6361/202039403},
archivePrefix = {arXiv},
       eprint = {2009.12112},
 primaryClass = {astro-ph.GA},
       adsurl = {https://ui.adsabs.harvard.edu/abs/2020A&A...644A..31E}
}

@ARTICLE{Scaramella-EP1,
       author = {{Euclid Collaboration: Scaramella}, R. and {Amiaux}, J. and {Mellier}, Y. and others},
        title = "{Euclid preparation. I. The Euclid Wide Survey}",
      journal = {\aap},
         year = 2022,
        month = jun,
       volume = {662},
          eid = {A112},
        pages = {A112},
          doi = {10.1051/0004-6361/202141938},
archivePrefix = {arXiv},
       eprint = {2108.01201},
 primaryClass = {astro-ph.CO},
       adsurl = {https://ui.adsabs.harvard.edu/abs/2022A&A...662A.112E}
}

@ARTICLE{2016arXiv161100036D,
       author = {{DESI Collaboration: Aghamousa}, Amir and {Aguilar}, Jessica and {Ahlen}, Steve and {Alam}, Shadab and {Allen}, Lori E. and {Allende Prieto}, Carlos and {Annis}, James and {Bailey}, Stephen and {Balland}, Christophe and {Ballester}, Otger and {Baltay}, Charles and {Beaufore}, Lucas and {Bebek}, Chris and {Beers}, Timothy C. and {Bell}, Eric F. and {Bernal}, Jos{\'e} Luis and {Besuner}, Robert and {Beutler}, Florian and {Blake}, Chris and {Bleuler}, Hannes and {Blomqvist}, Michael and {Blum}, Robert and {Bolton}, Adam S. and {Briceno}, Cesar and {Brooks}, David and {Brownstein}, Joel R. and {Buckley-Geer}, Elizabeth and {Burden}, Angela and {Burtin}, Etienne and {Busca}, Nicolas G. and {Cahn}, Robert N. and {Cai}, Yan-Chuan and {Cardiel-Sas}, Laia and {Carlberg}, Raymond G. and {Carton}, Pierre-Henri and {Casas}, Ricard and {Castander}, Francisco J. and {Cervantes-Cota}, Jorge L. and {Claybaugh}, Todd M. and {Close}, Madeline and {Coker}, Carl T. and {Cole}, Shaun and {Comparat}, Johan and {Cooper}, Andrew P. and {Cousinou}, M. -C. and {Crocce}, Martin and {Cuby}, Jean-Gabriel and {Cunningham}, Daniel P. and {Davis}, Tamara M. and {Dawson}, Kyle S. and {de la Macorra}, Axel and {De Vicente}, Juan and {Delubac}, Timoth{\'e}e and {Derwent}, Mark and {Dey}, Arjun and {Dhungana}, Govinda and {Ding}, Zhejie and {Doel}, Peter and {Duan}, Yutong T. and {Ealet}, Anne and {Edelstein}, Jerry and {Eftekharzadeh}, Sarah and {Eisenstein}, Daniel J. and {Elliott}, Ann and {Escoffier}, St{\'e}phanie and {Evatt}, Matthew and {Fagrelius}, Parker and {Fan}, Xiaohui and {Fanning}, Kevin and {Farahi}, Arya and {Farihi}, Jay and {Favole}, Ginevra and {Feng}, Yu and {Fernandez}, Enrique and {Findlay}, Joseph R. and {Finkbeiner}, Douglas P. and {Fitzpatrick}, Michael J. and {Flaugher}, Brenna and {Flender}, Samuel and {Font-Ribera}, Andreu and {Forero-Romero}, Jaime E. and {Fosalba}, Pablo and {Frenk}, Carlos S. and {Fumagalli}, Michele and {Gaensicke}, Boris T. and {Gallo}, Giuseppe and {Garcia-Bellido}, Juan and {Gaztanaga}, Enrique and {Pietro Gentile Fusillo}, Nicola and {Gerard}, Terry and {Gershkovich}, Irena and {Giannantonio}, Tommaso and {Gillet}, Denis and {Gonzalez-de-Rivera}, Guillermo and {Gonzalez-Perez}, Violeta and {Gott}, Shelby and {Graur}, Or and {Gutierrez}, Gaston and {Guy}, Julien and {Habib}, Salman and {Heetderks}, Henry and {Heetderks}, Ian and {Heitmann}, Katrin and {Hellwing}, Wojciech A. and {Herrera}, David A. and {Ho}, Shirley and {Holland}, Stephen and {Honscheid}, Klaus and {Huff}, Eric and {Hutchinson}, Timothy A. and {Huterer}, Dragan and {Hwang}, Ho Seong and {Illa Laguna}, Joseph Maria and {Ishikawa}, Yuzo and {Jacobs}, Dianna and {Jeffrey}, Niall and {Jelinsky}, Patrick and {Jennings}, Elise and {Jiang}, Linhua and {Jimenez}, Jorge and {Johnson}, Jennifer and {Joyce}, Richard and {Jullo}, Eric and {Juneau}, St{\'e}phanie and {Kama}, Sami and {Karcher}, Armin and {Karkar}, Sonia and {Kehoe}, Robert and {Kennamer}, Noble and {Kent}, Stephen and {Kilbinger}, Martin and {Kim}, Alex G. and {Kirkby}, David and {Kisner}, Theodore and {Kitanidis}, Ellie and {Kneib}, Jean-Paul and {Koposov}, Sergey and {Kovacs}, Eve and {Koyama}, Kazuya and {Kremin}, Anthony and {Kron}, Richard and {Kronig}, Luzius and {Kueter-Young}, Andrea and {Lacey}, Cedric G. and {Lafever}, Robin and {Lahav}, Ofer and {Lambert}, Andrew and {Lampton}, Michael and {Landriau}, Martin and {Lang}, Dustin and {Lauer}, Tod R. and {Le Goff}, Jean-Marc and {Le Guillou}, Laurent and {Le Van Suu}, Auguste and {Lee}, Jae Hyeon and {Lee}, Su-Jeong and {Leitner}, Daniela and {Lesser}, Michael and {Levi}, Michael E. and {L'Huillier}, Benjamin and {Li}, Baojiu and {Liang}, Ming and {Lin}, Huan and {Linder}, Eric and {Loebman}, Sarah R. and {Luki{\'c}}, Zarija and {Ma}, Jun and {MacCrann}, Niall and {Magneville}, Christophe and {Makarem}, Laleh and {Manera}, Marc and {Manser}, Christopher J. and {Marshall}, Robert and {Martini}, Paul and {Massey}, Richard and {Matheson}, Thomas and {McCauley}, Jeremy and {McDonald}, Patrick and {McGreer}, Ian D. and {Meisner}, Aaron and {Metcalfe}, Nigel and {Miller}, Timothy N. and {Miquel}, Ramon and {Moustakas}, John and {Myers}, Adam and {Naik}, Milind and {Newman}, Jeffrey A. and {Nichol}, Robert C. and {Nicola}, Andrina and {Nicolati da Costa}, Luiz and {Nie}, Jundan and {Niz}, Gustavo and {Norberg}, Peder and {Nord}, Brian and {Norman}, Dara and {Nugent}, Peter and {O'Brien}, Thomas and {Oh}, Minji and {Olsen}, Knut A.~G. and {Padilla}, Cristobal and {Padmanabhan}, Hamsa and {Padmanabhan}, Nikhil and {Palanque-Delabrouille}, Nathalie and {Palmese}, Antonella and {Pappalardo}, Daniel and {P{\^a}ris}, Isabelle and {Park}, Changbom and {Patej}, Anna and {Peacock}, John A. and {Peiris}, Hiranya V. and {Peng}, Xiyan and {Percival}, Will J. and {Perruchot}, Sandrine and {Pieri}, Matthew M. and {Pogge}, Richard and {Pollack}, Jennifer E. and {Poppett}, Claire and {Prada}, Francisco and {Prakash}, Abhishek and {Probst}, Ronald G. and {Rabinowitz}, David and {Raichoor}, Anand and {Ree}, Chang Hee and {Refregier}, Alexandre and {Regal}, Xavier and {Reid}, Beth and {Reil}, Kevin and {Rezaie}, Mehdi and {Rockosi}, Constance M. and {Roe}, Natalie and {Ronayette}, Samuel and {Roodman}, Aaron and {Ross}, Ashley J. and {Ross}, Nicholas P. and {Rossi}, Graziano and {Rozo}, Eduardo and {Ruhlmann-Kleider}, Vanina and {Rykoff}, Eli S. and {Sabiu}, Cristiano and {Samushia}, Lado and {Sanchez}, Eusebio and {Sanchez}, Javier and {Schlegel}, David J. and {Schneider}, Michael and {Schubnell}, Michael and {Secroun}, Aur{\'e}lia and {Seljak}, Uros and {Seo}, Hee-Jong and {Serrano}, Santiago and {Shafieloo}, Arman and {Shan}, Huanyuan and {Sharples}, Ray and {Sholl}, Michael J. and {Shourt}, William V. and {Silber}, Joseph H. and {Silva}, David R. and {Sirk}, Martin M. and {Slosar}, Anze and {Smith}, Alex and {Smoot}, George F. and {Som}, Debopam and {Song}, Yong-Seon and {Sprayberry}, David and {Staten}, Ryan and {Stefanik}, Andy and {Tarle}, Gregory and {Sien Tie}, Suk and {Tinker}, Jeremy L. and {Tojeiro}, Rita and {Valdes}, Francisco and {Valenzuela}, Octavio and {Valluri}, Monica and {Vargas-Magana}, Mariana and {Verde}, Licia and {Walker}, Alistair R. and {Wang}, Jiali and {Wang}, Yuting and {Weaver}, Benjamin A. and {Weaverdyck}, Curtis and {Wechsler}, Risa H. and {Weinberg}, David H. and {White}, Martin and {Yang}, Qian and {Yeche}, Christophe and {Zhang}, Tianmeng and {Zhao}, Gong-Bo and {Zheng}, Yi and {Zhou}, Xu and {Zhou}, Zhimin and {Zhu}, Yaling and {Zou}, Hu and {Zu}, Ying},
        title = "{The DESI Experiment Part I: Science,Targeting, and Survey Design}",
         year = 2016,
        month = oct,
          eid = {arXiv:1611.00036},
        pages = {arXiv:1611.00036},
          doi = {10.48550/arXiv.1611.00036},
archivePrefix = {arXiv},
       eprint = {1611.00036},
 primaryClass = {astro-ph.IM},
       adsurl = {https://ui.adsabs.harvard.edu/abs/2016arXiv161100036D}
}

@ARTICLE{2012ApJ...758L..17B,
       author = {{Brammer}, Gabriel B. and {S{\'a}nchez-Janssen}, Rub{\'e}n and {Labb{\'e}}, Ivo and {da Cunha}, Elisabete and {Erb}, Dawn K. and {Franx}, Marijn and {Fumagalli}, Mattia and {Lundgren}, Britt and {Marchesini}, Danilo and {Momcheva}, Ivelina and {Nelson}, Erica and {Patel}, Shannon and {Quadri}, Ryan and {Rix}, Hans-Walter and {Skelton}, Rosalind E. and {Schmidt}, Kasper B. and {van der Wel}, Arjen and {van Dokkum}, Pieter G. and {Wake}, David A. and {Whitaker}, Katherine E.},
        title = "{3D-HST Grism Spectroscopy of a Gravitationally Lensed, Low-metallicity Starburst Galaxy at z = 1.847}",
      journal = {\apjl},
         year = 2012,
        month = oct,
       volume = {758},
       number = {1},
          eid = {L17},
        pages = {L17},
          doi = {10.1088/2041-8205/758/1/L17},
archivePrefix = {arXiv},
       eprint = {1207.3795},
 primaryClass = {astro-ph.CO},
       adsurl = {https://ui.adsabs.harvard.edu/abs/2012ApJ...758L..17B}
}

@ARTICLE{2023MNRAS.519.1526P,
       author = {{Popesso}, P. and {Concas}, A. and {Cresci}, G. and {Belli}, S. and {Rodighiero}, G. and {Inami}, H. and {Dickinson}, M. and {Ilbert}, O. and {Pannella}, M. and {Elbaz}, D.},
        title = "{The main sequence of star-forming galaxies across cosmic times}",
      journal = {\mnras},
         year = 2023,
        month = feb,
       volume = {519},
       number = {1},
        pages = {1526-1544},
          doi = {10.1093/mnras/stac3214},
archivePrefix = {arXiv},
       eprint = {2203.10487},
 primaryClass = {astro-ph.GA},
       adsurl = {https://ui.adsabs.harvard.edu/abs/2023MNRAS.519.1526P}
}

@ARTICLE{2022A&A...661L...7D,
       author = {{Daddi}, E. and {Delvecchio}, I. and {Dimauro}, P. and {Magnelli}, B. and {Gomez-Guijarro}, C. and {Coogan}, R. and {Elbaz}, D. and {Kalita}, B.~S. and {Le Bail}, A. and {Rich}, R.~M. and {Tan}, Q.},
        title = "{The bending of the star-forming main sequence traces the cold- to hot-accretion transition mass over 0 < z < 4}",
      journal = {\aap},
         year = 2022,
        month = may,
       volume = {661},
          eid = {L7},
        pages = {L7},
          doi = {10.1051/0004-6361/202243574},
archivePrefix = {arXiv},
       eprint = {2203.10880},
 primaryClass = {astro-ph.CO},
       adsurl = {https://ui.adsabs.harvard.edu/abs/2022A&A...661L...7D}
}

@ARTICLE{2021A&A...647A.123D,
       author = {{Delvecchio}, I. and {Daddi}, E. and {Sargent}, M.~T. and {Jarvis}, M.~J. and {Elbaz}, D. and {Jin}, S. and {Liu}, D. and {Whittam}, I.~H. and {Algera}, H. and {Carraro}, R. and {D'Eugenio}, C. and {Delhaize}, J. and {Kalita}, B.~S. and {Leslie}, S. and {Moln{\'a}r}, D. Cs. and {Novak}, M. and {Prandoni}, I. and {Smol{\v{c}}i{\'c}}, V. and {Ao}, Y. and {Aravena}, M. and {Bournaud}, F. and {Collier}, J.~D. and {Randriamampandry}, S.~M. and {Randriamanakoto}, Z. and {Rodighiero}, G. and {Schober}, J. and {White}, S.~V. and {Zamorani}, G.},
        title = "{The infrared-radio correlation of star-forming galaxies is strongly M$_{{\ensuremath{\star}}}$-dependent but nearly redshift-invariant since z {\ensuremath{\sim}} 4}",
      journal = {\aap},
         year = 2021,
        month = mar,
       volume = {647},
          eid = {A123},
        pages = {A123},
          doi = {10.1051/0004-6361/202039647},
archivePrefix = {arXiv},
       eprint = {2010.05510},
 primaryClass = {astro-ph.GA},
       adsurl = {https://ui.adsabs.harvard.edu/abs/2021A&A...647A.123D}
}

@ARTICLE{1998ApJ...498..541K,
       author = {{Kennicutt}, Robert C., Jr.},
        title = "{The Global Schmidt Law in Star-forming Galaxies}",
      journal = {\apj},
         year = "1998",
        month = "May",
       volume = {498},
       number = {2},
        pages = {541-552},
          doi = {10.1086/305588},
archivePrefix = {arXiv},
       eprint = {astro-ph/9712213},
 primaryClass = {astro-ph},
       adsurl = {https://ui.adsabs.harvard.edu/abs/1998ApJ...498..541K}
}

@ARTICLE{2010ApJ...713..738W,
       author = {{Williams}, Rik J. and {Quadri}, Ryan F. and {Franx}, Marijn and {van Dokkum}, Pieter and {Toft}, Sune and {Kriek}, Mariska and {Labb{\'e}}, Ivo},
        title = "{The Evolving Relations Between Size, Mass, Surface Density, and Star Formation in 3 {\texttimes} {}10$^{4}$ Galaxies Since z = 2}",
      journal = {\apj},
         year = 2010,
        month = apr,
       volume = {713},
       number = {2},
        pages = {738-750},
          doi = {10.1088/0004-637X/713/2/738},
archivePrefix = {arXiv},
       eprint = {0906.4786},
 primaryClass = {astro-ph.CO},
       adsurl = {https://ui.adsabs.harvard.edu/abs/2010ApJ...713..738W}
}

@ARTICLE{2024A&A...691A.308S,
       author = {{Siudek}, M. and {Pucha}, R. and {Mezcua}, M. and {Juneau}, S. and {Aguilar}, J. and {Ahlen}, S. and {Brooks}, D. and {Circosta}, C. and {Claybaugh}, T. and {Cole}, S. and {Dawson}, K. and {de la Macorra}, A. and {Dey}, A. and {Dey}, B. and {Doel}, P. and {Font-Ribera}, A. and {Forero-Romero}, J.~E. and {Gazta{\~n}aga}, E. and {Gontcho A Gontcho}, S. and {Gutierrez}, G. and {Honscheid}, K. and {Howlett}, C. and {Ishak}, M. and {Kehoe}, R. and {Kirkby}, D. and {Kisner}, T. and {Kremin}, A. and {Lambert}, A. and {Landriau}, M. and {Le Guillou}, L. and {Manera}, M. and {Martini}, P. and {Meisner}, A. and {Miquel}, R. and {Moustakas}, J. and {Newman}, J.~A. and {Niz}, G. and {Pan}, Z. and {Percival}, W.~J. and {Poppett}, C. and {Prada}, F. and {Rossi}, G. and {Saintonge}, A. and {Sanchez}, E. and {Schlegel}, D. and {Scholte}, D. and {Schubnell}, M. and {Seo}, H. and {Speranza}, F. and {Sprayberry}, D. and {Tarl{\'e}}, G. and {Weaver}, B.~A. and {Zou}, H.},
        title = "{Value-added catalog of physical properties for more than 1.3 million galaxies from the DESI survey}",
      journal = {\aap},
         year = 2024,
        month = nov,
       volume = {691},
          eid = {A308},
        pages = {A308},
          doi = {10.1051/0004-6361/202451761},
archivePrefix = {arXiv},
       eprint = {2409.19066},
 primaryClass = {astro-ph.GA},
       adsurl = {https://ui.adsabs.harvard.edu/abs/2024A&A...691A.308S}
}

@ARTICLE{2007ApJ...655...51W,
       author = {{Wuyts}, Stijn and {Labb{\'e}}, Ivo and {Franx}, Marijn and {Rudnick}, Gregory and {van Dokkum}, Pieter G. and {Fazio}, Giovanni G. and {F{\"o}rster Schreiber}, Natascha M. and {Huang}, Jiasheng and {Moorwood}, Alan F.~M. and {Rix}, Hans-Walter and {R{\"o}ttgering}, Huub and {van der Werf}, Paul},
        title = "{What Do We Learn from IRAC Observations of Galaxies at 2 < z < 3.5?}",
      journal = {\apj},
         year = 2007,
        month = jan,
       volume = {655},
       number = {1},
        pages = {51-65},
          doi = {10.1086/509708},
archivePrefix = {arXiv},
       eprint = {astro-ph/0609548},
 primaryClass = {astro-ph},
       adsurl = {https://ui.adsabs.harvard.edu/abs/2007ApJ...655...51W}
}

@ARTICLE{2005ApJ...624L..81L,
       author = {{Labb{\'e}}, Ivo and {Huang}, Jiasheng and {Franx}, Marijn and {Rudnick}, Gregory and {Barmby}, Pauline and {Daddi}, Emanuele and {van Dokkum}, Pieter G. and {Fazio}, Giovanni G. and {F{\"o}rster Schreiber}, Natascha M. and {Moorwood}, Alan F.~M. and {Rix}, Hans-Walter and {R{\"o}ttgering}, Huub and {Trujillo}, Ignacio and {van der Werf}, Paul},
        title = "{IRAC Mid-Infrared Imaging of the Hubble Deep Field-South: Star Formation Histories and Stellar Masses of Red Galaxies at z>2}",
      journal = {\apjl},
         year = 2005,
        month = may,
       volume = {624},
       number = {2},
        pages = {L81-L84},
          doi = {10.1086/430700},
archivePrefix = {arXiv},
       eprint = {astro-ph/0504219},
 primaryClass = {astro-ph},
       adsurl = {https://ui.adsabs.harvard.edu/abs/2005ApJ...624L..81L}
}

@ARTICLE{2003MNRAS.344.1000B,
       author = {{Bruzual}, G. and {Charlot}, S.},
        title = "{Stellar population synthesis at the resolution of 2003}",
      journal = {\mnras},
         year = 2003,
        month = oct,
       volume = {344},
       number = {4},
        pages = {1000-1028},
          doi = {10.1046/j.1365-8711.2003.06897.x},
archivePrefix = {arXiv},
       eprint = {astro-ph/0309134},
 primaryClass = {astro-ph},
       adsurl = {https://ui.adsabs.harvard.edu/abs/2003MNRAS.344.1000B}
}

@ARTICLE{2018PASJ...70S...9T,
       author = {{Tanaka}, Masayuki and {Coupon}, Jean and {Hsieh}, Bau-Ching and {Mineo}, Sogo and {Nishizawa}, Atsushi J. and {Speagle}, Joshua and {Furusawa}, Hisanori and {Miyazaki}, Satoshi and {Murayama}, Hitoshi},
        title = "{Photometric redshifts for Hyper Suprime-Cam Subaru Strategic Program Data Release 1}",
      journal = {\pasj},
         year = 2018,
        month = jan,
       volume = {70},
          eid = {S9},
        pages = {S9},
          doi = {10.1093/pasj/psx077},
archivePrefix = {arXiv},
       eprint = {1704.05988},
 primaryClass = {astro-ph.GA},
       adsurl = {https://ui.adsabs.harvard.edu/abs/2018PASJ...70S...9T}
}

@ARTICLE{2008SerAJ.176....1I,
       author = {{Ivezic}, Z. and {Axelrod}, T. and {Brandt}, W.~N. and {Burke}, D.~L. and {Claver}, C.~F. and {Connolly}, A. and {Cook}, K.~H. and {Gee}, P. and {Gilmore}, D.~K. and {Jacoby}, S.~H. and {Jones}, R.~L. and {Kahn}, S.~M. and {Kantor}, J.~P. and {Krabbendam}, V.~V. and {Lupton}, R.~H. and {Monet}, D.~G. and {Pinto}, P.~A. and {Saha}, A. and {Schalk}, T.~L. and {Schneider}, D.~P. and {Strauss}, M.~A. and {Stubbs}, C.~W. and {Sweeney}, D. and {Szalay}, A. and {Thaler}, J.~J. and {Tyson}, J.~A. and {LSST Collaboration}},
        title = "{Large Synoptic Survey Telescope: From Science Drivers To Reference Design}",
      journal = {Serbian Astron. J.},
         year = 2008,
        month = jun,
       volume = {176},
        pages = {1-13},
          doi = {10.2298/SAJ0876001I},
       adsurl = {https://ui.adsabs.harvard.edu/abs/2008SerAJ.176....1I}
}

@ARTICLE{2010A&A...523A..13P,
       author = {{Pozzetti}, L. and {Bolzonella}, M. and {Zucca}, E. and {Zamorani}, G. and {Lilly}, S. and {Renzini}, A. and {Moresco}, M. and {Mignoli}, M. and {Cassata}, P. and {Tasca}, L. and {Lamareille}, F. and {Maier}, C. and {Meneux}, B. and {Halliday}, C. and {Oesch}, P. and {Vergani}, D. and {Caputi}, K. and {Kova{\v{c}}}, K. and {Cimatti}, A. and {Cucciati}, O. and {Iovino}, A. and {Peng}, Y. and {Carollo}, M. and {Contini}, T. and {Kneib}, J. -P. and {Le F{\'e}vre}, O. and {Mainieri}, V. and {Scodeggio}, M. and {Bardelli}, S. and {Bongiorno}, A. and {Coppa}, G. and {de la Torre}, S. and {de Ravel}, L. and {Franzetti}, P. and {Garilli}, B. and {Kampczyk}, P. and {Knobel}, C. and {Le Borgne}, J. -F. and {Le Brun}, V. and {Pell{\`o}}, R. and {Perez Montero}, E. and {Ricciardelli}, E. and {Silverman}, J.~D. and {Tanaka}, M. and {Tresse}, L. and {Abbas}, U. and {Bottini}, D. and {Cappi}, A. and {Guzzo}, L. and {Koekemoer}, A.~M. and {Leauthaud}, A. and {Maccagni}, D. and {Marinoni}, C. and {McCracken}, H.~J. and {Memeo}, P. and {Porciani}, C. and {Scaramella}, R. and {Scarlata}, C. and {Scoville}, N.},
        title = "{zCOSMOS - 10k-bright spectroscopic sample. The bimodality in the galaxy stellar mass function: exploring its evolution with redshift}",
      journal = {\aap},
         year = 2010,
        month = nov,
       volume = {523},
          eid = {A13},
        pages = {A13},
          doi = {10.1051/0004-6361/200913020},
archivePrefix = {arXiv},
       eprint = {0907.5416},
 primaryClass = {astro-ph.CO},
       adsurl = {https://ui.adsabs.harvard.edu/abs/2010A&A...523A..13P}
}

@ARTICLE{2022ApJS..258...11W,
       author = {{Weaver}, J.~R. and {Kauffmann}, O.~B. and {Ilbert}, O. and {McCracken}, H.~J. and {Moneti}, A. and {Toft}, S. and {Brammer}, G. and {Shuntov}, M. and {Davidzon}, I. and {Hsieh}, B.~C. and {Laigle}, C. and {Anastasiou}, A. and {Jespersen}, C.~K. and {Vinther}, J. and {Capak}, P. and {Casey}, C.~M. and {McPartland}, C.~J.~R. and {Milvang-Jensen}, B. and {Mobasher}, B. and {Sanders}, D.~B. and {Zalesky}, L. and {Arnouts}, S. and {Aussel}, H. and {Dunlop}, J.~S. and {Faisst}, A. and {Franx}, M. and {Furtak}, L.~J. and {Fynbo}, J.~P.~U. and {Gould}, K.~M.~L. and {Greve}, T.~R. and {Gwyn}, S. and {Kartaltepe}, J.~S. and {Kashino}, D. and {Koekemoer}, A.~M. and {Kokorev}, V. and {Le F{\`e}vre}, O. and {Lilly}, S. and {Masters}, D. and {Magdis}, G. and {Mehta}, V. and {Peng}, Y. and {Riechers}, D.~A. and {Salvato}, M. and {Sawicki}, M. and {Scarlata}, C. and {Scoville}, N. and {Shirley}, R. and {Silverman}, J.~D. and {Sneppen}, A. and {Smolc̆i{\'c}}, V. and {Steinhardt}, C. and {Stern}, D. and {Tanaka}, M. and {Taniguchi}, Y. and {Teplitz}, H.~I. and {Vaccari}, M. and {Wang}, W. -H. and {Zamorani}, G.},
        title = "{COSMOS2020: A Panchromatic View of the Universe to z{\ensuremath{\sim}}10 from Two Complementary Catalogs}",
      journal = {\apjs},
         year = 2022,
        month = jan,
       volume = {258},
       number = {1},
          eid = {11},
        pages = {11},
          doi = {10.3847/1538-4365/ac3078},
archivePrefix = {arXiv},
       eprint = {2110.13923},
 primaryClass = {astro-ph.GA},
       adsurl = {https://ui.adsabs.harvard.edu/abs/2022ApJS..258...11W}
}

@ARTICLE{2003PASP..115..763C,
       author = {{Chabrier}, Gilles},
        title = "{Galactic Stellar and Substellar Initial Mass Function}",
      journal = {\pasp},
         year = 2003,
        month = jul,
       volume = {115},
       number = {809},
        pages = {763-795},
          doi = {10.1086/376392},
archivePrefix = {arXiv},
       eprint = {astro-ph/0304382},
 primaryClass = {astro-ph},
       adsurl = {https://ui.adsabs.harvard.edu/abs/2003PASP..115..763C}
}

@ARTICLE{2015AJ....150..150F,
       author = {{Flaugher}, B. and {Diehl}, H.~T. and {Honscheid}, K. and {Abbott}, T.~M.~C. and {Alvarez}, O. and {Angstadt}, R. and {Annis}, J.~T. and {Antonik}, M. and {Ballester}, O. and {Beaufore}, L. and {Bernstein}, G.~M. and {Bernstein}, R.~A. and {Bigelow}, B. and {Bonati}, M. and {Boprie}, D. and {Brooks}, D. and {Buckley-Geer}, E.~J. and {Campa}, J. and {Cardiel-Sas}, L. and {Castander}, F.~J. and {Castilla}, J. and {Cease}, H. and {Cela-Ruiz}, J.~M. and {Chappa}, S. and {Chi}, E. and {Cooper}, C. and {da Costa}, L.~N. and {Dede}, E. and {Derylo}, G. and {DePoy}, D.~L. and {de Vicente}, J. and {Doel}, P. and {Drlica-Wagner}, A. and {Eiting}, J. and {Elliott}, A.~E. and {Emes}, J. and {Estrada}, J. and {Fausti Neto}, A. and {Finley}, D.~A. and {Flores}, R. and {Frieman}, J. and {Gerdes}, D. and {Gladders}, M.~D. and {Gregory}, B. and {Gutierrez}, G.~R. and {Hao}, J. and {Holland}, S.~E. and {Holm}, S. and {Huffman}, D. and {Jackson}, C. and {James}, D.~J. and {Jonas}, M. and {Karcher}, A. and {Karliner}, I. and {Kent}, S. and {Kessler}, R. and {Kozlovsky}, M. and {Kron}, R.~G. and {Kubik}, D. and {Kuehn}, K. and {Kuhlmann}, S. and {Kuk}, K. and {Lahav}, O. and {Lathrop}, A. and {Lee}, J. and {Levi}, M.~E. and {Lewis}, P. and {Li}, T.~S. and {Mandrichenko}, I. and {Marshall}, J.~L. and {Martinez}, G. and {Merritt}, K.~W. and {Miquel}, R. and {Mu{\~n}oz}, F. and {Neilsen}, E.~H. and {Nichol}, R.~C. and {Nord}, B. and {Ogando}, R. and {Olsen}, J. and {Palaio}, N. and {Patton}, K. and {Peoples}, J. and {Plazas}, A.~A. and {Rauch}, J. and {Reil}, K. and {Rheault}, J. -P. and {Roe}, N.~A. and {Rogers}, H. and {Roodman}, A. and {Sanchez}, E. and {Scarpine}, V. and {Schindler}, R.~H. and {Schmidt}, R. and {Schmitt}, R. and {Schubnell}, M. and {Schultz}, K. and {Schurter}, P. and {Scott}, L. and {Serrano}, S. and {Shaw}, T.~M. and {Smith}, R.~C. and {Soares-Santos}, M. and {Stefanik}, A. and {Stuermer}, W. and {Suchyta}, E. and {Sypniewski}, A. and {Tarle}, G. and {Thaler}, J. and {Tighe}, R. and {Tran}, C. and {Tucker}, D. and {Walker}, A.~R. and {Wang}, G. and {Watson}, M. and {Weaverdyck}, C. and {Wester}, W. and {Woods}, R. and {Yanny}, B. and {DES Collaboration}},
        title = "{The Dark Energy Camera}",
      journal = {\aj},
         year = 2015,
        month = nov,
       volume = {150},
       number = {5},
          eid = {150},
        pages = {150},
          doi = {10.1088/0004-6256/150/5/150},
archivePrefix = {arXiv},
       eprint = {1504.02900},
 primaryClass = {astro-ph.IM},
       adsurl = {https://ui.adsabs.harvard.edu/abs/2015AJ....150..150F}
}

@ARTICLE{2016MNRAS.460.1270D,
       author = {{DES Collaboration: Abbott}, T. and {Abdalla}, F.~B. and {Aleksi{\'c}}, J. and {Allam}, S. and {Amara}, A. and {Bacon}, D. and {Balbinot}, E. and {Banerji}, M. and {Bechtol}, K. and {Benoit-L{\'e}vy}, A. and {Bernstein}, G.~M. and {Bertin}, E. and {Blazek}, J. and {Bonnett}, C. and {Bridle}, S. and {Brooks}, D. and {Brunner}, R.~J. and {Buckley-Geer}, E. and {Burke}, D.~L. and {Caminha}, G.~B. and {Capozzi}, D. and {Carlsen}, J. and {Carnero-Rosell}, A. and {Carollo}, M. and {Carrasco-Kind}, M. and {Carretero}, J. and {Castander}, F.~J. and {Clerkin}, L. and {Collett}, T. and {Conselice}, C. and {Crocce}, M. and {Cunha}, C.~E. and {D'Andrea}, C.~B. and {da Costa}, L.~N. and {Davis}, T.~M. and {Desai}, S. and {Diehl}, H.~T. and {Dietrich}, J.~P. and {Dodelson}, S. and {Doel}, P. and {Drlica-Wagner}, A. and {Estrada}, J. and {Etherington}, J. and {Evrard}, A.~E. and {Fabbri}, J. and {Finley}, D.~A. and {Flaugher}, B. and {Foley}, R.~J. and {Fosalba}, P. and {Frieman}, J. and {Garc{\'\i}a-Bellido}, J. and {Gaztanaga}, E. and {Gerdes}, D.~W. and {Giannantonio}, T. and {Goldstein}, D.~A. and {Gruen}, D. and {Gruendl}, R.~A. and {Guarnieri}, P. and {Gutierrez}, G. and {Hartley}, W. and {Honscheid}, K. and {Jain}, B. and {James}, D.~J. and {Jeltema}, T. and {Jouvel}, S. and {Kessler}, R. and {King}, A. and {Kirk}, D. and {Kron}, R. and {Kuehn}, K. and {Kuropatkin}, N. and {Lahav}, O. and {Li}, T.~S. and {Lima}, M. and {Lin}, H. and {Maia}, M.~A.~G. and {Makler}, M. and {Manera}, M. and {Maraston}, C. and {Marshall}, J.~L. and {Martini}, P. and {McMahon}, R.~G. and {Melchior}, P. and {Merson}, A. and {Miller}, C.~J. and {Miquel}, R. and {Mohr}, J.~J. and {Morice-Atkinson}, X. and {Naidoo}, K. and {Neilsen}, E. and {Nichol}, R.~C. and {Nord}, B. and {Ogando}, R. and {Ostrovski}, F. and {Palmese}, A. and {Papadopoulos}, A. and {Peiris}, H.~V. and {Peoples}, J. and {Percival}, W.~J. and {Plazas}, A.~A. and {Reed}, S.~L. and {Refregier}, A. and {Romer}, A.~K. and {Roodman}, A. and {Ross}, A. and {Rozo}, E. and {Rykoff}, E.~S. and {Sadeh}, I. and {Sako}, M. and {S{\'a}nchez}, C. and {Sanchez}, E. and {Santiago}, B. and {Scarpine}, V. and {Schubnell}, M. and {Sevilla-Noarbe}, I. and {Sheldon}, E. and {Smith}, M. and {Smith}, R.~C. and {Soares-Santos}, M. and {Sobreira}, F. and {Soumagnac}, M. and {Suchyta}, E. and {Sullivan}, M. and {Swanson}, M. and {Tarle}, G. and {Thaler}, J. and {Thomas}, D. and {Thomas}, R.~C. and {Tucker}, D. and {Vieira}, J.~D. and {Vikram}, V. and {Walker}, A.~R. and {Wechsler}, R.~H. and {Weller}, J. and {Wester}, W. and {Whiteway}, L. and {Wilcox}, H. and {Yanny}, B. and {Zhang}, Y. and {Zuntz}, J.},
        title = "{The Dark Energy Survey: more than dark energy - an overview}",
      journal = {\mnras},
         year = 2016,
        month = aug,
       volume = {460},
       number = {2},
        pages = {1270-1299},
          doi = {10.1093/mnras/stw641},
archivePrefix = {arXiv},
       eprint = {1601.00329},
 primaryClass = {astro-ph.CO},
       adsurl = {https://ui.adsabs.harvard.edu/abs/2016MNRAS.460.1270D}
}

@ARTICLE{1983ApJ...266..713O,
       author = {{Oke}, J.~B. and {Gunn}, J.~E.},
        title = "{Secondary standard stars for absolute spectrophotometry.}",
      journal = {\apj},
         year = 1983,
        month = mar,
       volume = {266},
        pages = {713-717},
          doi = {10.1086/160817},
       adsurl = {https://ui.adsabs.harvard.edu/abs/1983ApJ...266..713O}
}

@ARTICLE{2018A&A...609A.111D,
       author = {{D'Isanto}, A. and {Polsterer}, K.~L.},
        title = "{Photometric redshift estimation via deep learning. Generalized and pre-classification-less, image based, fully probabilistic redshifts}",
      journal = {\aap},
         year = 2018,
        month = jan,
       volume = {609},
          eid = {A111},
        pages = {A111},
          doi = {10.1051/0004-6361/201731326},
archivePrefix = {arXiv},
       eprint = {1706.02467},
 primaryClass = {astro-ph.IM},
       adsurl = {https://ui.adsabs.harvard.edu/abs/2018A&A...609A.111D}
}

@ARTICLE{2016A&A...594A..13P,
       author = {{Planck Collaboration} and {Ade}, P.~A.~R. and {Aghanim}, N. and {Arnaud}, M. and {Ashdown}, M. and {Aumont}, J. and {Baccigalupi}, C. and {Banday}, A.~J. and {Barreiro}, R.~B. and {Bartlett}, J.~G. and {Bartolo}, N. and {Battaner}, E. and {Battye}, R. and {Benabed}, K. and {Beno{\^\i}t}, A. and {Benoit-L{\'e}vy}, A. and {Bernard}, J.-P. and {Bersanelli}, M. and {Bielewicz}, P. and {Bock}, J.~J. and {Bonaldi}, A. and {Bonavera}, L. and {Bond}, J.~R. and {Borrill}, J. and {Bouchet}, F.~R. and {Boulanger}, F. and {Bucher}, M. and {Burigana}, C. and {Butler}, R.~C. and {Calabrese}, E. and {Cardoso}, J.-F. and {Catalano}, A. and {Challinor}, A. and {Chamballu}, A. and {Chary}, R.-R. and {Chiang}, H.~C. and {Chluba}, J. and {Christensen}, P.~R. and {Church}, S. and {Clements}, D.~L. and {Colombi}, S. and {Colombo}, L.~P.~L. and {Combet}, C. and {Coulais}, A. and {Crill}, B.~P. and {Curto}, A. and {Cuttaia}, F. and {Danese}, L. and {Davies}, R.~D. and {Davis}, R.~J. and {de Bernardis}, P. and {de Rosa}, A. and {de Zotti}, G. and {Delabrouille}, J. and {D{\'e}sert}, F.-X. and {Di Valentino}, E. and {Dickinson}, C. and {Diego}, J.~M. and {Dolag}, K. and {Dole}, H. and {Donzelli}, S. and {Dor{\'e}}, O. and {Douspis}, M. and {Ducout}, A. and {Dunkley}, J. and {Dupac}, X. and {Efstathiou}, G. and {Elsner}, F. and {En{\ss}lin}, T.~A. and {Eriksen}, H.~K. and {Farhang}, M. and {Fergusson}, J. and {Finelli}, F. and {Forni}, O. and {Frailis}, M. and {Fraisse}, A.~A. and {Franceschi}, E. and {Frejsel}, A. and {Galeotta}, S. and {Galli}, S. and {Ganga}, K. and {Gauthier}, C. and {Gerbino}, M. and {Ghosh}, T. and {Giard}, M. and {Giraud-H{\'e}raud}, Y. and {Giusarma}, E. and {Gjerl{\o}w}, E. and {Gonz{\'a}lez-Nuevo}, J. and {G{\'o}rski}, K.~M. and {Gratton}, S. and {Gregorio}, A. and {Gruppuso}, A. and {Gudmundsson}, J.~E. and {Hamann}, J. and {Hansen}, F.~K. and {Hanson}, D. and {Harrison}, D.~L. and {Helou}, G. and {Henrot-Versill{\'e}}, S. and {Hern{\'a}ndez-Monteagudo}, C. and {Herranz}, D. and {Hildebrandt}, S.~R. and {Hivon}, E. and {Hobson}, M. and {Holmes}, W.~A. and {Hornstrup}, A. and {Hovest}, W. and {Huang}, Z. and {Huffenberger}, K.~M. and {Hurier}, G. and {Jaffe}, A.~H. and {Jaffe}, T.~R. and {Jones}, W.~C. and {Juvela}, M. and {Keih{\"a}nen}, E. and {Keskitalo}, R. and {Kisner}, T.~S. and {Kneissl}, R. and {Knoche}, J. and {Knox}, L. and {Kunz}, M. and {Kurki-Suonio}, H. and {Lagache}, G. and {L{\"a}hteenm{\"a}ki}, A. and {Lamarre}, J.-M. and {Lasenby}, A. and {Lattanzi}, M. and {Lawrence}, C.~R. and {Leahy}, J.~P. and {Leonardi}, R. and {Lesgourgues}, J. and {Levrier}, F. and {Lewis}, A. and {Liguori}, M. and {Lilje}, P.~B. and {Linden-V{\o}rnle}, M. and {L{\'o}pez-Caniego}, M. and {Lubin}, P.~M. and {Mac{\'\i}as-P{\'e}rez}, J.~F. and {Maggio}, G. and {Maino}, D. and {Mandolesi}, N. and {Mangilli}, A. and {Marchini}, A. and {Maris}, M. and {Martin}, P.~G. and {Martinelli}, M. and {Mart{\'\i}nez-Gonz{\'a}lez}, E. and {Masi}, S. and {Matarrese}, S. and {McGehee}, P. and {Meinhold}, P.~R. and {Melchiorri}, A. and {Melin}, J.-B. and {Mendes}, L. and {Mennella}, A. and {Migliaccio}, M. and {Millea}, M. and {Mitra}, S. and {Miville-Desch{\^e}nes}, M.-A. and {Moneti}, A. and {Montier}, L. and {Morgante}, G. and {Mortlock}, D. and {Moss}, A. and {Munshi}, D. and {Murphy}, J.~A. and {Naselsky}, P. and {Nati}, F. and {Natoli}, P. and {Netterfield}, C.~B. and {N{\o}rgaard-Nielsen}, H.~U. and {Noviello}, F. and {Novikov}, D. and {Novikov}, I. and {Oxborrow}, C.~A. and {Paci}, F. and {Pagano}, L. and {Pajot}, F. and {Paladini}, R. and {Paoletti}, D. and {Partridge}, B. and {Pasian}, F. and {Patanchon}, G. and {Pearson}, T.~J. and {Perdereau}, O. and {Perotto}, L. and {Perrotta}, F. and {Pettorino}, V. and {Piacentini}, F. and {Piat}, M. and {Pierpaoli}, E. and {Pietrobon}, D. and {Plaszczynski}, S. and {Pointecouteau}, E. and {Polenta}, G. and {Popa}, L. and {Pratt}, G.~W. and {Pr{\'e}zeau}, G.},
        title = "{Planck 2015 results. XIII. Cosmological parameters}",
      journal = {\aap},
         year = 2016,
        month = sep,
       volume = {594},
          eid = {A13},
        pages = {A13},
          doi = {10.1051/0004-6361/201525830},
archivePrefix = {arXiv},
       eprint = {1502.01589},
 primaryClass = {astro-ph.CO},
       adsurl = {https://ui.adsabs.harvard.edu/abs/2016A&A...594A..13P}
}

@article{10.2307/2981683,
    author = {Dawid, A. P.},
    title = {Present Position and Potential Developments: Some Personal Views Statistical Theory the Prequential Approach},
    journal = {Royal Statistical Society. Journal. Series A: General},
    volume = {147},
    number = {2},
    pages = {278-290},
    year = {1984},
    month = {12},
    issn = {0035-9238},
    doi = {10.2307/2981683},
    url = {https://doi.org/10.2307/2981683},
    eprint = {https://academic.oup.com/jrsssa/article-pdf/147/2/278/49758146/jrsssa_147_2_278.pdf},
}

@ARTICLE{2025AJ....170..324G,
       author = {{Gwyn}, Stephen and {McConnachie}, Alan W. and {Cuillandre}, Jean-Charles and {Chambers}, Kenneth C. and {Magnier}, Eugene A. and {de Boer}, Thomas and {Hudson}, Michael J. and {Oguri}, Masamune and {Furusawa}, Hisanori and {Hildebrandt}, Hendrik and {Carlberg}, Raymond and {Ellison}, Sara L. and {Furusawa}, Junko and {Gavazzi}, Rapha{\"e}l and {Ibata}, Rodrigo and {Mellier}, Yannick and {Osato}, Ken and {Aussel}, H. and {Baumont}, Lucie and {Bayer}, Manuel and {Boulade}, Olivier and {C{\^o}t{\'e}}, Patrick and {Chemaly}, David and {Daley}, Cail and {Duc}, Pierre-Alain and {Durret}, Florence and {Ellien}, A. and {Fabbro}, S{\'e}bastien and {Ferreira}, Leonardo and {Fitriana}, Itsna K. and {Le Floc'h}, Emeric and {Fudamoto}, Yoshinobu and {Gao}, Hua and {Goh}, L.~W.~K. and {Goto}, Tomotsugu and {Guerrini}, Sacha and {Guinot}, Axel and {H{\'e}nault-Brunet}, Vincent and {Hammer}, Francois and {Harikane}, Yuichi and {Hayashi}, Kohei and {Heesters}, Nick and {Ichikawa}, Kohei and {Kilbinger}, Martin and {Kuzma}, P.~B. and {Li}, Qinxun and {Liaudat}, Tob{\'\i}as I. and {Lin}, Chien-Cheng and {M{\"u}ller}, Oliver and {Martin}, Nicolas F. and {Matsuoka}, Yoshiki and {Medina}, Gustavo E. and {Miyatake}, Hironao and {Miyazaki}, Satoshi and {Mpetha}, Charlie T. and {Nagao}, Tohru and {Navarro}, Julio F. and {Niwano}, Masafumi and {Ogami}, Itsuki and {Okabe}, Nobuhiro and {Onoue}, Masafusa and {Paek}, Gregory S.~H. and {Parker}, Laura C. and {Patton}, David R. and {Peters}, Fabian Hervas and {Prunet}, Simon and {S{\'a}nchez-Janssen}, Rub{\'e}n and {Schultheis}, M. and {Sestito}, Federico and {Smith}, Simon E.~T. and {Starck}, J.-L. and {Starkenburg}, Else and {Stone}, Connor and {Storfer}, Christopher and {Suzuki}, Yoshihisa and {Erben}, T. and {Taibi}, Salvatore and {Thomas}, G.~F. and {Toba}, Yoshiki and {Uchiyama}, Hisakazu and {Valls-Gabaud}, David and {Venn}, Kim A. and {Van Waerbeke}, Ludovic and {Wainscoat}, Richard J. and {Wilkinson}, Scott and {Wittje}, Anna and {Yoshida}, Taketo and {Zhang}, TianFang and {Zhong}, Yuxing},
        title = "{UNIONS: The Ultraviolet Near-infrared Optical Northern Survey}",
      journal = {\aj},
         year = 2025,
        month = dec,
       volume = {170},
       number = {6},
          eid = {324},
        pages = {324},
          doi = {10.3847/1538-3881/ae03ab},
archivePrefix = {arXiv},
       eprint = {2503.13783},
 primaryClass = {astro-ph.GA},
       adsurl = {https://ui.adsabs.harvard.edu/abs/2025AJ....170..324G}
}

@ARTICLE{2016A&A...595A..97M,
       author = {{Merlin}, E. and {Bourne}, N. and {Castellano}, M. and {Ferguson}, H.~C. and {Wang}, T. and {Derriere}, S. and {Dunlop}, J.~S. and {Elbaz}, D. and {Fontana}, A.},
        title = "{T-PHOT version 2.0: Improved algorithms for background subtraction, local convolution, kernel registration, and new options}",
      journal = {\aap},
         year = 2016,
        month = nov,
       volume = {595},
          eid = {A97},
        pages = {A97},
          doi = {10.1051/0004-6361/201628751},
archivePrefix = {arXiv},
       eprint = {1609.00146},
 primaryClass = {astro-ph.IM},
       adsurl = {https://ui.adsabs.harvard.edu/abs/2016A&A...595A..97M}
}

@ARTICLE{2015A&A...582A..15M,
       author = {{Merlin}, E. and {Fontana}, A. and {Ferguson}, H.~C. and {Dunlop}, J.~S. and {Elbaz}, D. and {Bourne}, N. and {Bruce}, V.~A. and {Buitrago}, F. and {Castellano}, M. and {Schreiber}, C. and {Grazian}, A. and {McLure}, R.~J. and {Okumura}, K. and {Shu}, X. and {Wang}, T. and {Amor{\'\i}n}, R. and {Boutsia}, K. and {Cappelluti}, N. and {Comastri}, A. and {Derriere}, S. and {Faber}, S.~M. and {Santini}, P.},
        title = "{T-PHOT: A new code for PSF-matched, prior-based, multiwavelength extragalactic deconfusion photometry}",
      journal = {\aap},
         year = 2015,
        month = oct,
       volume = {582},
          eid = {A15},
        pages = {A15},
          doi = {10.1051/0004-6361/201526471},
archivePrefix = {arXiv},
       eprint = {1505.02516},
 primaryClass = {astro-ph.IM},
       adsurl = {https://ui.adsabs.harvard.edu/abs/2015A&A...582A..15M}
}

@ARTICLE{2026ApJS..282....6K,
       author = {{Khostovan}, Ali Ahmad and {Kartaltepe}, Jeyhan S. and {Salvato}, Mara and {Ilbert}, Olivier and {Casey}, Caitlin M. and {Algera}, Hiddo and {Antwi-Danso}, Jacqueline and {Battisti}, Andrew and {Brinch}, Malte and {Brusa}, Marcella and {Calabr{\`o}}, Antonello and {Capak}, Peter L. and {Chartab}, Nima and {Cooper}, Olivia R. and {Cox}, Isa G. and {Darvish}, Behnam and {Drakos}, Nicole E. and {Faisst}, Andreas L. and {George}, Matthew R. and {Gozaliasl}, Ghassem and {Harish}, Santosh and {Hasinger}, G{\"u}nther and {Hatamnia}, Hossein and {Iovino}, Angela and {Jin}, Shuowen and {Kashino}, Daichi and {Koekemoer}, Anton M. and {Laishram}, Ronaldo and {Lee}, Khee-Gan and {Lertprasertpong}, Jitrapon and {Lilly}, Simon J. and {Liu}, Daizhong and {Masters}, Daniel C. and {Mobasher}, Bahram and {Nagao}, Tohru and {Onodera}, Masato and {Peng}, Yingjie and {Sanders}, David B. and {Sanders}, Ryan L. and {Sattari}, Zahra and {Scoville}, Nick and {Shah}, Ekta A. and {Silverman}, John D. and {Suzuki}, Nao and {Taamoli}, Sina and {Tanaka}, Masayuki and {Tasca}, Lidia A.~M. and {Toft}, Sune and {Toni}, Greta and {Trakhtenbrot}, Benny and {Trump}, Jonathan R. and {Vaccari}, Mattia and {Valentino}, Francesco and {Vanderhoof}, Brittany N. and {Weaver}, John R. and {Yun}, Min S. and {Zavala}, Jorge A.},
        title = "{COSMOS Spectroscopic Redshift Compilation (First Data Release): 488,000 Redshifts Encompassing Two Decades of Spectroscopy}",
      journal = {\apjs},
         year = 2026,
        month = jan,
       volume = {282},
       number = {1},
          eid = {6},
        pages = {6},
          doi = {10.3847/1538-4365/ae1cb9},
archivePrefix = {arXiv},
       eprint = {2503.00120},
 primaryClass = {astro-ph.GA},
       adsurl = {https://ui.adsabs.harvard.edu/abs/2026ApJS..282....6K}
}

@ARTICLE{2013arXiv1305.2140V,
       author = {{van Dokkum}, Pieter and {Brammer}, Gabriel and {Momcheva}, Ivelina and {Skelton}, Rosalind E. and {Whitaker}, Katherine E.},
        title = "{3D-HST Data Release v3.0: Extremely Deep Spectra in the UDF and WFC3 Mosaics in the 3D-HST/CANDELS Fields}",
         year = 2013,
        month = may,
          eid = {arXiv:1305.2140},
        pages = {arXiv:1305.2140},
          doi = {10.48550/arXiv.1305.2140},
archivePrefix = {arXiv},
       eprint = {1305.2140},
 primaryClass = {astro-ph.CO},
       adsurl = {https://ui.adsabs.harvard.edu/abs/2013arXiv1305.2140V}
}

@ARTICLE{2016ApJS..225...27M,
       author = {{Momcheva}, Ivelina G. and {Brammer}, Gabriel B. and {van Dokkum}, Pieter G. and {Skelton}, Rosalind E. and {Whitaker}, Katherine E. and {Nelson}, Erica J. and {Fumagalli}, Mattia and {Maseda}, Michael V. and {Leja}, Joel and {Franx}, Marijn and {Rix}, Hans-Walter and {Bezanson}, Rachel and {Da Cunha}, Elisabete and {Dickey}, Claire and {F{\"o}rster Schreiber}, Natascha M. and {Illingworth}, Garth and {Kriek}, Mariska and {Labb{\'e}}, Ivo and {Ulf Lange}, Johannes and {Lundgren}, Britt F. and {Magee}, Daniel and {Marchesini}, Danilo and {Oesch}, Pascal and {Pacifici}, Camilla and {Patel}, Shannon G. and {Price}, Sedona and {Tal}, Tomer and {Wake}, David A. and {van der Wel}, Arjen and {Wuyts}, Stijn},
        title = "{The 3D-HST Survey: Hubble Space Telescope WFC3/G141 Grism Spectra, Redshifts, and Emission Line Measurements for \raisebox{-0.5ex}\textasciitilde 100,000 Galaxies}",
      journal = {\apjs},
         year = 2016,
        month = aug,
       volume = {225},
       number = {2},
          eid = {27},
        pages = {27},
          doi = {10.3847/0067-0049/225/2/27},
archivePrefix = {arXiv},
       eprint = {1510.02106},
 primaryClass = {astro-ph.GA},
       adsurl = {https://ui.adsabs.harvard.edu/abs/2016ApJS..225...27M}
}

\begin{appendix}
\end{appendix}

\end{document}